\documentclass[useAMS,usenatbib]{mn2e}

\usepackage{natbib}
\usepackage{graphicx}
\usepackage{multirow}
\usepackage{subfig}

\title[A mid-infrared spectroscopic atlas of local
AGN]{A mid-infrared spectroscopic atlas of local active galactic nuclei on sub-arcsecond resolution using GTC/CanariCam}
\author[Almudena Alonso-Herrero et al.]
{\parbox{\textwidth}{A.
  Alonso-Herrero,$^{1,2,3}$\thanks{E-mail: aalonso@ifca.unican.es}
 P. Esquej,$^{4}$
P. F. Roche,$^2$
  C. Ramos Almeida,$^{5,6}$\thanks{Marie Curie Fellow.}
O. Gonz\'alez-Mart\'{\i}n,$^{7}$
C. Packham,$^{3}$ N. A. Levenson,$^{8}$ R. E. Mason,$^{9}$
A. Hern\'an-Caballero,$^{1}$ 
 M. Pereira-Santaella,$^{10,11}$ C. Alvarez,$^{5,6}$ I. Aretxaga$^{12}$, 
E. L\'opez-Rodr\'{\i}guez,$^{3}$\thanks{Research
  Affiliate-Postdoctoral, Department of Astronomy, University of Texas
  at Austin}  L. Colina,$^{10,11}$ T. D\'{\i}az-Santos$^{13}$,
M. Imanishi,$^{14,15,16}$ J. M. Rodr\'{\i}guez
Espinosa,$^{5,6}$ E. Perlman$^{17}$}
\vspace{0.2cm}\\ 
$^{1}$Instituto de F\'{\i}sica de Cantabria, CSIC-UC, E-39005 Santander,
Spain\\
$^{2}$Department of Physics, University of Oxford, Oxford OX1 3RH,
UK\\
$^{3}$Department of Physics and Astronomy, University of Texas at San
Antonio, San Antonio, TX 78249, USA\\
$^{4}$Departamento de Astrof\'{\i}sica, Universidad Complutense de
Madrid, E-28040 Madrid, Spain\\ 
$^5$Instituto de Astrof\'{\i}sica de Canarias (IAC), E-38205 La Laguna,
Tenerife, Spain\\
$^6$Departamento de Astrof\'{\i}sica, Universidad de la Laguna (ULL),
E-38206 La Laguna, Tenerife, Spain\\ 
$^7$Centro de Radioastronom\'{\i}a y Astrof\'{\i}sica (CRyA-UNAM),
3-72 (Xangari), 8701, Morelia, Mexico\\ 
$^{8}$Gemini Observatory, Casilla 603, La Serena, Chile\\
$^{9}$Gemini Observatory, Northern Operations Center, Hilo,  HI 96720,
USA\\
$^{10}$Centro de Astrobiolog\'{\i}a, CSIC-INTA, E-28850 Torrej\'on de
Ardoz, Madrid, Spain\\
$^{11}$ASTRO-UAM, Universidad Aut\'onoma de Madrid,
  Unidad Asociada CSIC, Madrid, Spain\\
$^{12}$Instituto Nacional de Astrof\'{\i}sica, Optica y Electr\'onica
  (INAOE), 72000 Puebla, Mexico\\
$^{13}$N\'ucleo de Astronom\'{\i}a de la Facultad de Ingenier\'{\i}a, Universidad Diego Portales, Av. Ej\'ercito Libertador 441, Santiago, Chile\\
$^{14}$Subaru Telescope, 650 North A'ohoku Place, Hilo, Hawaii, 96720,
USA\\
$^{15}$National Astronomical Observatory of Japan, 2-21-1
Osawa, Mitaka, Tokyo 181-8588, Japan\\
$^{16}$Department of Astronomical Science,
The Graduate University for Advanced Studies (SOKENDAI),
Mitaka, Tokyo 181-8588, Japan\\
$^{17}$Florida Institute of Technology, Melbourne, FL 32901, USA
}

\begin{document}

\date{Accepted ---. Received ---; in original form ---}

\pagerange{\pageref{firstpage}--\pageref{lastpage}} \pubyear{2015}

\maketitle

\label{firstpage}
\vspace{-1cm}

\begin{abstract}
We present an atlas of mid-infrared (mid-IR) $\sim 7.5-13\,\mu$m
spectra of 45
local active galactic nuclei (AGN) obtained  
with CanariCam on the 10.4\,m
Gran Telescopio CANARIAS (GTC) as part of an ESO/GTC large
program. The sample includes Seyferts and other low 
luminosity AGN (LLAGN)
at a median distance of 35\,Mpc and  luminous AGN,
namely PG quasars,
(U)LIRGs, and radio galaxies (RG) at a median distance of 254\,Mpc.
To date, this is the
largest mid-IR spectroscopic catalog of local AGN at
sub-arcsecond resolution (median 0.3\,arcsec). The goal of this work is to
 give an overview of the spectroscopic properties of the
 sample. The nuclear $12\,\mu$m luminosities of the
AGN  span more than four orders of magnitude, $\nu
L_{12\mu{\rm m}}\sim  3\times 10^{41}-10^{46}\,{\rm 
  erg \,s}^{-1}$. In
a simple  mid-IR spectral index vs. strength of the $9.7\,\mu$m 
silicate feature diagram most LLAGN, Seyfert nuclei, PG quasars, and RGs lie
in the region occupied by clumpy torus model tracks. However,
the mid-IR spectra of some  might include contributions from other
mechanisms.  Most (U)LIRG nuclei in our sample
have deeper silicate features and flatter spectral
indices than predicted by  these models suggesting deeply embedded
dust heating sources and/or contribution from 
star formation. The $11.3\,\mu$m PAH feature
is clearly detected in approximately half of the Seyfert nuclei, LLAGN, and (U)LIRGs. 
While the RG, PG quasars, and (U)LIRGs in our sample have
similar nuclear $\nu L_{12\mu{\rm m}}$, we do not
detect nuclear PAH emission in the RGs and PG quasars. 

\end{abstract}

\vspace{-1cm}

\section{Introduction}\label{sec:intro}

The mid-infrared (mid-IR)  range has been proven to be
exceptionally rich in spectral features that can be  used to
characterise the properties of active galactic 
nuclei (AGN) and their host galaxies. 
In particular, {\it Spitzer} and previously ground-based and {\it ISO}
spectroscopy  have
provided excellent mid-IR spectroscopy of large samples of local AGN
\citep[see
e.g.][]{Roche1991,Laurent2000,Buchanan2006,Tommasin2008,Wu2009,Tommasin2010}. For
instance, 
these observations allowed the 
study of the silicate dust near the AGN and in their host galaxies using
the 10 and $18\,\mu$m spectral 
features \citep[][]{Sturm2005,Shi2006,Thompson2009,Mor2009,Goulding2012} and
define mid-IR features that trace star formation activity such as 
polycyclic aromatic hydrocarbon (PAH)
features and
the [Ne\,{\sc ii}]$12.81\,\mu$m line
\citep[][]{Shi2007,PereiraSantaella2010,DiamondStanic2012}. Moreover, a number of
different mid-IR indicators have been used to detect previously
unidentified AGN in local galaxies \citep{Goulding2009}, look for buried AGN
\citep{Imanishi2009,AAH2012} in local luminous and
ultraluminous IR galaxies (LIRGs and ULIRGs, respectively), and
provide a general activity class \citep{Genzel1998,Spoon2007,HernanCaballero2011}.

While space-based mid-IR observations provide excellent sensitivity
and access to large samples of AGN,  
ground-based mid-IR observations can take advantage of having telescopes with
an order of magnitude larger diameters and therefore much higher
angular resolution. Mid-IR instruments on 
8-10\,m telescopes deliver routinely imaging and spectroscopic
observations with angular resolutions of typically 
0.3-0.4\,arcsec. This is approximately a factor of 10 better than
what is achieved with {\it Spitzer}, albeit with limited sensitivity.
So far, these high angular resolution mid-IR spectroscopic
observations have been obtained for relatively small samples of AGN,
although they are producing interesting results. These include the study of the
properties of the  AGN dusty torus \citep[see  
e.g.][]{Hoenig2010,AAH2011,RamosAlmeida2014torus,RuschelDutra2014,Ichikawa2015},
the star formation activity in the nuclear regions of local AGN
\citep{Sales2013,Esquej2014,AAH2014}, 
the properties  of the obscuring material in the
nuclei of active galaxies
\citep{Mason2006,Roche2006,Roche2007,GonzalezMartin2013,Roche2015},
the nature of the nuclear dust heating source in local (U)LIRGs
\citep{Soifer2002,Lira2008,DiazSantos2010,AAH2013,Mori2014,MartinezParedes2015,PereiraSantaella2015}, and obscured super star clusters \citep{Snijders2006}. 

\begin{table*}
 \centering
 \begin{minipage}{175mm}
  \caption{The ESO/GTC large program sample of local AGN
  }\label{table:sample} 
  \begin{tabular}{lcccccccccc}
 \hline
Name           &   Redshift & Dist & scale & Type & IRAS$12\,\mu$m & Ref & Other\\ 
               &            & (Mpc) & (kpc/arcsec) & & (Jy) & & Name\\

\hline
3C273          & 0.158339 & 734. &2.647 &RG/PG Quasar              &0.417 &1  &PG~1226+023 \\   
3C382          & 0.05787  & 246. &1.068 &RG                &0.071
&2 & \\ 
3C390.3        & 0.056100 & 239. &1.041 &RG                &0.140 &2\\
IRAS~08572+3915 & 0.058350 & 254. &1.097 &(U)LIRG/Sy2:(NW)/Sy2:(SE)     &0.33  &3\\
IRAS~13197$-$1627 & 0.016541 & 73.2 &0.343 &(U)LIRG/Sy1.8       &0.94  &3         &MCG-03-34-064\\
IRAS~13349+2438 & 0.107641 & 483. &1.905 &(U)LIRG/Sy1       &0.631 &4\\
IRAS~14348$-$1447 & 0.08300  & 366. &1.512 &(U)LIRG/ cp:(NE)/cp::(SW)     &$<$0.10&  3\\  
IRAS~17208$-$0014 & 0.042810 & 181. &0.809 &(U)LIRG/HII     &0.22  &3\\
Mrk~3           & 0.013509 & 55.9 &0.264 &Sy2               &0.760 &4\\
Mrk~231         & 0.042170 & 181. &0.807 &(U)LIRG/Sy1         &1.83
&3 & IRAS~12540+5708 \\
Mrk~463         & 0.050355 & 219. &0.959 &(U)LIRG/Sy2(E)       &0.510
&4 & IRAS~13536+1836 \\  
Mrk~478         & 0.079055 & 347. &1.443 &PG Quasar            &0.098 &1  &PG~1440+356 \\
Mrk~841         & 0.036422 & 157. &0.706 &PG Quasar            &0.109 &1  &PG~1501+106 \\
Mrk~1014        & 0.163110 & 748  &2.684 &PG Quasar/(U)LIRG    &0.137 &1  &PG~0157+002\\
Mrk~1066        & 0.012025 & 47.2 &0.224 &Sy2         &0.460 &3
&UGC~02456 \\
Mrk~1073        & 0.023343 & 95.3 &0.442 &(U)LIRG/Sy2 &0.440 &3
&UGC~02608 \\
Mrk~1210        & 0.013496 & 59.5 &0.280 &Sy2           &0.496 &4\\
Mrk~1383        & 0.086570 & 383. &1.571 &PG Quasar            &0.124 &1  &PG~1426+015 \\
NGC~931         & 0.016652 & 66.1 &0.310 &Sy1         &0.610 &4 &
Mrk~1040\\
NGC~1194        & 0.013596 & 53.7 &0.254 &Sy1.9       &0.266 &4 \\ 
NGC~1275        & 0.017559 & 70.9 &0.332 &RG/Sy1.5     &1.06
&3 & 3C84\\
NGC~1320        & 0.008883 & 34.5 &0.164 &Sy2         &1.069 &4\\
NGC~1614        & 0.015938 & 65.5 &0.308 &(U)LIRG/cp         &1.38
&3 & IRAS~04315-0840 \\
NGC~2273        & 0.006138 & 25.8 &0.124 &Sy2         &0.44  &3\\
NGC~2992        & 0.007710 & 36.6 &0.174 &Sy1.8       &0.63  &3\\
NGC~3227        & 0.003859 & 20.4 &0.098 &Sy1.5       &0.94  &3\\
NGC~4051        & 0.002336 & 12.7 &0.061 &Sy1         &1.35  &3\\
NGC~4253        & 0.01292  & 57.6 &0.272 &Sy1         &0.386 &4\\
NGC~4258        & 0.001494 & 8.97 &0.043 &LLAGN/Sy1.9 &2.25  &5 
& M106\\
NGC~4388        & 0.008419 & 17.0$^*$ &0.082 &Sy1.9       &0.29  &3\\    
NGC~4419        & 0.000871 & 17.0$^*$ &0.082 &LLAGN/T2             &0.67  &3\\
NGC~4569        & -0.000784& 17.0$^*$ &0.082 &LLAGN/T2             &1.27
&3 & M90\\
NGC~4579        & 0.005060 & 17.0$^*$ &0.082&LLAGN/Sy1.9
&1.12  &3 & M58\\
NGC~5347        & 0.007789 & 35.1 &0.167 &Sy2         &0.309 &4\\
NGC~5548        & 0.017175 & 74.5 &0.348 &Sy1.5       &0.401 &4\\ 
NGC~5793        & 0.011645 & 51.1 &0.242 &Sy2         &0.17  &3\\
NGC~6240        & 0.024480 & 103. &0.475 &(U)LIRG/LINER     &0.59  &3
& IRAS~16504+0228 \\
NGC~7465        & 0.006538 & 21.9 &0.105 &Sy2         &0.26  &3\\
OQ208          & 0.076576 & 336. &1.406 &RG                &$<$0.185
&4     & Mrk~668  \\
PG~0804+761     & 0.100000 & 443. &1.773 &PG Quasar            &0.190 &1 \\
PG~0844+349     & 0.064000 & 279. &1.194 &PG Quasar            &0.126 &1\\
PG~1211+143     & 0.080900 & 358. &1.484 &PG Quasar            &0.362 &1\\
PG~1229+204     & 0.063010 & 276. &1.183 &PG Quasar            &0.417 &1\\
PG~1411+442     & 0.089600 & 396. &1.615 &PG Quasar            &0.115 &1\\ 
UGC~5101        & 0.039367 & 168. &0.755 &(U)LIRG/Sy2:     &0.25  &3
& IRAS~09320+6134 \\

\hline
\end{tabular}

{\it Notes.} Redshifts,  luminosity distances, and
projected 1\,arcsec\, scales are from NED for $H_0= 73\,{\rm km \,
  s}^{-1}\,{\rm Mpc}^{-1}$, $\Omega_{\rm M} =   0.27$, and
$\Omega_{\Lambda} =   0.73$. $^*$These  galaxies are  in
the Virgo Cluster (Binggeli et al. 1985).
In the column of ''Type'', the single colons mean 
that the nucleus does not have the same classification in
the three optical diagrams used and double colons mean 
that the classification using different optical line ratios does not
agree on any of the diagrams (see 
Yuan et al. 2010 for details). If two optical classes are given, they 
refer to the individual nuclei of the system.
The {\it IRAS} $12\,\mu$m flux densities are taken from the reference 
listed in the column to the right. References. 1. Sanders et
al. (1989).  2. Golombek et al. (1988). 3. {\it IRAS} Revised Bright
Galaxy Sample (RBGS), Sanders 
et al. (2003). 4.  {\it IRAS} Faint Catalog (IRASFC), Moshir et al. (1990). 
5. Rice et al. (1988). 
\end{minipage}
\end{table*}

In this paper we present a mid-IR spectroscopic atlas of 45 local AGN
with accompanying imaging obtained with the CanariCam
instrument \citep{Telesco2003, Packham2005}
on the 10.4\,m Gran Telescopio CANARIAS (GTC) in El Roque de los
Muchachos Observatory. The nearly diffraction limited 
(median 0.3\,arcsec) imaging and spectroscopic observations were
taken as part of an ESO/GTC large program (ID 182.B-2005, PI
Alonso-Herrero). The main goal of this work is to provide a brief overview of the
mid-IR spectroscopic properties of the nuclear regions of local
AGN. The paper is organized as 
follows. In Section~\ref{sec:sample} 
we present the main goals of our mid-IR survey of local AGN and the
sample. Section~\ref{sec:observations} describes the data reduction and analysis of
the mid-IR spectroscopic observations 
and accompanying mid-IR imaging
observations. In Section~\ref{sec:results} we discuss the
main spectroscopic properties of the AGN sample 
and Section~\ref{sec:summary} summarizes our results.

\section{The GTC/CanariCam mid-IR survey of  local AGN}\label{sec:sample}
The main objective of our CanariCam mid-IR survey of local AGN is to
understand the properties of the obscuring material around active
nuclei, including 
the so-called torus of the AGN Unified Model \citep{Antonucci1993, Netzer2015}.
 In particular this survey was designed to address a number of open
questions such as, (1) the nature of the torus material and its
connection with the interstellar material in the  host galaxy, (2)
the dependence of the torus properties (e.g., torus physical and
angular size, number of clouds, covering
factor) on the AGN luminosity and/or
activity class, (3) the relation  between the 
dust properties (for instance, composition, grain size) and the AGN
luminosity/type, and  (4) the role of nuclear 
($< 100\,$pc)  starbursts
in feeding and/or obscuring the active nuclei of galaxies.

\begin{figure}

\hspace{0.2cm}
\resizebox{0.95\hsize}{!}{\rotatebox[]{-90}{\includegraphics{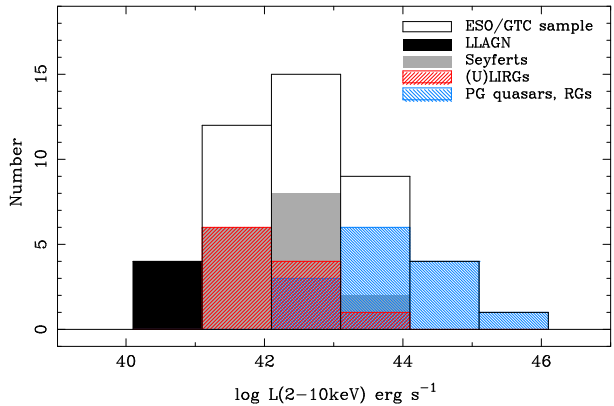}}}

\vspace{-1cm}
\caption{Hard X-ray luminosities of the AGN in the ESO/GTC sample. The
filled and hatched histograms represent different AGN classes in the sample.}
\label{fig:sample}
    \end{figure}

Although our team and others have already addressed some of the open
questions posed 
above, this was done for relatively small samples of local 
AGN (see references in the Introduction). We thus decided to exploit the unique
combination of the diffraction limited 
($\simeq 0.3\,$arcsec) angular resolution and the imaging and spectroscopy
capabilities of the GTC/CanariCam system to address these questions.   We were awarded 
total of  180\,hours of observing 
time through an ESO/GTC large program (see Section~\ref{sec:observations}
for full details of these observations). In future papers we will combine the ESO/GTC
time with   the   on-going (at the time of writing this paper) observations of 
$\sim 100\,$hours of CanariCam guaranteed time (GT)
devoted to AGN science  observations. The  final goal is 
to build a statistically significant sample of local AGN   in
terms of AGN classes and AGN luminosity bins (see below). As part of the
ESO/GTC and GT
  observations we are also obtaining mid-IR imaging and spectropolarimetry
  observations of some of the brightest targets in our sample 
\citep[][L\'opez-Rodr\'{\i}guez et al. 2016, in
  preparation]{LopezRodriguez2014,LopezRodriguez2015}.

For  the ESO/GTC large program  we selected a sample of
45 local active galaxies with the purpose of covering a broad range of 
AGN luminosities and different AGN classes.  We chose the hard (2-10\,keV) X-ray 
luminosities as a proxy for the AGN
luminosity.  Finally, to ensure a spectroscopic detection
with reasonable integration times we also imposed
sufficiently bright arcsecond resolution literature $N$-band
fluxes. This limit was  approximately 20\,mJy. The ESO/GTC sample of
local AGN  (see Table~\ref{table:sample}) includes transition (T) or 
  composite (cp) objects (that is, nuclei with contributions from
AGN and star formation activity), Seyfert galaxies, radio galaxies (RG), LIRGs and ULIRGs, and 
quasars from the Bright Quasar Survey \citep{Schmidt1983}, which
are selected from the Palomar-Green (PG) Survey. 
All galaxies in the  
(U)LIRG class, except IRAS~17208-0014, 
have been spectroscopically identified as 
AGN  or composite \citep[see][]{Wu2009, Yuan2010}. For those (U)LIRGs with double nuclei (see
  Section~\ref{sec:imaging}) at 
  least one nucleus is spectroscopically classified as an AGN, usually
the mid-IR bright one. The ULIRG sample was chosen to match approximately the PG
quasars in AGN bolometric luminosities (see also Section~\ref{sec:12micronemission}).

It has been suggested that the torus disappears at AGN
bolometric luminosities $<~10^{42}\,{\rm erg \, s}^{-1}$ \citep{Elitzur2006}. 
To draw particular attention to AGN below this possible threshold, we identify objects with
hard X-ray luminosities $<10^{41}\,{\rm erg \,s}^{-1}$
as low-luminosity AGN (LLAGN). In summary, our sample contains 4 LLAGN, 16 Seyfert nuclei,
11 (U)LIRG, 5 RGs, and 9 PG quasars. We note that some (U)LIRG nuclei, and RG galaxies are also
classified optically as Seyferts.  

Fig.~\ref{fig:sample} shows the distribution of hard X-ray
luminosities of the AGN in the ESO/GTC
sample. As can be seen from this figure, the
AGN luminosities   span more than  
four orders of magnitude with the 
PG quasars and RGs at the high luminosity end. 
For the assumed cosmology ($H_0= 73\,{\rm km \,
  s}^{-1}\,{\rm Mpc}^{-1}$, $\Omega_{\rm M} =   0.27$, and
$\Omega_{\Lambda} =   0.73$), the Seyferts and LLAGN are at a median
distance of 35\,Mpc, whereas
the rest of the sample is at a median distance of 254\,Mpc. 
The requirement that
  the nuclei for the galaxies in our sample be classified as AGN
  or composite using optical spectroscopy might
  exclude highly obscured AGN. We also note that the small-aperture
  mid-IR flux limit provides a mid-IR flux limited AGN sample,
  although some of the ULIRG nuclei and LLAGN also have a significant star formation contribution
in the mid-IR \citep[see e.g.,][and Alonso-Herrero et al. 2016, in 
preparation]{Mason2012,Mori2014,MartinezParedes2015}.

\begin{table*}
 \centering
  \caption{Log of the GTC/CanariCam Si-2 filter imaging
    observations.}\label{table:imaging_log} 

  \begin{tabular}{lccccccc}
 \hline
Galaxy & Date & t$_{\rm on}$ $\times$ rep &  PA & Star & FWHM & Comment\\
       & (yyyy-mm-dd) & (s)       & (degree) & & (arcsec)\\ 
 \hline

3C273      &2014-03-17&    $139 \times  3$    &0   &HD~107328    &0.31\\ 
3C382      &2013-08-27&   $348\times   3$  &0   &HD~176670    &0.23\\ 
3C390.3    &2013-07-22&   $348\times   3$  &0   &HD~158986    &0.29\\ 
IRAS~08572+3915  &2014-03-16&    $139\times   3$  &0   &HD~83787     &0.28\\ 
IRAS~13197$-$1627   &2013-01-07&    $81 \times  4$    &0   &HD~116870    &0.25 & Pattern Noise\\ 
IRAS~13349+2438  &2013-07-22&    $139\times   3$    &0   &HD~121710    &0.31\\ 
IRAS~14348$-$1447  &2014-06-10&   $348\times    3$    &0   &HD~130157    &0.40 \\ 
IRAS~17208$-$0014  &2013-06-07&   $348\times    3$  &0   &HD~157999    &0.27\\ 
Mrk~3       &2013-08-27&    $139\times   3$  &0   &HD~34450     &0.49\\ 
Mrk~231     &2013-01-07&    $81\times   4$  &0   &HD~111335    &0.34 &Pattern Noise\\ 
Mrk~463     &2014-02-09&    $139  \times 1$    &0   &HD~125560
&0.37& Distorted PSF\\ 
Mrk~478     &2014-03-16&    $209\times    3$    &0   &HD~128902    &0.25\\ 
Mrk~841     &2013-08-30&    $209\times   4$    &0   &HD~140573    &0.26\\ 
Mrk~1014    &2013-01-01&   $242\times  3$    &0   &HD~10550     &0.59 &Pattern Noise\\ 
Mrk~1066    &2013-08-27&    $139\times   3$   &45   &HD~18449
&0.24 & References 1, 2\\ 
Mrk~1073    &2013-08-27&    $209\times   3$  &345   &HD~19476
&0.26 & Reference 1\\ 
Mrk~1210    &2012-12-28&    $147\times   3$    &0   &HD~66141     &0.31 &Pattern Noise\\ 
Mrk~1383    &2012-03-09&    $220\times  3$    &0   &HD~126927    &0.42 &Pattern Noise\\ 
NGC~931     &2013-08-26&    $139 \times   3$  &350   &HD~14146     &0.28\\ 
NGC~1194    &2013-08-28&    $209\times  3$    &40  &HD~20356     &0.32\\ 
NGC~1275    &2013-08-27&    $139\times  4$    &0   &HD~19476     &0.37
\\ 
NGC~1320    &2013-01-03&    $161\times   3$    &0   &HD~20356     &0.30
& Distorted PSF\\ 
NGC~1614    &2013-09-18&   $139 \times 1$    &0   &HD~28749      & 0.35
& Acquisition image \\  
NGC~2273    &2013-09-24&    $209\times  3$   &20   &HD~42633     &0.26
& Reference 1\\  
NGC~2992    &2014-02-13&    $348\times   2$   &30   &HD~82660     &0.32
& Distorted PSF  \\ 
NGC~3227    &2014-03-17&    $209\times  3$   &55   &HD~85503     &0.31\\ 
NGC~4051    &2014-02-09&    $139\times  3$   &40   &HD~95212     &0.32
& Distorted PSF \\ 
NGC~4253    &2014-03-17&    $139\times   3$   &15   &HD~108381    &0.32\\ 
NGC~4388    & 2015-02-01&   $348\times   3$   &90   &HD~111067    &0.39 &\\  
NGC~4419    & 2014-05-23&    $209\times   3$   &40   &HD~109511    &0.33\\ 
NGC~4569    & 2014-03-16&    $209\times   3$  &300   &HD~111067    &0.28\\ 
NGC~5347    &2014-03-16&    $139\times   3$   &15   &HD~121710    &0.27\\ 
NGC~5548    &2014-03-17&    $139 \times  3$  &0   &HD~127093    &0.44& \\ 
NGC~5793    &2014-05-17&    $209\times  3$   &45   &HD~133774    &0.30\\ 
NGC~6240    &2013-08-27&    $209\times  3$  &286   &HD~151217    &0.38
& References 1, 3\\ 
NGC~7465    &2013-08-27&   $348\times   3$   &60   &HD~220363    &0.38
& Distorted PSF\\
OQ208      &2014-03-17&    $209\times   3$  &0   &HD~127093    &0.43\\ 
PG~0804+761 &2014-01-03&    $209\times  3$  &0   &HD~64307     &0.34
&Distorted PSF\\
PG~0844+349 &2014-01-06&    $216\times   2$  &0   &HD~81146     &0.38&
Distorted PSF \\
PG~1211+143 &2014-03-14&    $209\times  3$    &0   &HD~107328    &0.29\\ 
PG~1229+204 &2014-06-08&   $1251 \times 1$  &0   &HD~111067    &0.27\\ 
PG~1411+442 &2014-03-16&    $209\times  3$  &0   &HD~128902    &0.27\\ 
UGC~5101    &2014-01-06&   $1224 \times  1$  &0   &HD~79354
&0.41 & Reference 4\\  
\hline
\end{tabular}

{\it Notes}. The references listed in the last column indicate
previous works where the galaxies have been presented. 1. Alonso-Herrero et al. (2014). 2. Ramos Almeida et
al. (2014b). 3. Mori et 
al. (2014). 4. Mart\'{\i}nez-Paredes et al. (2015). 
\end{table*}

\begin{figure}

\hspace{0.2cm}
\resizebox{0.95\hsize}{!}{\rotatebox[]{-90}{\includegraphics{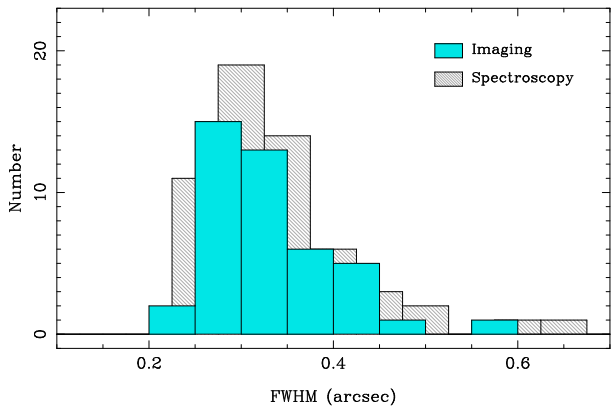}}}

\vspace{-1cm}
\caption{Image quality of the GTC/CanariCam imaging and spectroscopic
  observations in the ESO/GTC large program as 
  measured from the FWHM of the standard stars at $8.7\,\mu$m. }
\label{fig:FWHM}
    \end{figure}

\section{Observations,  Data Reduction, and
  Analysis}\label{sec:observations} 
In this section we describe the CanariCam imaging and spectroscopic
observations taken within the ESO/GTC large program as well as the data
reduction and analysis.  
The ESO/GTC
CanariCam imaging and spectroscopic observations were obtained between
March 2012 and March 2015. As  required by 
our approved ESO/GTC large program, all the data were observed in
queue mode under photometric conditions
and image quality  better than 0.6\,arcsec measured at
mid-IR wavelengths from the full width half maximum (FWHM)
of the standard stars (see 
Sections~\ref{sec:imaging} and \ref{sec:spectroscopy}).

\subsection{Imaging observations and data reduction}\label{sec:imaging}
We obtained imaging observations using the Si-2
filter, which has a central wavelength of $\lambda_{\rm 
  c}=8.7\,\mu$m and a width at 50\%
cut-on/off of $\Delta \lambda_{\rm cut} =1.1\,\mu$m. We chose the 
Si-2 filter because it gives the optimal sensitivity with the best
FWHM among the CanariCam filters.   We observed all AGN in
our sample except for two LLAGN
(NGC~4258 and 
NGC~4579) and the LIRG NGC~1614\footnote{We note that we used the
NGC~1614  acquisition image for the CanariCam spectroscopy (see next section)
  for the aperture photometry in
  Section~\ref{sec:aperturephotometry}.}, which had been  previously
observed  
by \cite{Mason2012} and \cite{DiazSantos2008}, respectively. The plate
scale of the CanariCam $320 \times 240$ Si:As detector is
0.0798\,arcsec pixel$^{-1}$ (0.08\,arcsec pixel$^{-1}$ hereafter),
which provides a field of 
view in imaging 
mode of $26\,{\rm arcsec} \times 19\,{\rm arcsec}$.

All the observations were carried out in observing blocks (OB) using
the standard mid-IR chop-nod technique. The typical
imaging sequence included an OB for the galaxy target with one or
several repetitions together with an OB for the standard star, which
was observed immediately before or after the galaxy observation. We
used the observations of the standard stars to both perform the
photometric calibration of the galaxy images and measure the angular resolution
 of the observations. The
chop and nod throws were 15\,arcsec, whereas the chop and nod 
angles, which were the same for the galaxy and the star,
were chosen for each target to avoid extended galaxy emission
in the sky images.

Table~\ref{table:imaging_log} summarizes the details of the imaging
observations including the date of the observation, the on-source
integration time, number of repetitions, position angle (PA) of the
detector on the sky, and the name of the standard star.
Observations taken prior to March 2013 were obtained with the S1R1-CR
readout mode which greatly reduced the vertical "level drop" pattern
\citep{Sako2003}
although it also produced a pattern noise resembling horizontal stripes in the  
images\footnote{The horizontal stripes are clearly seen in  the
  commissioning images of the CanariCam polarimetric mode: 
  http://www.gtc.iac.es/instruments/canaricam/data-commissioning.php\#Pol\_readout\_mode}.
After March 2013 all
the observations were taken with 
the S1R3 mode which is now the default mode for CanariCam\footnote{We refer the reader to the GTC/CanariCam
  webpage for further information on the different readout modes:
  http://www.gtc.iac.es/instruments/canaricam/
canaricam.php\#ReadoutModes. 
}. This level drop effect did not affect our CanariCam science observations and the new readout model resulted in a signal-to-noise gain. The observations taken with the S1R1-CR readout mode are marked in the 
last column of Table~\ref{table:imaging_log}. 

We reduced the CanariCam imaging data using the {\sc redcan} pipeline
\citep[see][]{GonzalezMartin2013}. The reduction process of 
the imaging data includes sky subtraction, stacking of the individual
images, and rejection of bad images. The flux calibration of the
galaxy images is done using the observation of the standard star taken
immediately before or after the galaxy OB. In those cases with
several repetitions for the same galaxy, we 
shifted the individual images to a common position and then combined them 
using the average. We finally rotated them to the usual orientation of
north up, east to the left.

To estimate the angular resolution of
the imaging observations we measured the FWHM of the standard star using a moffat  function. These are listed in Table~\ref{table:imaging_log} and
plotted in Fig.~\ref{fig:FWHM}. As can be seen from this figure, the
measured FWHM of the imaging observations are between 0.23 and
0.59\,arcsec with a median value of 0.31\,arcsec. At the median
distances of our sample, this corresponds to a median physical resolution of
approximately
51\,pc for the Seyfert galaxies and LLAGN and 382\,pc for the rest of the sample.

\begin{figure*}

\resizebox{0.25\hsize}{!}{\rotatebox[]{-90}{\includegraphics{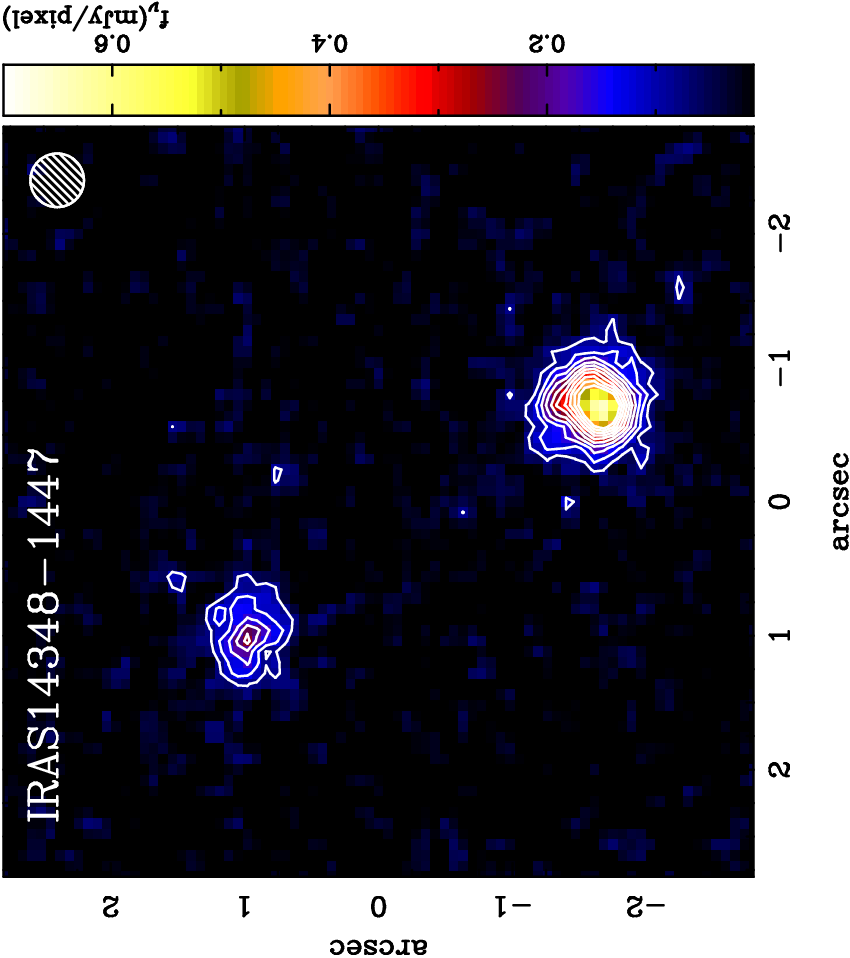}}}
\hspace{0.3cm}
\resizebox{0.25\hsize}{!}{\rotatebox[]{-90}{\includegraphics{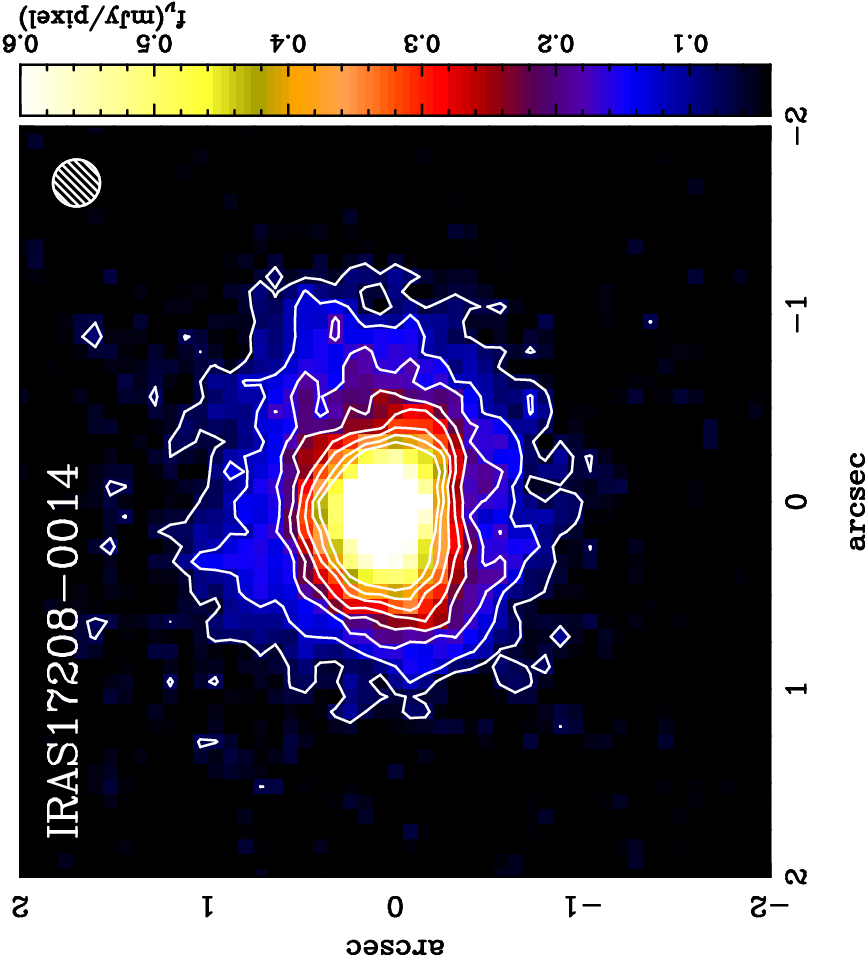}}}
\hspace{0.3cm}
\resizebox{0.25\hsize}{!}{\rotatebox[]{-90}{\includegraphics{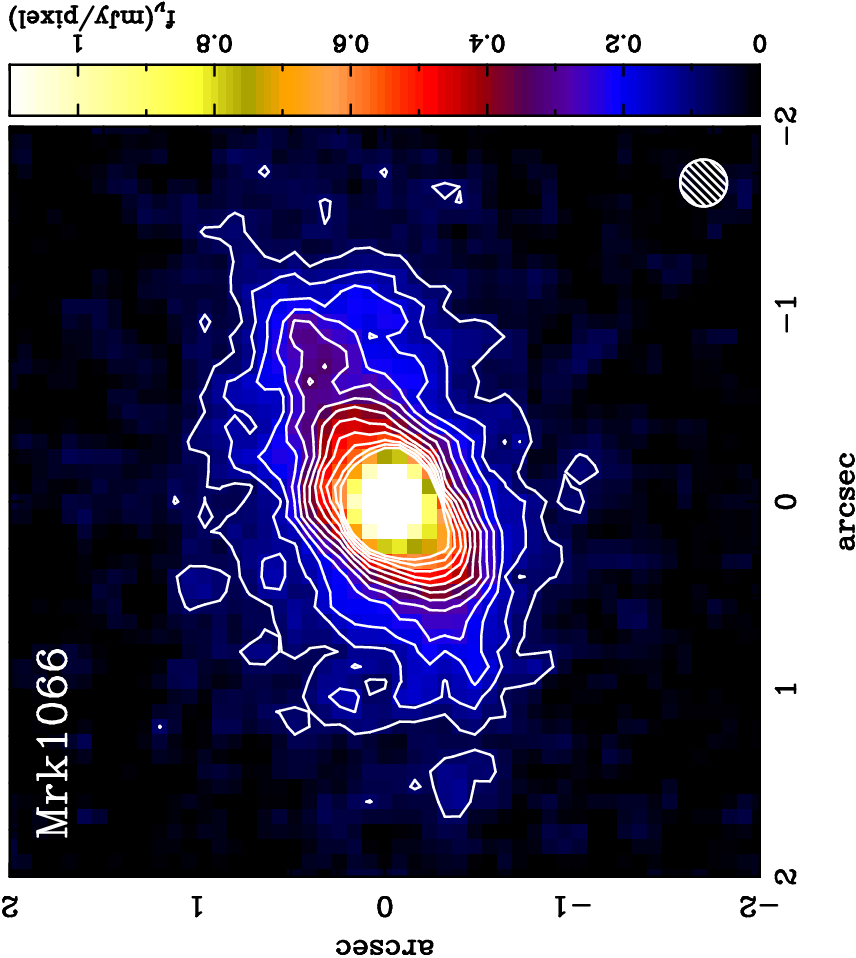}}}

\resizebox{0.25\hsize}{!}{\rotatebox[]{-90}{\includegraphics{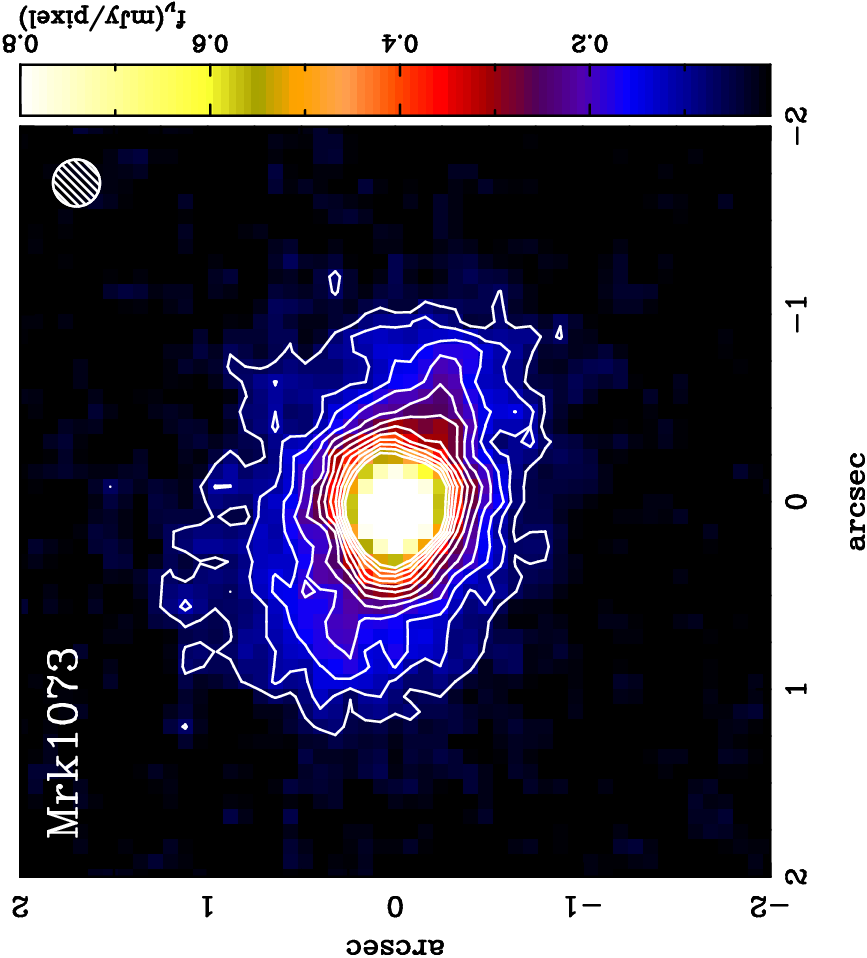}}}
\hspace{0.3cm}
\resizebox{0.25\hsize}{!}{\rotatebox[]{-90}{\includegraphics{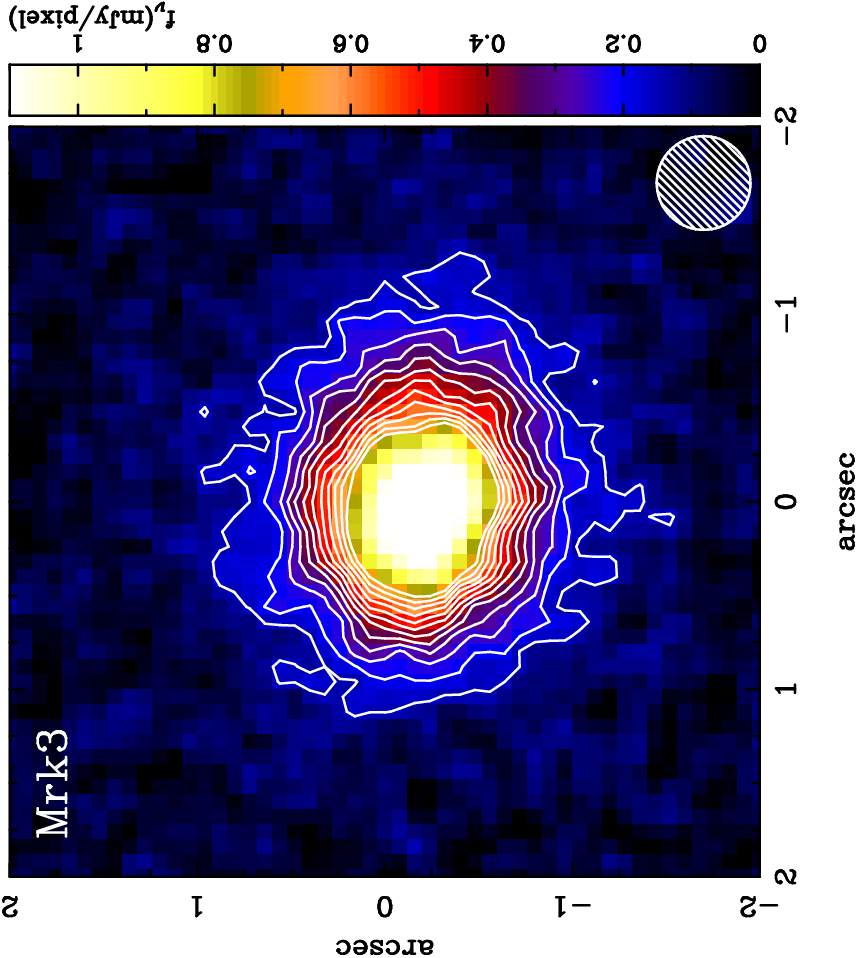}}}
\hspace{0.3cm}
\resizebox{0.25\hsize}{!}{\rotatebox[]{-90}{\includegraphics{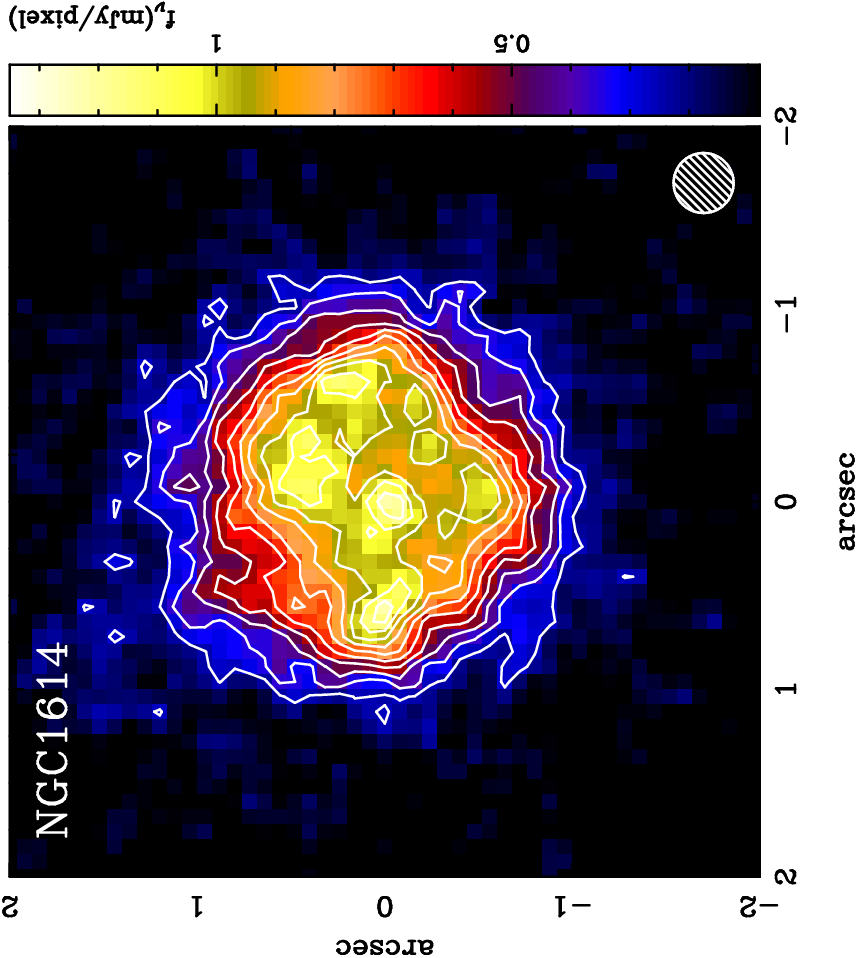}}}

\resizebox{0.25\hsize}{!}{\rotatebox[]{-90}{\includegraphics{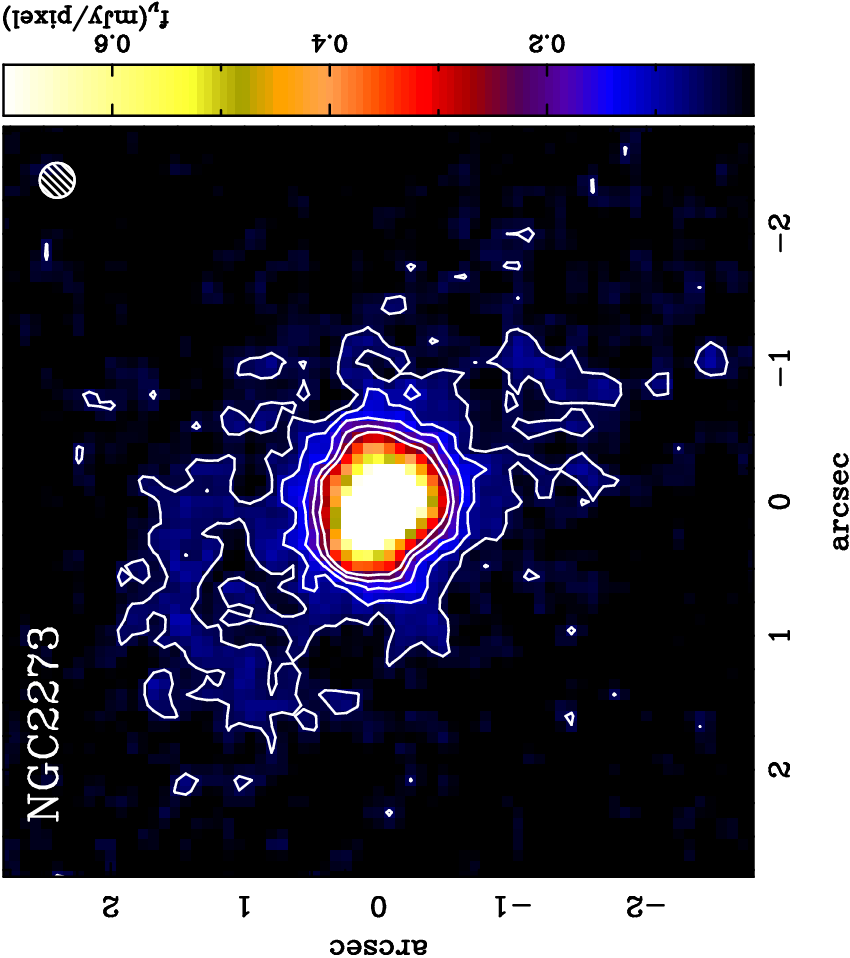}}}
\hspace{0.3cm}
\resizebox{0.25\hsize}{!}{\rotatebox[]{-90}{\includegraphics{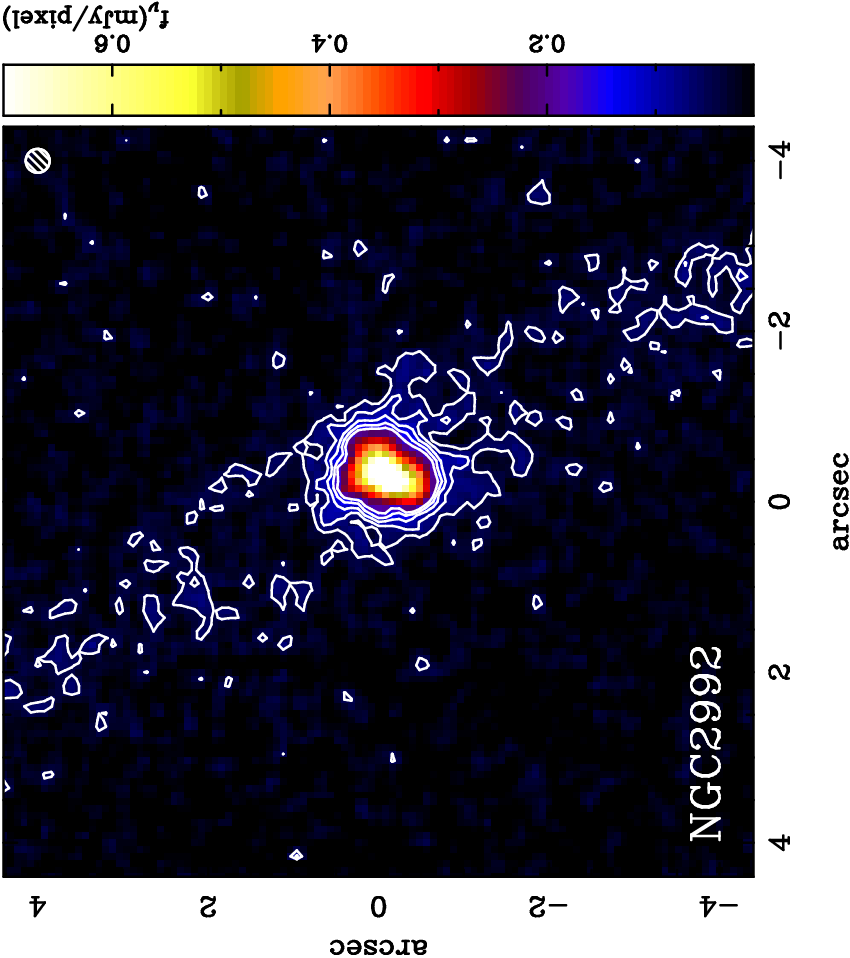}}}
\hspace{0.3cm}
\resizebox{0.25\hsize}{!}{\rotatebox[]{-90}{\includegraphics{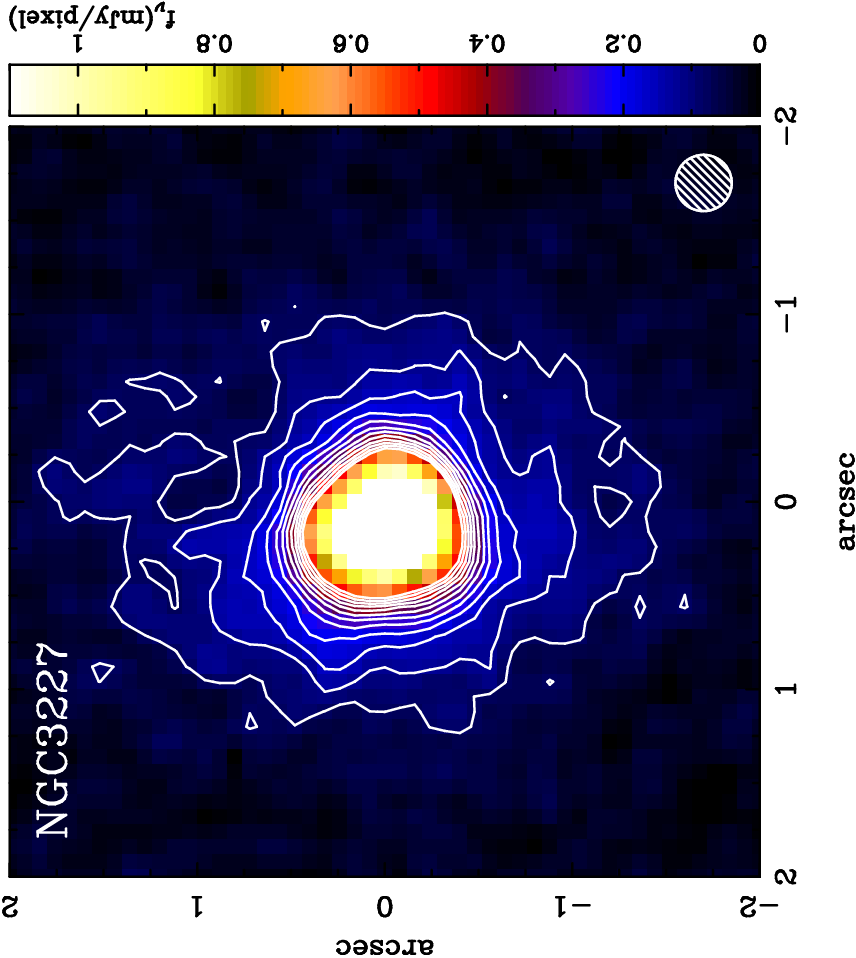}}}

\resizebox{0.25\hsize}{!}{\rotatebox[]{-90}{\includegraphics{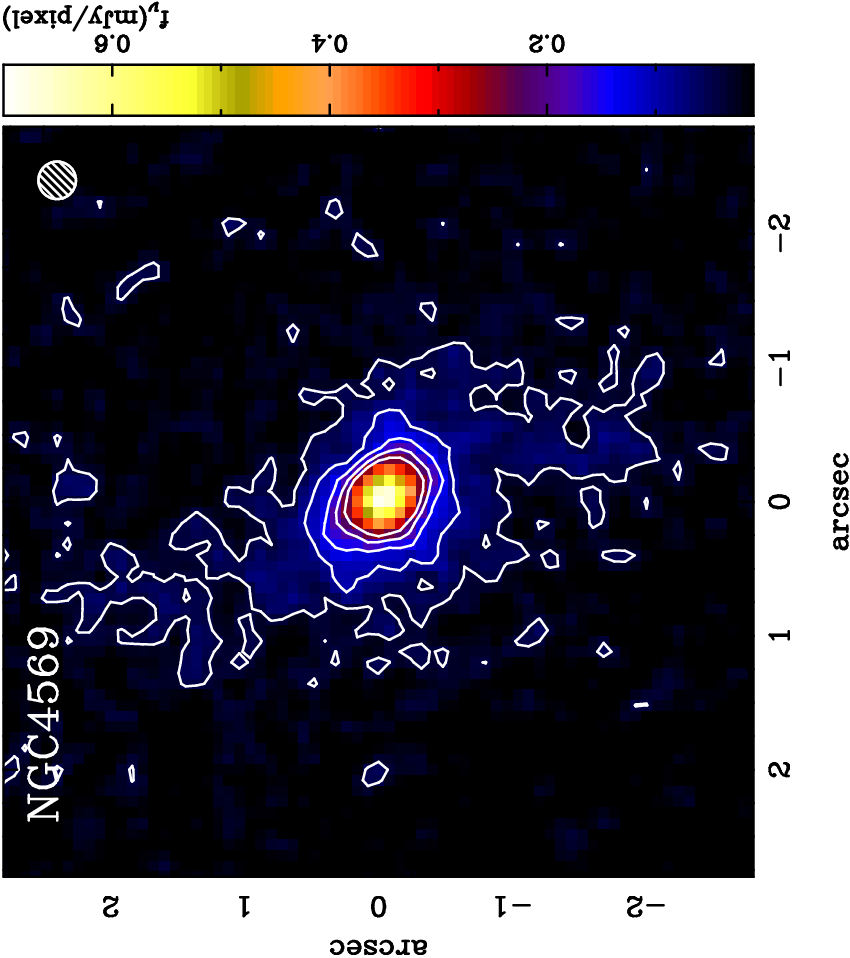}}}
\hspace{0.3cm}
\resizebox{0.25\hsize}{!}{\rotatebox[]{-90}{\includegraphics{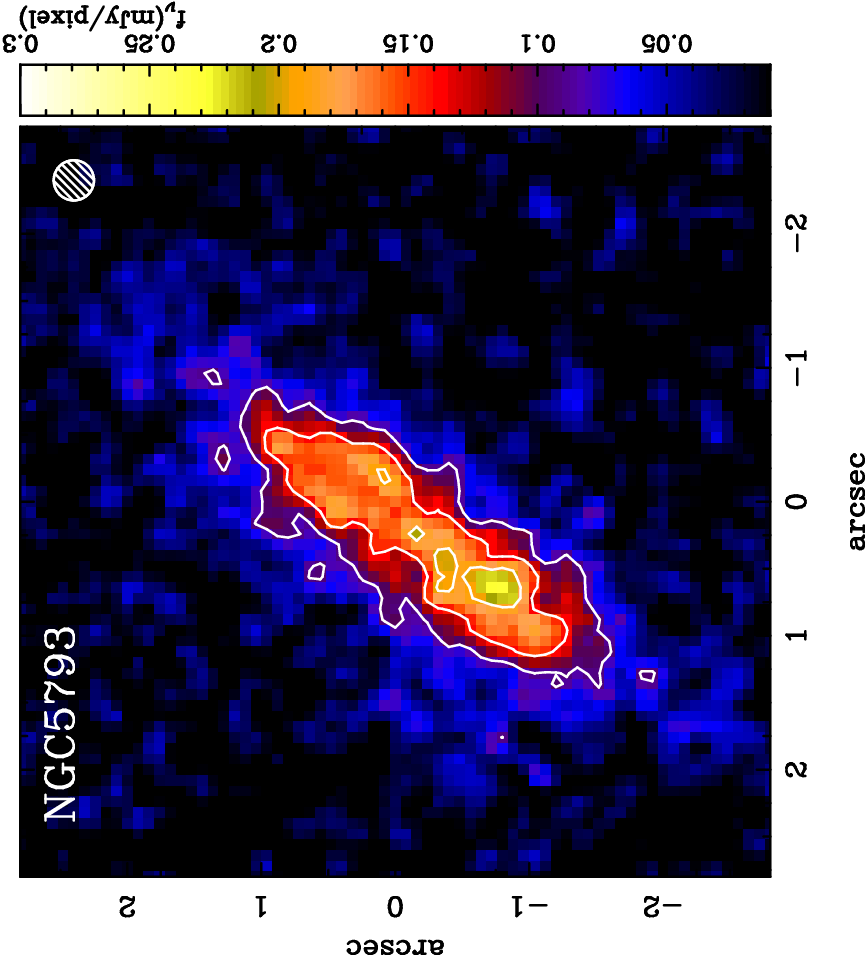}}}
\hspace{0.3cm}
\resizebox{0.25\hsize}{!}{\rotatebox[]{-90}{\includegraphics{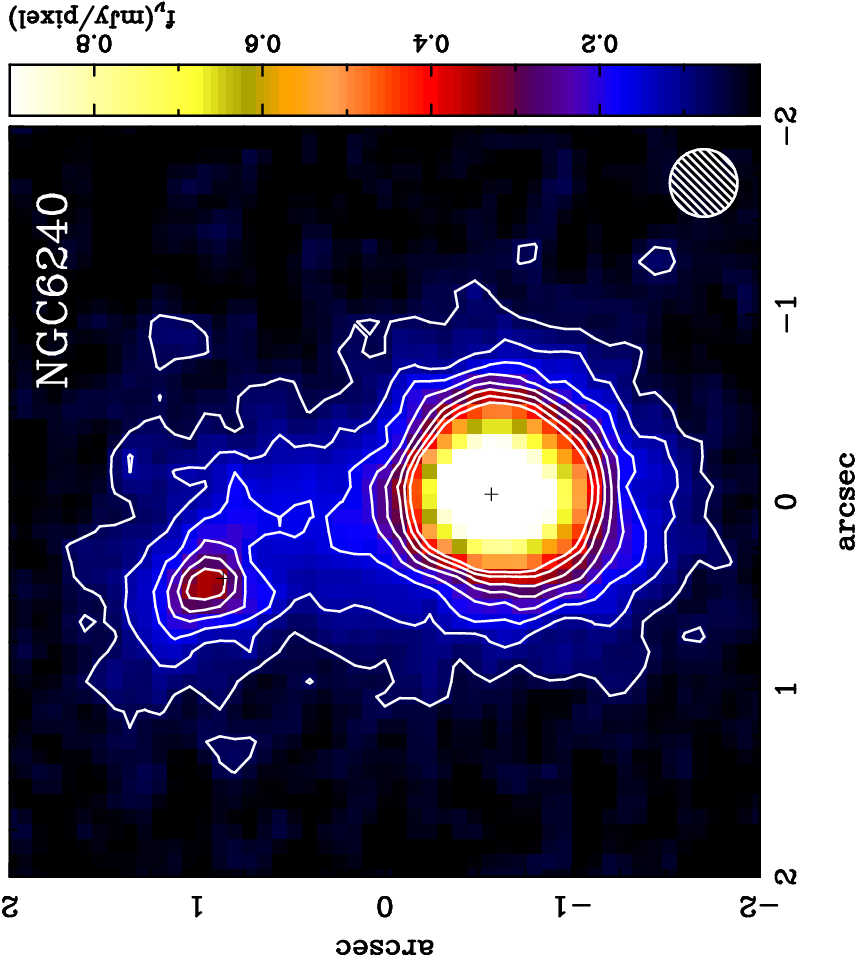}}}
\resizebox{0.25\hsize}{!}{\rotatebox[]{-90}{\includegraphics{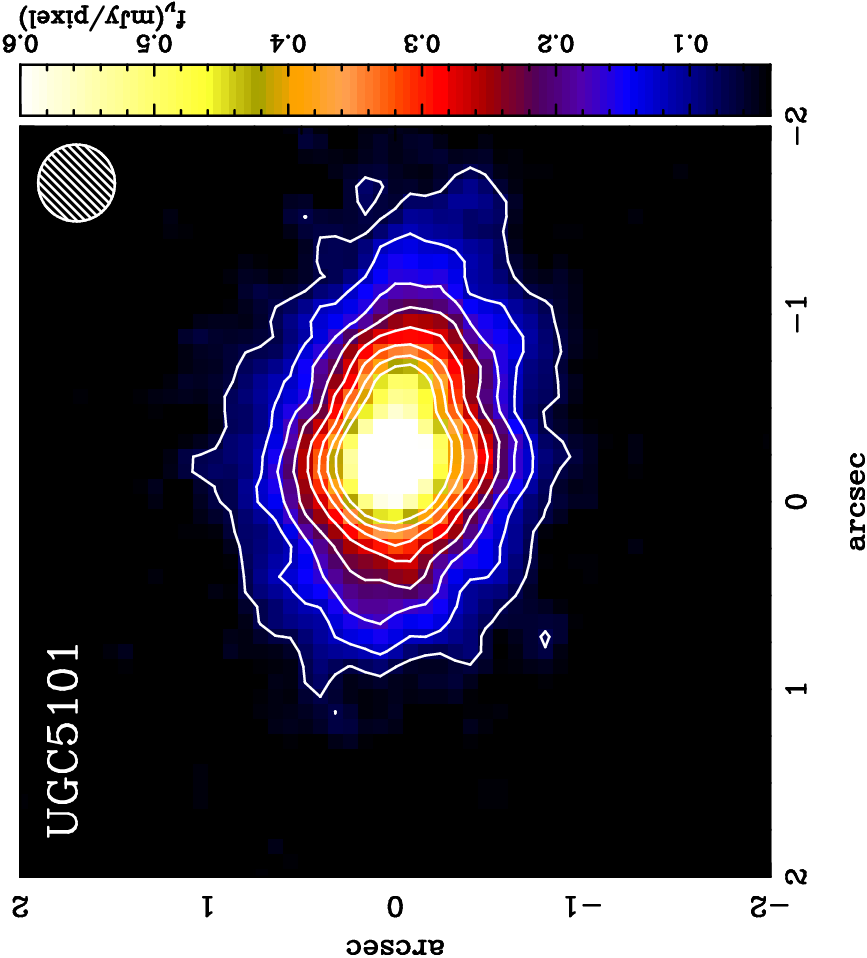}}}

\vspace{-0.1cm}
\caption{Examples of flux-calibrated GTC/CanariCam Si-2 ($\lambda_{\rm
    c}=8.7\,\mu$m) images of 
  galaxies in the ESO/GTC large program which are clearly extended
  in the mid-IR. Orientation is north up, east to the left. We
  smoothed the CanariCam images with a
Gaussian function with a width (sigma) between 0.6 and 0.7 pixels. The hatched
circle represents the angular resolution of the image 
(FWHM) as measured from the
corresponding standard star of each galaxy.}
\label{fig:Images}
    \end{figure*}

Apart from the FWHM,
we also measured the Strehl ratios of the standard stars. The median
value for the imaging observations was 0.19, although we note that at
the time the observations the GTC did not have fast tip/tilt guiding. A small
fraction of the images show distorted shapes for the core of the
point spread function (PSF). These images have lower Strehl ratios than expected
for the  measured FWHM of the star. For reference we marked
those in the last column of Table~\ref{table:imaging_log}. 

Four (U)LIRGs in our sample have double nuclei, namely,
IRAS~08572+3915, IRAS~14348$-$1447, Mrk~463 \citep[see e.g.,][]{GarciaMarin2009},
and NGC~6240. The CanariCam 
$8.7\,\mu$m image of IRAS~08572+3915 only shows emission
from one bright nucleus \citep[see also][]{Soifer2000}. A comparison with archival {\it Spitzer}/IRAC
images of this system at $3.4\,\mu$m and $8\,\mu$m reveals that the
bright source detected in the mid-IR corresponds to the nucleus of the
northern galaxy, which appears unresolved (FWHM=0.3\,arcsec$\simeq 330\,$pc) in our
CanariCam $8.7\,\mu$m image. The CanariCam image of IRAS~14348$-$1447 (see
Fig.~\ref{fig:Images}) reveals $8.7\,\mu$m emission arising from both
nuclei, with the southern one being brighter than the northern one
(see Section~\ref{sec:aperturephotometry} and Table~\ref{table:aperture_photometry}). The CanariCam
$8.7\,\mu$m imaging data of NGC~6240 has been extensively discussed by
\cite{Mori2014} and \cite{AAH2014}. \cite{Soifer2002} obtained Keck
mid-IR imaging and spectroscopy of Mrk~463 and showed that the
emission comes from the eastern nucleus.

A large fraction of the AGN observed in the ESO/GTC large program
show compact emission in the mid-IR, as 
already found in the large mid-IR imaging compilation of \cite{Asmus2014}. 
In Fig.~\ref{fig:Images} we show the CanariCam images of those AGN in
our sample that are  
clearly extended in the mid-IR. For completeness we also show galaxies
already studied in some of our previous work
\citep[][see also
Table~\ref{table:imaging_log}]{AAH2014,RamosAlmeida2014,Mori2014,GarciaBernete2015,MartinezParedes2015}. In
a series of forthcoming papers 
we will present  
detailed analyses of the mid-IR nuclear and extended emission of Seyfert galaxies
(Garc\'{\i}a-Bernete et al. 2016 in preparation) and of PG quasars
(Mart\'{\i}nez-Paredes et al. 2016 in preparation).

\begin{table*}
 \centering
  \caption{Log of the GTC/CanariCam spectroscopic
    observations.}\label{table:spectroscopy_log} 

  \begin{tabular}{lccccccc}
 \hline
Galaxy & Date & t$_{\rm on}$ $\times$ rep &  PA & Star & FWHM & Ref\\
       & (yyyy-mm-dd) & (s)       & (degree) & & (arcsec)\\ 
 \hline

3C273       &2014-03-17& $354\times 3$&   0  &HD~107328 &0.42\\
3C382       &2013-08-29& $943\times 1$&   0  &HD~176670 &0.28\\  
3C390.3     &2013-08-26& $354\times 2$& 0  &HD~158996 &0.49\\  
IRAS~08572+3915   &2013-03-16& $354\times 3$& 0  &HD~83787  &0.24\\
IRAS~13197$-$1627    &2015-03-02& $943\times 1$&  90   &HD~116879 &0.31\\
IRAS~13349+2438   &2014-06-15& $943\times 1$& 0  &HD~121710 &0.31\\  
IRAS~14348$-$1447   &2015-03-16& $1179\times 2$&  25   &HD~130157 &0.47\\
IRAS~17208$-$0014   &2013-09-09& $1238\times 1$&  90  &HD~157999 &0.30\\
Mrk~3       &2014-11-06& $943\times 1$& 30   &HD~34450  &0.25\\
Mrk~231     &2014-03-13& $354\times 2$& 290  &HD~111335 &0.28\\  
Mrk~463     &2014-03-18& $177\times 3$& 82   &HD~125560 &0.50\\
Mrk~478     &2014-06-07& $1238\times 1$& 0  &HD~128902 &0.30\\ 
            &2014-06-07& $1238\times 1$& 0  &HD~128902 &0.26\\ 
Mrk~841     &2014-05-27& $943\times 1$& 0  &HD~133165 &0.37\\ 
            &2014-05-01& $943\times 1$& 0  &HD~133165 &0.28\\ 
Mrk~1014    &2013-09-09& $1238\times 2$&   0  &HD~10550  &0.28\\  
Mrk~1066    &2013-08-31& $354\times 3$& 315  &HD~18449  &0.28 &
1, 2\\
Mrk~1073    &2013-09-10& $413\times 3$&  75  &HD~14146  & 0.34 & 1 \\
Mrk~1210    &2014-12-04& $295\times 3$& 0    &HD~66141  &0.27\\  
Mrk~1383    &2014-05-19& $943\times 1$& 0    &HD~126927 &0.34\\
            &2014-06-09& $943\times 1$& 0    &HD~126927 &0.30\\  
NGC~931     &2013-09-16& $354\times 3$& 80    &HD~14146  &0.33\\
NGC~1194    &2013-09-06& $295\times 3$& 310   &HD~20356  &0.34\\ 
NGC~1275    &2013-09-09& $354\times 3$& 0  &HD~19476  &0.27\\  
NGC~1320    &2013-09-06& $295\times 3$& 315   &HD~20356  &0.34\\  
NGC~1614    &2014-01-05& $1242\times 1$& 90   &HD~28749  &0.64 & 3\\ 
            &2013-09-08& $1242\times 1$ &  0   &HD~28749  &0.45\\ 
NGC~2273    &2013-09-22& $354\times 1$& 290  &HD~42633  &0.32 &
1\\ 
            &2013-09-23& $295\times 3$& 290  &HD~42633  &0.26 &
            1\\ 
NGC~2992    &2014-02-14& $943\times 1$& 30   &HD~82660  &0.30 &
4\\
NGC~3227    &2014-12-03& $943\times 1$&   0  &HD~85503  &0.35\\  
NGC~4051    &2014-02-09& $354\times 3$& 310  &HD~95212  &0.42\\  
NGC~4253    &2014-03-17& $354\times 3$& 285  &HD~108381 &0.27\\  
NGC~4258    &2015-02-04& $1238\times 1$& 325  &HD~107274 &0.26\\
NGC~4388    &2015-03-07& $943\times 1$&  90  &HD~111067 &0.58\\
NGC~4419    &2015-02-17& $1238 \times 1$& 310   &HD~109511 &0.41\\
NGC~4569    &2015-02-06& $1238\times 2$& 30    &HD~111067 &0.43\\  
NGC~4579    &2015-02-12& $1238\times 1$& 55    &HD~111067 &0.29\\
NGC~5347    &2014-06-11& $943\times 1$& 283   &HD~121710 &0.32\\
NGC~5548    &2014-06-10& $1061\times 1$& 0   &HD~127093 &0.32\\
NGC~5793    &2015-03-02& $943\times 1$& 315   &HD~133774 &0.30\\ 
NGC~6240    &2013-09-15& $1238\times 1$& 16    &HD~157999 &0.40 &
1, 5\\
NGC~7465    &2013-08-30& $943\times 1$& 330   &HD~220363 &0.29\\ 
            &2013-08-29& $295\times 3$& 330   &HD~220363 &0.26\\ 
OQ208       &2014-06-09& $943\times 1$& 0   &HD~127093 &0.28\\ 
            &2014-06-06& $354\times 3$& 0   &HD~127093 &0.25\\ 
PG~0804+761  &2014-01-03& $354\times 3$& 0  &HD~64307  &0.33\\ 
            &2014-03-15& $354\times 3$& 0  &HD~64307  &0.33\\ 
PG~0844+349  &2014-12-05& $1238\times 1$& 0   &HD~81146  &0.35\\ 
            &2014-12-03& $1238\times 1$& 0   &HD~81146  &0.28\\ 
PG~1211+143  &2014-03-14& $294\times 3$& 0  &HD~113996 &0.34\\ 
            &2014-06-18& $943\times 1$& 0   &HD~109511 &0.24\\ 
PG~1229+204  &2014-06-09& $1238 \times 1$& 0   &HD~111067 &0.35\\ 
            &2014-06-20& $1238 \times 1$& 0   &HD~111067 &0.42\\ 
PG~1411+442  &2014-05-30& $1238 \times 1$& 0   &HD~128902 &0.42\\ 
            &2014-05-31& $1238 \times 1$& 0   &HD~128902 &0.36\\ 
UGC~5101     &2014-01-06& $1242\times 1$& 90  &HD~79354 & 0.33 &
6\\ 
\hline
\end{tabular}

{\it Notes}. The references listed in the last column indicate
previous works where the galaxies have been presented. 1. Alonso-Herrero et al. (2014). 2. Ramos Almeida et
al. (2014b). 3. Pereira-Santaella et al. (2015). 4. Garc\'{\i}a-Bernete
et al. (2015).  5. Mori et 
al. (2014). 6. Mart\'{\i}nez-Paredes et al. (2015). 

\end{table*}

\subsection{Spectroscopic observations and data reduction}\label{sec:spectroscopy}

We used the CanariCam low spectral resolution ($R=\lambda / \Delta
\lambda \sim 175$) $N$-band grating to obtain $\sim 7.5-13\,\mu$m
spectroscopy of the nuclear regions of 45 local AGN using the  
0.52\,arcsec  wide slit. As for the imaging data, all the
spectroscopic observations
were taken using separate OBs for the galaxy and the standard star.
Within a spectroscopy OB the observing sequence was to rotate the
detector to the requested PA of the slit, 
take an
acquisition image of the target (either galaxy or star) using 
 the Si-2 filter, then place the slit, and finally
 integrate for the requested on-source time and number of
 repetitions. We chose the PA of the slit based on the extension of
 the mid-IR emission if that information existed or otherwise along
 the major axis of the galaxy.  The
chop-nod parameters were the same as for the imaging observations. The OB of
the corresponding standard star was executed right before or after the
galaxy OB. We used the observations of the standard stars to derive
the photometric calibration, the telluric correction,  the slit  
loss correction, and the FWHM of the
observations. 

Table~\ref{table:spectroscopy_log} summarizes the
details of the spectroscopic observations including the date of the
observation, on-source time and number of repetitions, PA of the slit,
and the name of the standard star and measured FWHM (see below). For
reference the 
last column of this table specifies if the spectroscopic observations
of the galaxy have already been published by us. Since all the
spectroscopic observations in the ESO/GTC large program were
obtained from 2013 March onwards, they are not affected by the noise
introduced by the S1R1-CR readout mode (see Section~\ref{sec:imaging}).

We also used {\sc redcan} \citep{GonzalezMartin2013}
to reduce the CanariCam spectroscopy. The
first three steps of the data  reduction are the same as for the
imaging, then  {\sc redcan} performs 
 the two-dimensional  wavelength calibration of the galaxy and standard star
spectra using sky lines. Finally, the trace determination needed for the
extracting the one-dimensional (1D) spectra is done using the standard
star data.  The last steps of the data reduction are the spectral
extraction and the correction for slit losses for point
sources. For the final steps it is necessary to
determine whether the nuclear emission is unresolved or extended. This
analysis is done in the next section and the extraction of the 1D
spectra and analysis are presented in Section~\ref{sec:spectraanalysis}.

\begin{figure}

\hspace{0.2cm}
\resizebox{0.9\hsize}{!}{\rotatebox[]{-90}{\includegraphics{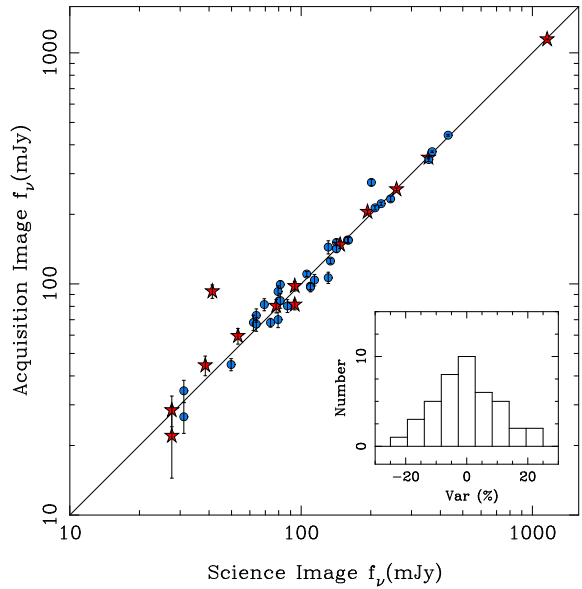}}}

\caption{Comparison between the $8.7\,\mu$m nuclear aperture
  photometry on the CanariCam
  science images and the acquisition images for the spectroscopy 
using a 2\,arcsec-diameter
aperture. The star symbols represent galaxies for which 
the CanariCam science images were affected by pattern noise or had
distorted PSF (see Table~\ref{table:imaging_log}), whereas the circles 
are the rest of the galaxies in the sample.
The solid line is not a fit but the 1:1 relation. For
clarity we only show the
errors on the acquisition image photometry, which are larger than on
the science image photometry,  as 
calculated using Equation~3 (see text for details). The inset shows
the variation computed as the difference for a given galaxy of the two flux
densities divided by the average of the two.}
\label{fig:comparisonphot}
    \end{figure}

For the spectroscopic observations we estimated the  angular
resolution of the data from
the FWHM of the standard star acquisition images. The measured values
for the CanariCam spectroscopy are given Table~\ref{table:spectroscopy_log} and
plotted in Fig.~\ref{fig:FWHM}. The median value of the FWHM of the
spectroscopic observations is 0.32\,arcsec, which is similar to that
of the imaging data.

\begin{table*}
 \centering
  \caption{Aperture photometry on the GTC/CanariCam $8.7\,\mu$m
    images. }\label{table:aperture_photometry}  

  \begin{tabular}{lcccccc}
 \hline
Galaxy & FWHM & \multicolumn{3}{c}{Flux densities (mJy)}\\
       & (arcsec) & 1arcsec & 2arcsec & Unresolved\\
 \hline
3C273               &  0.30&  $ 188.0\pm 0.6$ & $  221.7\pm1.5$ & 236\\
3C382               &  0.23&  $  60.3\pm 0.8$ & $   69.3\pm1.8$ &  74\\
3C390.3             &  0.30&   $  80.4\pm 0.6$ & $   95.9\pm1.3$ & 102\\
IRAS~08572+3915 North         &  0.28  & $ 358.9\pm 0.8$ & $  431.8\pm1.7$ & 460\\
IRAS~13197$-$1627           &  0.36  & $ 274.7\pm 0.9$ & $  355.1\pm2.0$ &  \\
IRAS~13349+2438           &  0.28  & $ 314.2\pm 0.8$ & $  356.3\pm1.9$ & 379\\
IRAS~14348$-$1447 North         & ...&   $  12.1\pm 0.8$ & $   17.7\pm1.9$ &  \\
IRAS~14348$-$1447 South   &  0.43&   $  28.2\pm 0.8$ & $   40.5\pm1.9$ &  \\
IRAS~17208$-$0014           &  0.53&   $  57.9\pm 0.5$ & $  109.7\pm1.1$ &  \\
Mrk~3                &  0.74&   $ 110.4\pm 0.8$ & $  201.2\pm1.8$ & 227\\
Mrk~231              &  0.42&   $ 802.4\pm 0.8$ & $ 1158.0\pm1.9$ &1273\\
Mrk~463 East             &  0.46&   $ 186.9\pm 1.3$ & $  258.5\pm3.0$ & 284\\
Mrk~478              &  0.28&   $  50.4\pm 0.5$ & $   62.2\pm1.2$ &  65\\
Mrk~841              &  0.30&   $  72.8\pm 0.6$ & $   87.2\pm1.3$ &  93\\
Mrk~1014             &  0.51&   $  27.8\pm 0.5$ & $   41.3\pm1.1$ &  47\\
Mrk~1066             &  0.30&   $  82.3\pm 0.7$ & $  142.2\pm1.7$ &  \\
Mrk~1073             &  0.33&   $  62.1\pm 0.6$ & $  109.8\pm1.4$ &  \\
Mrk~1210             &  0.32&   $ 158.5\pm 0.6$ & $  193.6\pm1.5$ & 210\\
Mrk~1383             &  0.46&   $  40.0\pm 0.6$ & $   53.3\pm1.3$ &  60\\
NGC~931              &  0.39&   $ 182.9\pm 0.9$ & $  243.7\pm2.0$ & 268\\
NGC~1194             &  0.33&   $ 102.4\pm 0.6$ & $  131.0\pm1.4$ & 141\\
NGC~1275             &  0.27&   $ 314.4\pm 0.6$ & $  368.6\pm1.4$ & 396\\
NGC~1320             &  0.32&   $ 113.7\pm 0.6$ & $  147.4\pm1.5$ & 158\\
NGC~2273             &  0.30&   $  96.1\pm 0.7$ & $  133.9\pm1.7$ & 142\\
NGC~2992             & 0.40&   $  51.3\pm 0.5$ & $   78.0\pm1.1$ &  \\
NGC~3227             &  0.32&   $ 150.1\pm 0.5$ & $  208.8\pm1.2$ & 225\\
NGC~4051             &  0.36&   $ 138.4\pm 0.6$ & $  206.2\pm1.4$ & 222\\
NGC~4253             &  0.31&   $ 128.8\pm 0.7$ & $  158.8\pm1.5$ & 169\\
NGC~4388             &  0.37&   $  76.3\pm 0.7$ & $  108.6\pm1.6$ & 119\\
NGC~4419             &  0.38&   $  45.9\pm 0.7$ & $   73.7\pm1.5$ &  \\
NGC~4569             &  0.36&   $  28.7\pm 0.5$ & $   49.8\pm1.1$ &  \\
NGC~5347             &  0.29&   $  91.9\pm 0.6$ & $  114.2\pm1.4$ & 121\\
NGC~5548             &  0.29&   $ 120.3\pm 0.7$ & $  142.2\pm1.7$ & 151\\
NGC~5793             & ...&   $  18.2\pm 0.7$ & $   47.6\pm1.6$ &  \\
NGC~6240 South            &  0.38&   $ 101.1\pm 0.6$ & $  160.2\pm1.3$ & 178\\
NGC~7465             &  0.31&   $  25.9\pm 0.5$ & $   38.5\pm1.1$ &  41\\
OQ~208             &  0.36&   $  58.5\pm 0.6$ & $   79.5\pm1.4$ &  86\\
PG0804+761          &  0.31&   $  69.1\pm 0.6$ & $   93.8\pm1.4$ & 108\\
PG0844+349          &  0.35&   $  20.4\pm 0.6$ & $   27.6\pm1.4$ &  32\\
PG1211+143          &  0.29&   $  67.9\pm 0.5$ & $   81.3\pm1.1$ &  86\\
PG1229+204          &  0.27&   $  25.8\pm 0.5$ & $   31.1\pm1.2$ &  33\\
PG1411+442          &  0.27&   $  55.0\pm 0.6$ & $   64.0\pm1.3$ &  67\\
UGC~5101             &  0.57&   $  54.0\pm 0.4$ & $  105.8\pm0.8$ &  \\

 \hline
\end{tabular}

  {\it Notes.---} The apertures are diameters. The quoted errors are calculated
    according to Equation~3 and do not include the $\sim 10\%$ uncertainty associated with
  the photometric calibration (see Section~3.3). 
\end{table*}

\subsection{Aperture Photometry}\label{sec:aperturephotometry}

We performed aperture photometry on the CanariCam $8.7\,\mu$m images using {\sc
  iraf}\footnote{IRAF is distributed by the National Optical Astronomy
  Observatory, which is operated by the Association of Universities
  for Research in Astronomy (AURA) under cooperative agreement with
  the National Science Foundation.} routines. The  
nuclear $8.7\,\mu$m flux densities (without the correction for point source
emission) measured through different apertures are given in
Table~\ref{table:aperture_photometry}.   The background in the
sky-subtracted images was measured in an annulus with an inner radius
of 30\,pixels ($2.4\,$arcsec) and a 5\,pixel width, except for the
very extended sources where we used an annulus with a radius of
45\,pixels ($3.6\,$arcsec).

We list in Table~\ref{table:aperture_photometry}  
the errors in the photometry
due to the background subtraction uncertainty, which 
are computed as follows \citep[see e.g.][]{Reach2005}. The first term is 
  associated to the annulus used for the background
subtraction:
\begin{equation}
\sigma_{\rm sky} = S_{\rm sky} N_{\rm on}/\sqrt{N_{\rm sky}}
\end{equation}

\noindent where $S_{\rm sky}$ is the standard deviation of the
sky annulus, $N_{\rm on}$ is the number of pixels in the on-source photometric
aperture, and $N_{\rm sky}$ is the number of pixels in the sky
annulus. The second term accounts for noise due to sky variations
within the on-source aperture which can be expressed as:

\begin{equation}
\sigma_{\rm sky,on} = S_{\rm sky} \sqrt{N_{\rm on}}
\end{equation}

\noindent The total error is then calculated by summing these two
terms in quadrature:

\begin{equation}
\sigma_{\rm tot} = \sqrt{\sigma^2_{\rm sky} + \sigma^2_{\rm sky, on}}
\end{equation}

\begin{figure*}
\resizebox{0.35\hsize}{!}{\rotatebox[]{-90}{\includegraphics{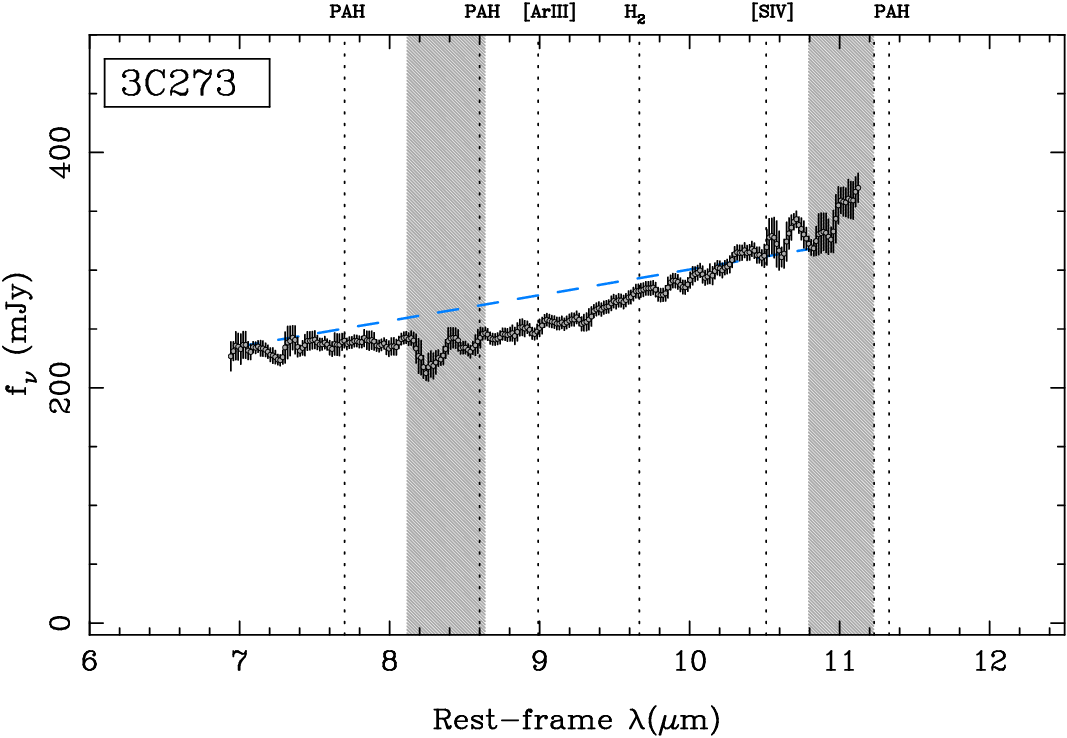}}}
\hspace{0.5cm}
\resizebox{0.35\hsize}{!}{\rotatebox[]{-90}{\includegraphics{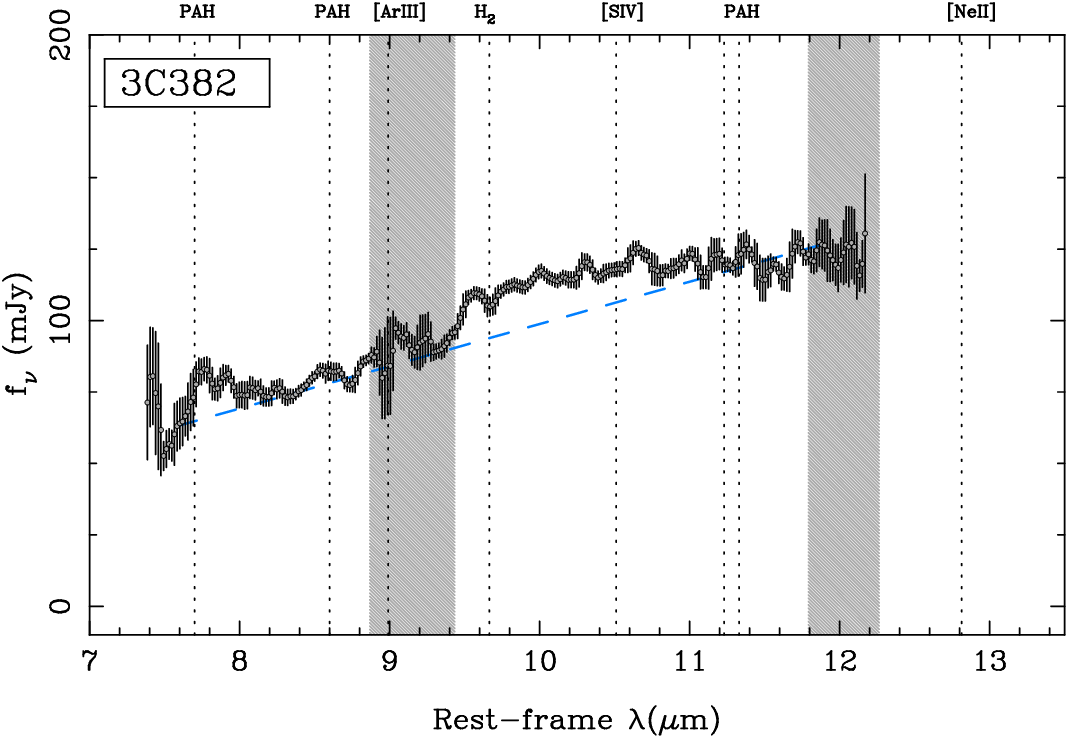}}}

\vspace{-0.8cm}
\resizebox{0.35\hsize}{!}{\rotatebox[]{-90}{\includegraphics{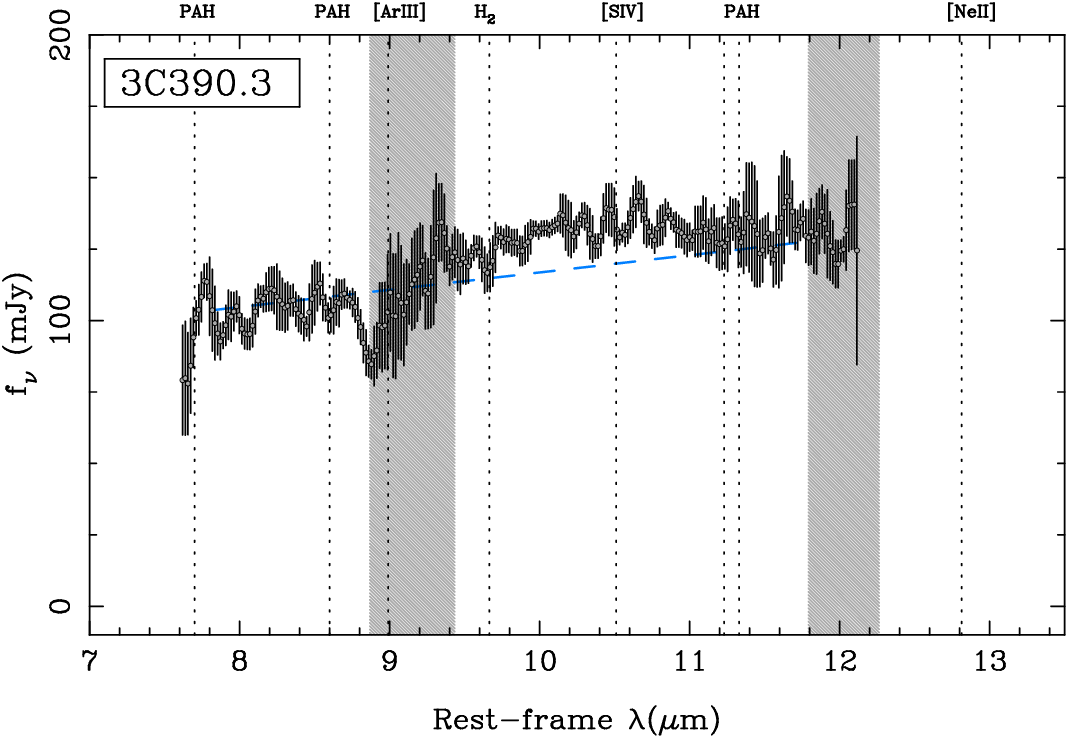}}}
\hspace{0.5cm}
\resizebox{0.35\hsize}{!}{\rotatebox[]{-90}{\includegraphics{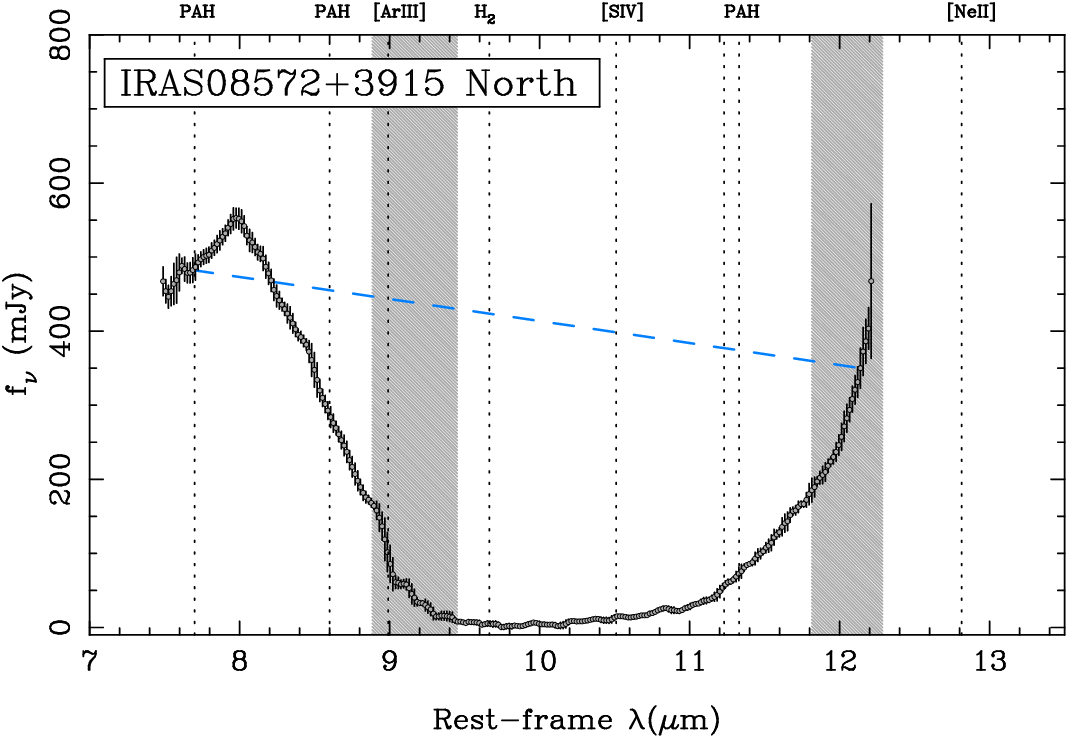}}}

\vspace{-0.8cm}
\resizebox{0.35\hsize}{!}{\rotatebox[]{-90}{\includegraphics{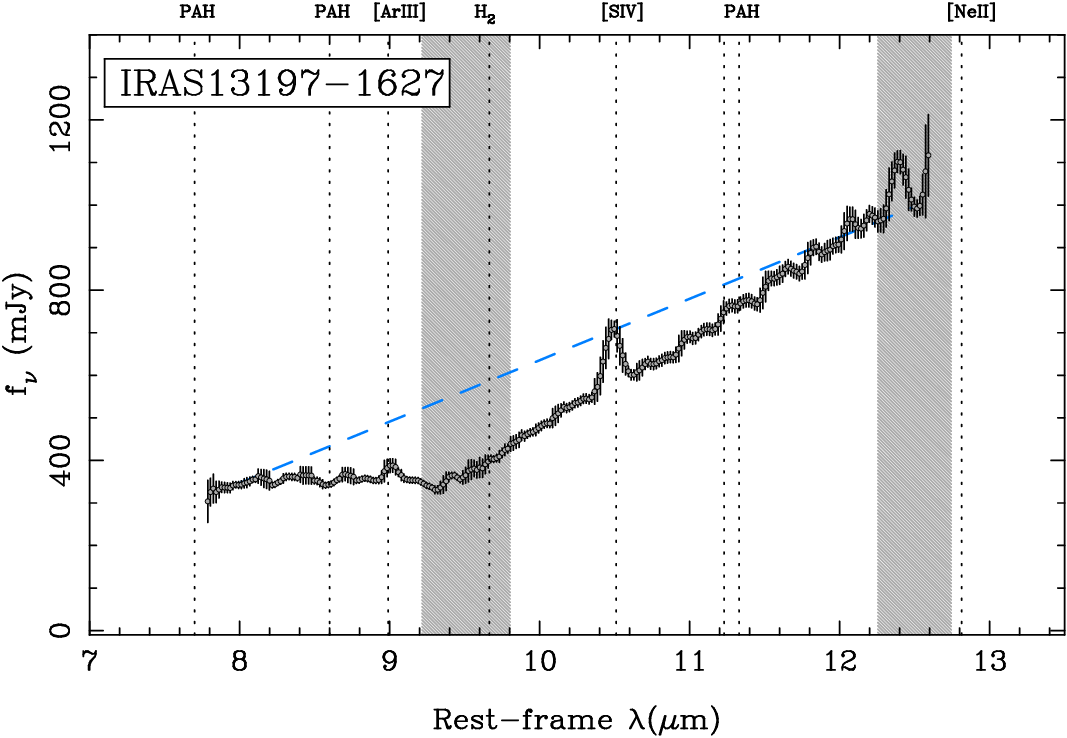}}}
\hspace{0.5cm}
\resizebox{0.35\hsize}{!}{\rotatebox[]{-90}{\includegraphics{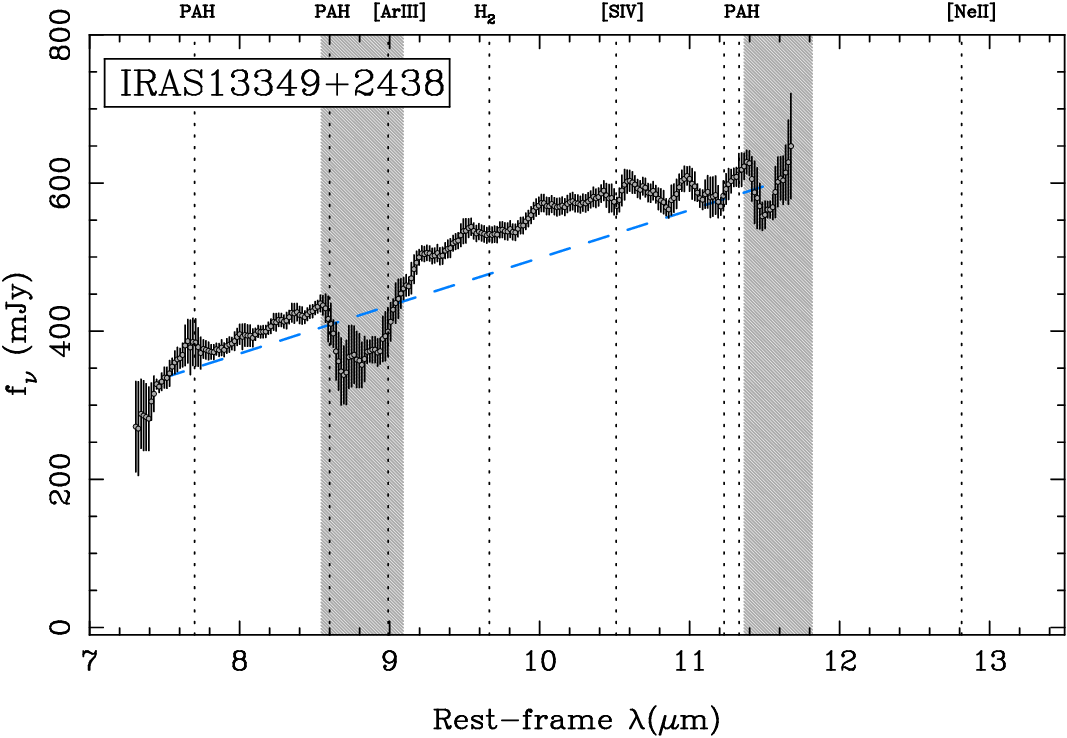}}}

\vspace{-0.8cm}
\resizebox{0.35\hsize}{!}{\rotatebox[]{-90}{\includegraphics{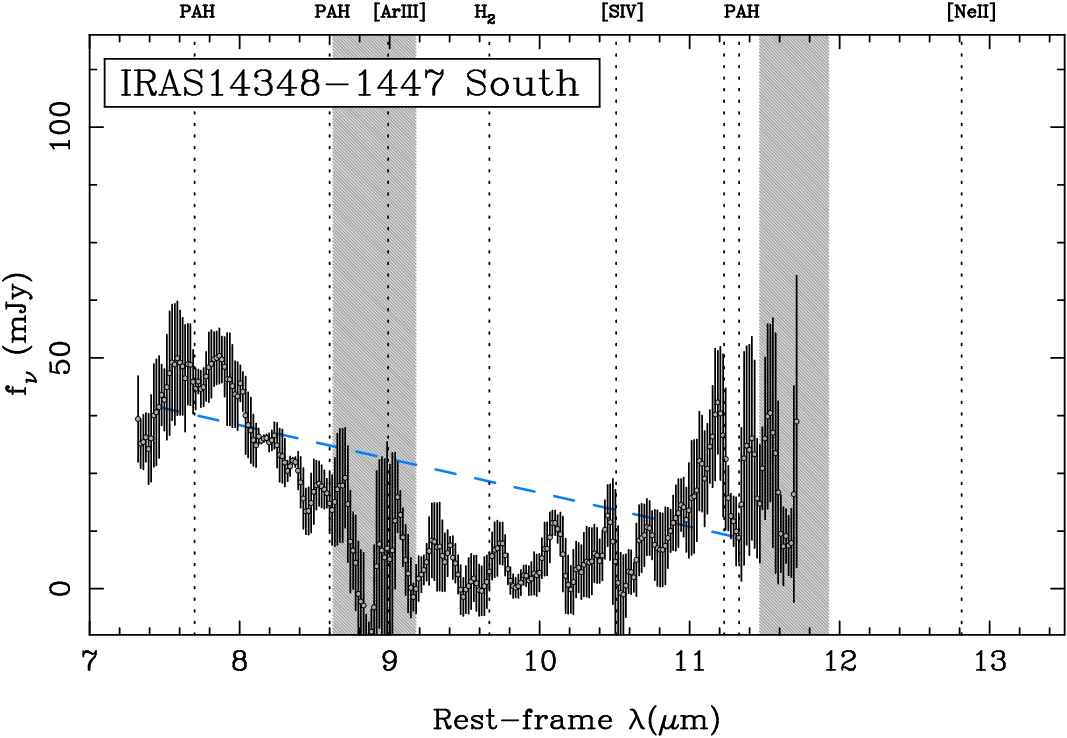}}}
\hspace{0.5cm}
\resizebox{0.35\hsize}{!}{\rotatebox[]{-90}{\includegraphics{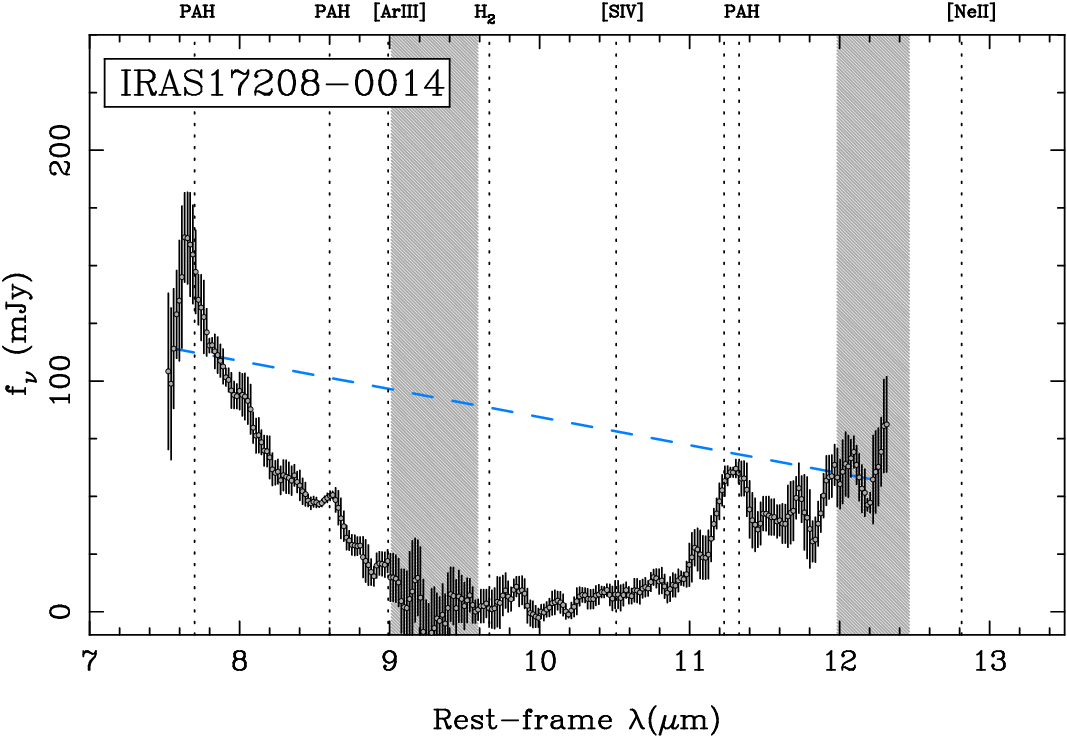}}}

\vspace{-0.8cm}
\resizebox{0.35\hsize}{!}{\rotatebox[]{-90}{\includegraphics{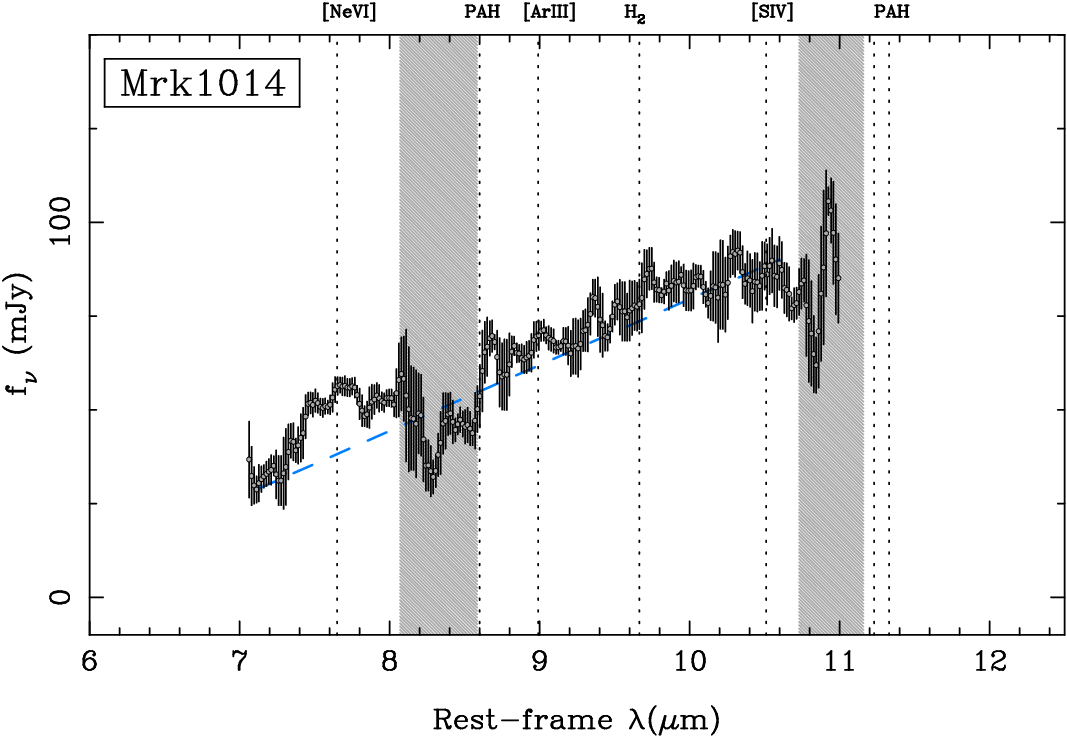}}}
\hspace{0.5cm}
\resizebox{0.35\hsize}{!}{\rotatebox[]{-90}{\includegraphics{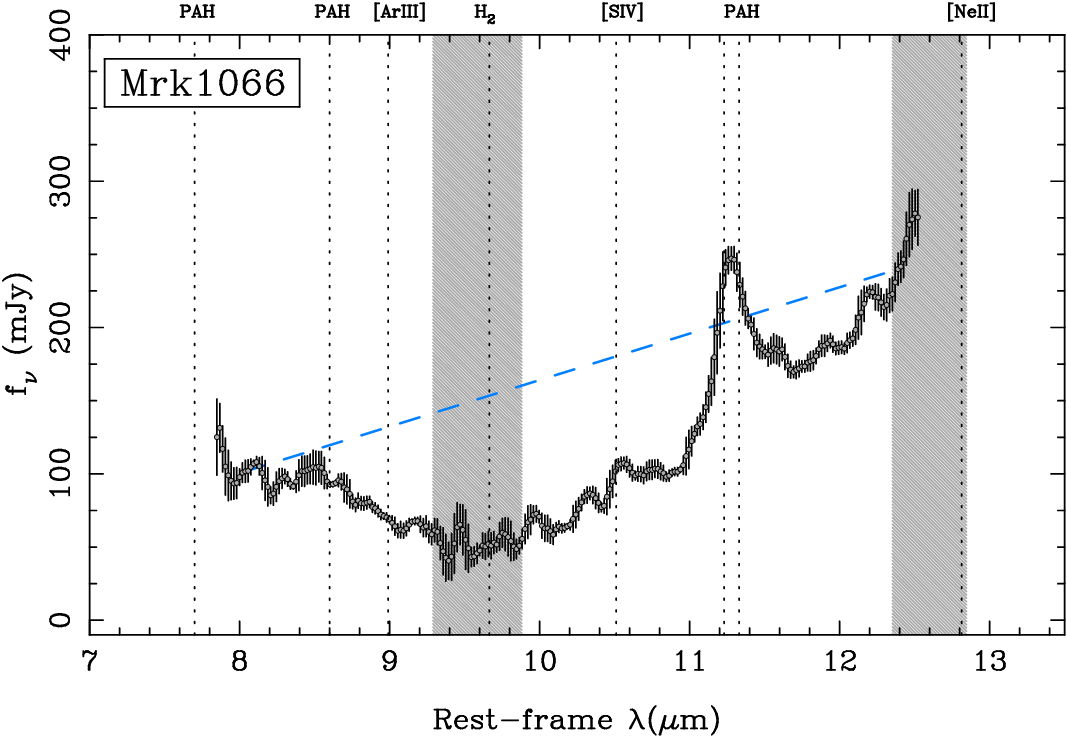}}}

\vspace{-1.1cm}
\caption{Flux-calibrated GTC/CanariCam 1D spectra (see
  Section~\ref{sec:spectraextraction}  for details) of the 45 AGN in the
  ESO/GTC sample. The dashed blue
  line is the   continuum fit (see text for details). The shaded areas represent
  spectral ranges of low atmospheric transmission and highly variable emission especially
  near $9.5\,\mu$m due the ozone  at El Roque de los
  Muchachos observatory.}
\label{fig:Spectra}
    \end{figure*}

\setcounter{figure}{4}
\begin{figure*}
\resizebox{0.35\hsize}{!}{\rotatebox[]{-90}{\includegraphics{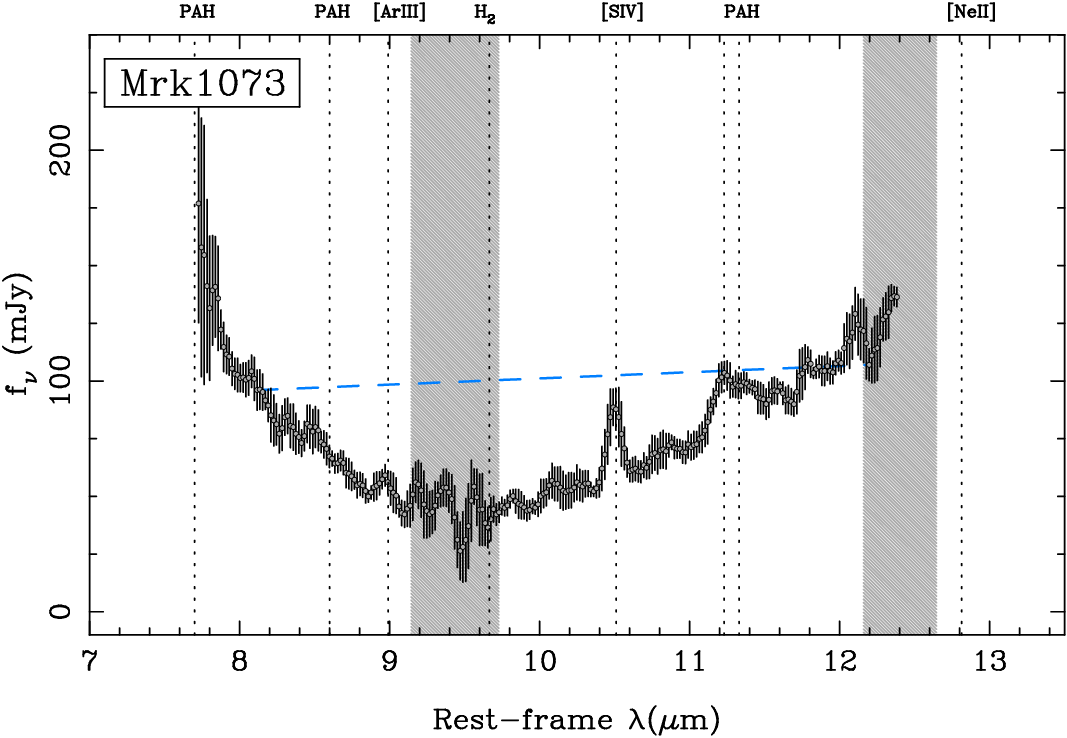}}}
\hspace{0.5cm}
\resizebox{0.35\hsize}{!}{\rotatebox[]{-90}{\includegraphics{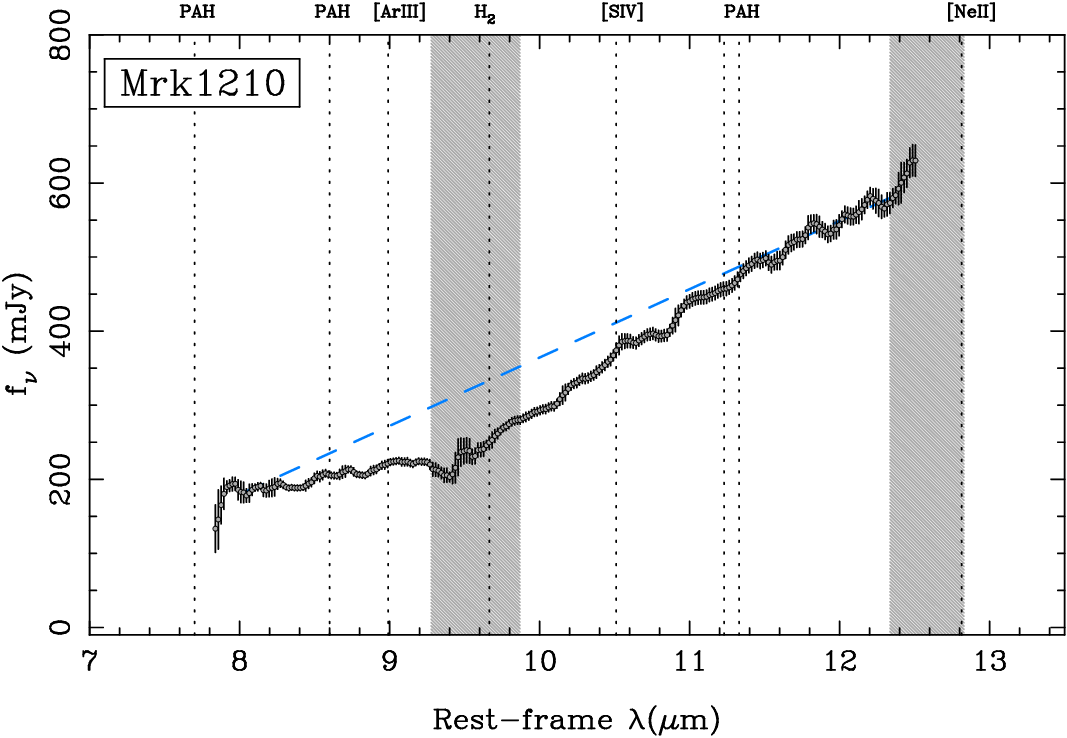}}}

\vspace{-0.8cm}
\resizebox{0.35\hsize}{!}{\rotatebox[]{-90}{\includegraphics{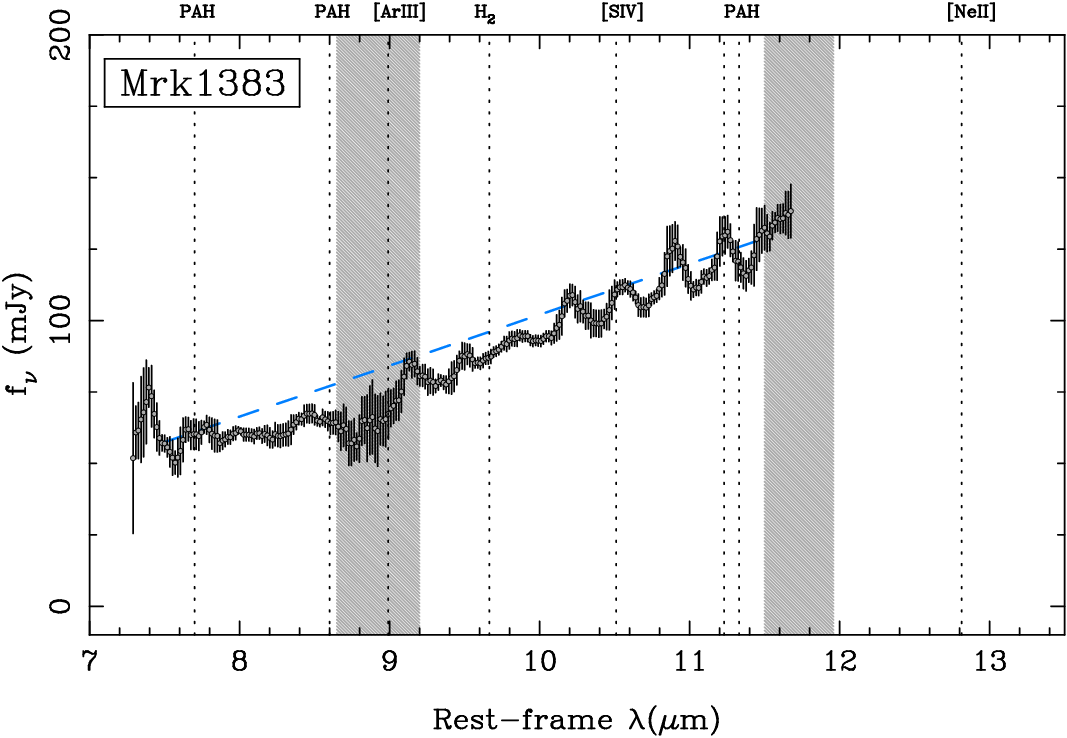}}}
\hspace{0.5cm}
\resizebox{0.35\hsize}{!}{\rotatebox[]{-90}{\includegraphics{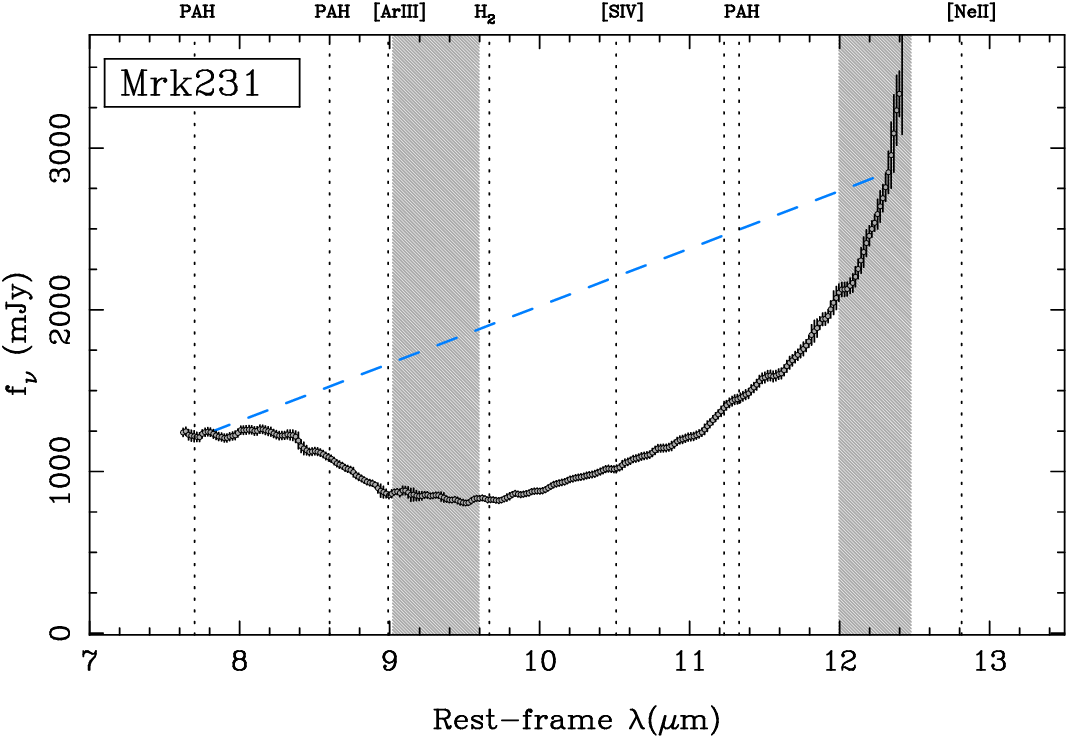}}}

\vspace{-0.8cm}
\resizebox{0.35\hsize}{!}{\rotatebox[]{-90}{\includegraphics{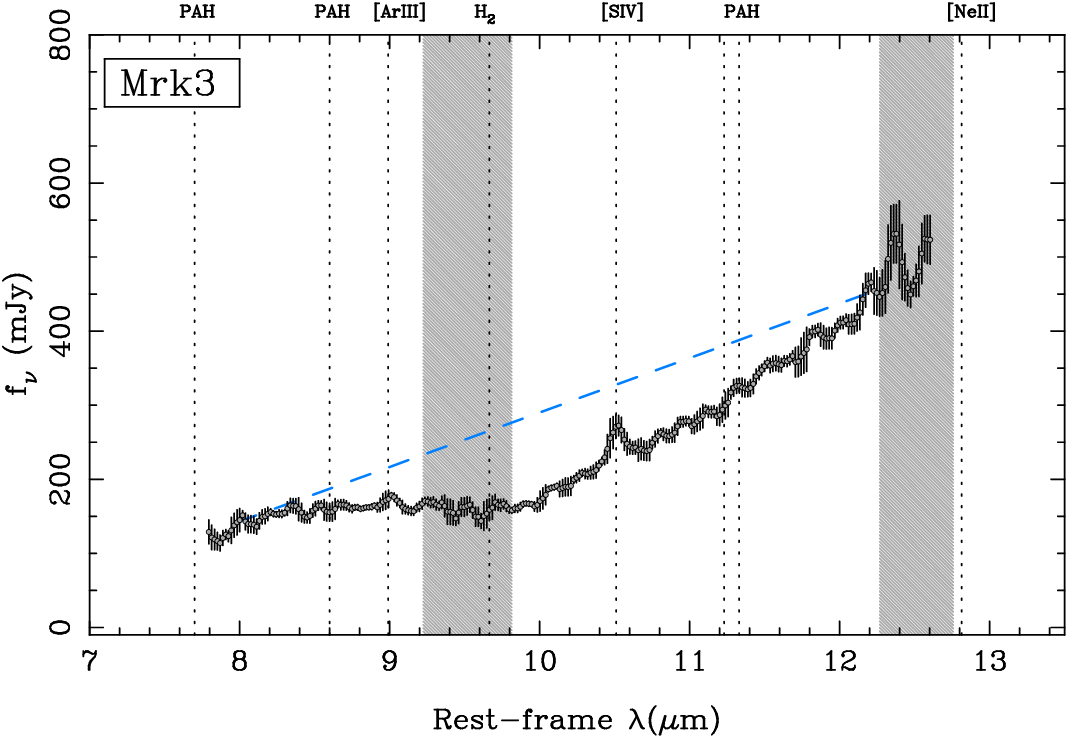}}}
\hspace{0.5cm}
\resizebox{0.35\hsize}{!}{\rotatebox[]{-90}{\includegraphics{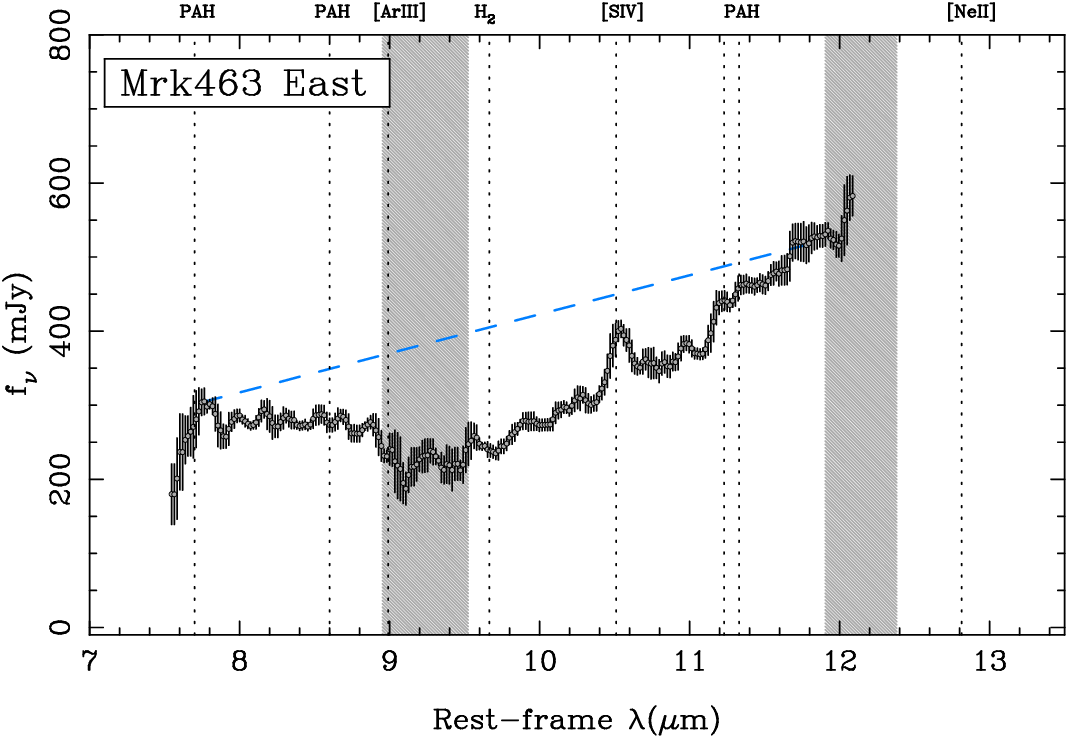}}}

\vspace{-0.8cm}
\resizebox{0.35\hsize}{!}{\rotatebox[]{-90}{\includegraphics{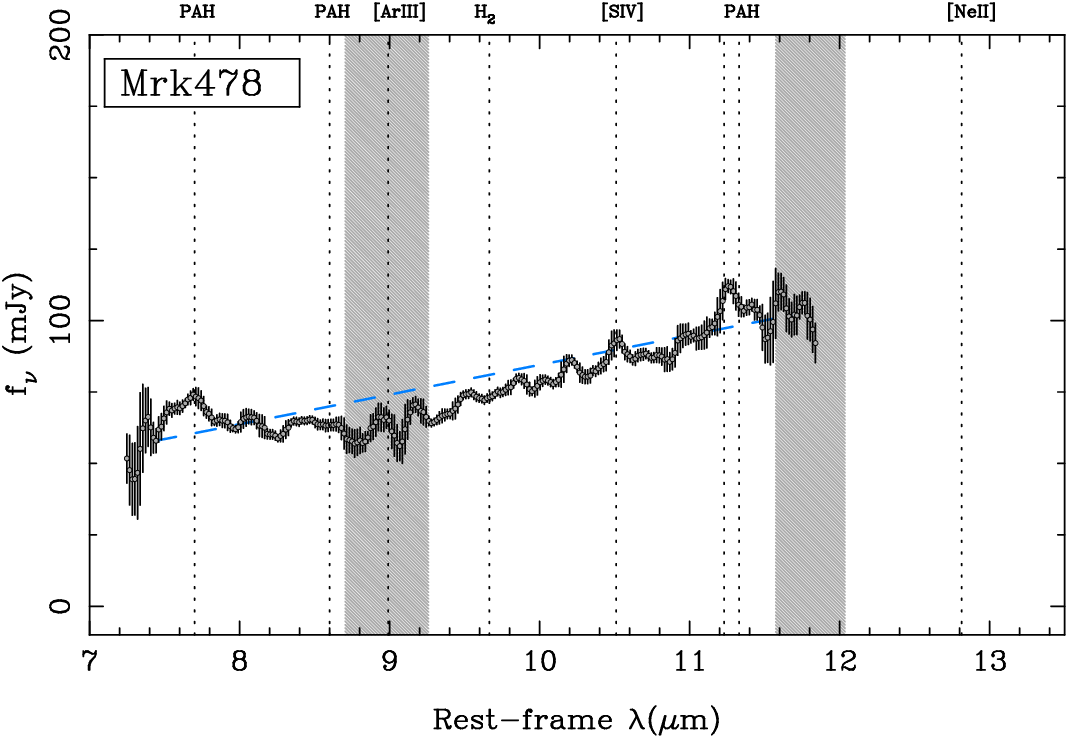}}}
\hspace{0.5cm}
\resizebox{0.35\hsize}{!}{\rotatebox[]{-90}{\includegraphics{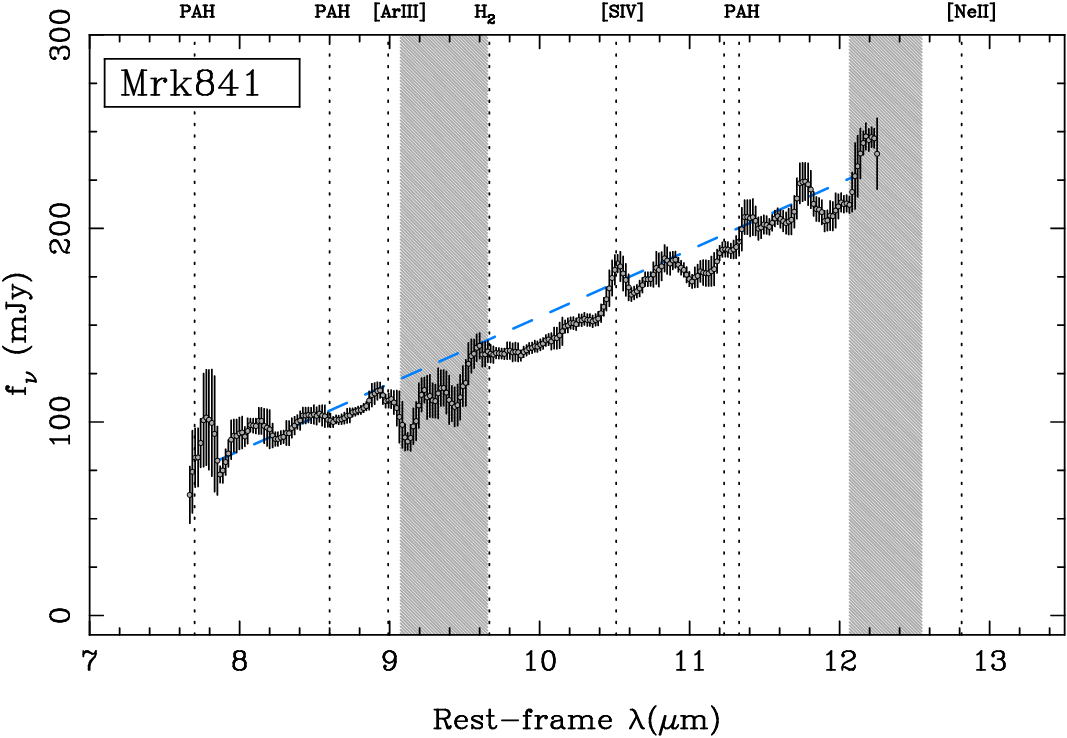}}}

\vspace{-0.8cm}
\resizebox{0.35\hsize}{!}{\rotatebox[]{-90}{\includegraphics{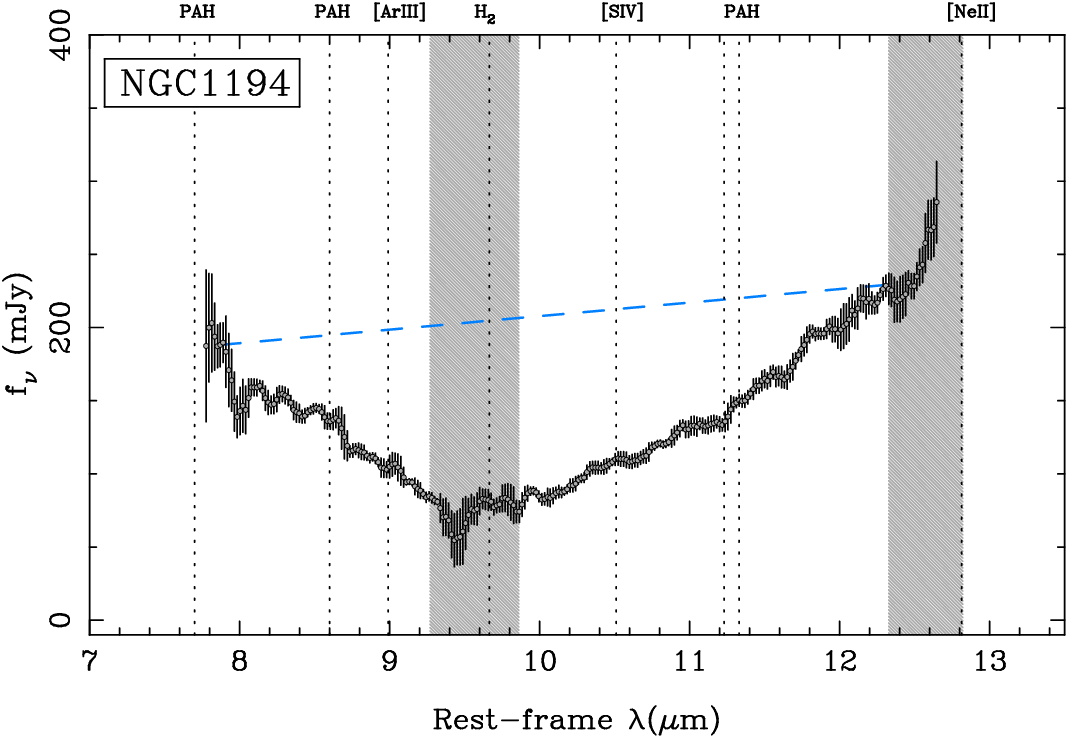}}}
\hspace{0.5cm}
\resizebox{0.35\hsize}{!}{\rotatebox[]{-90}{\includegraphics{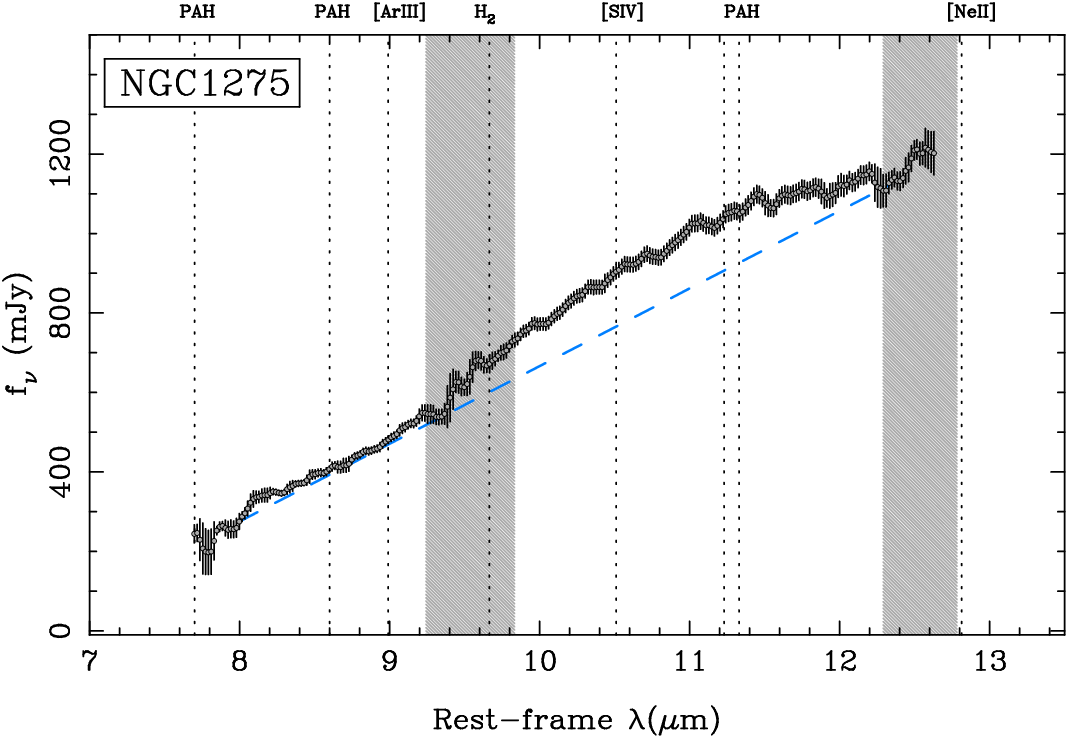}}}

\vspace{-0.9cm}
\caption{Continued.}
    \end{figure*}

\setcounter{figure}{4}
\begin{figure*}
\resizebox{0.35\hsize}{!}{\rotatebox[]{-90}{\includegraphics{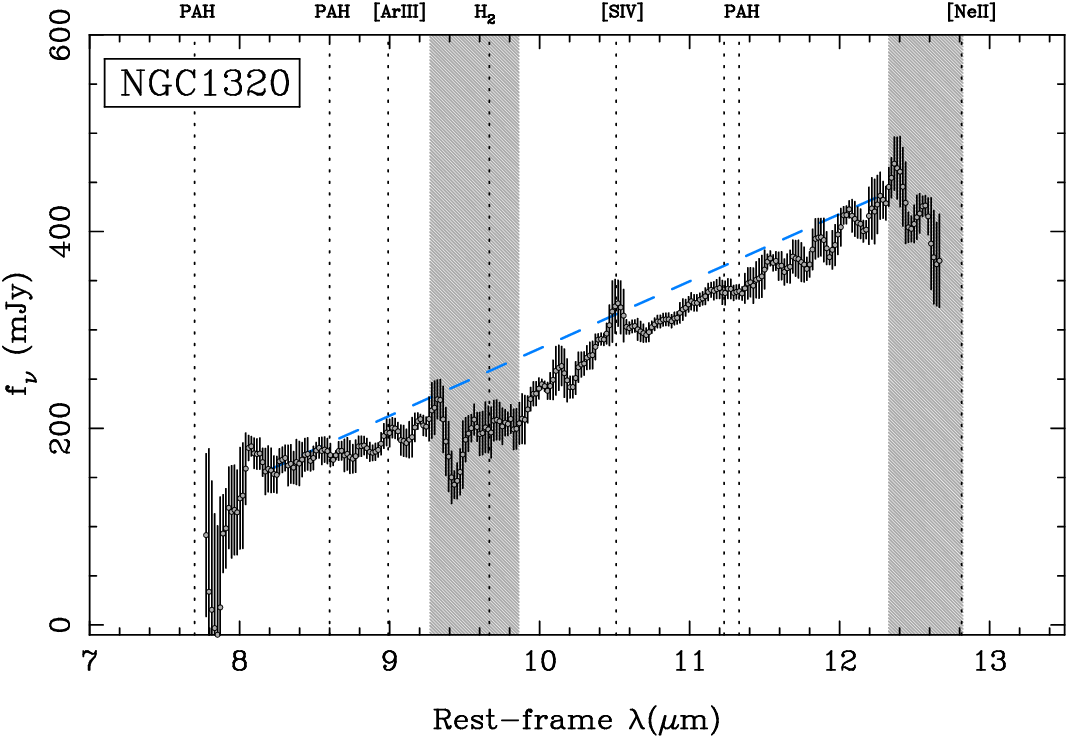}}}
\hspace{0.5cm}
\resizebox{0.35\hsize}{!}{\rotatebox[]{-90}{\includegraphics{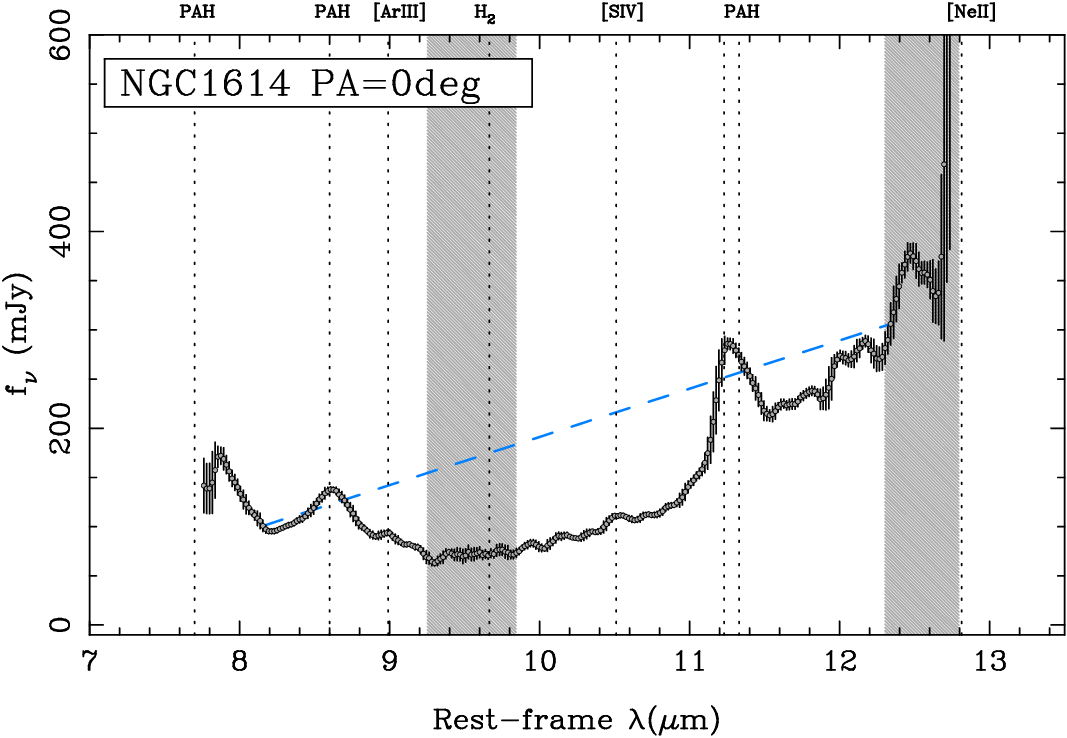}}}

\vspace{-0.8cm}
\resizebox{0.35\hsize}{!}{\rotatebox[]{-90}{\includegraphics{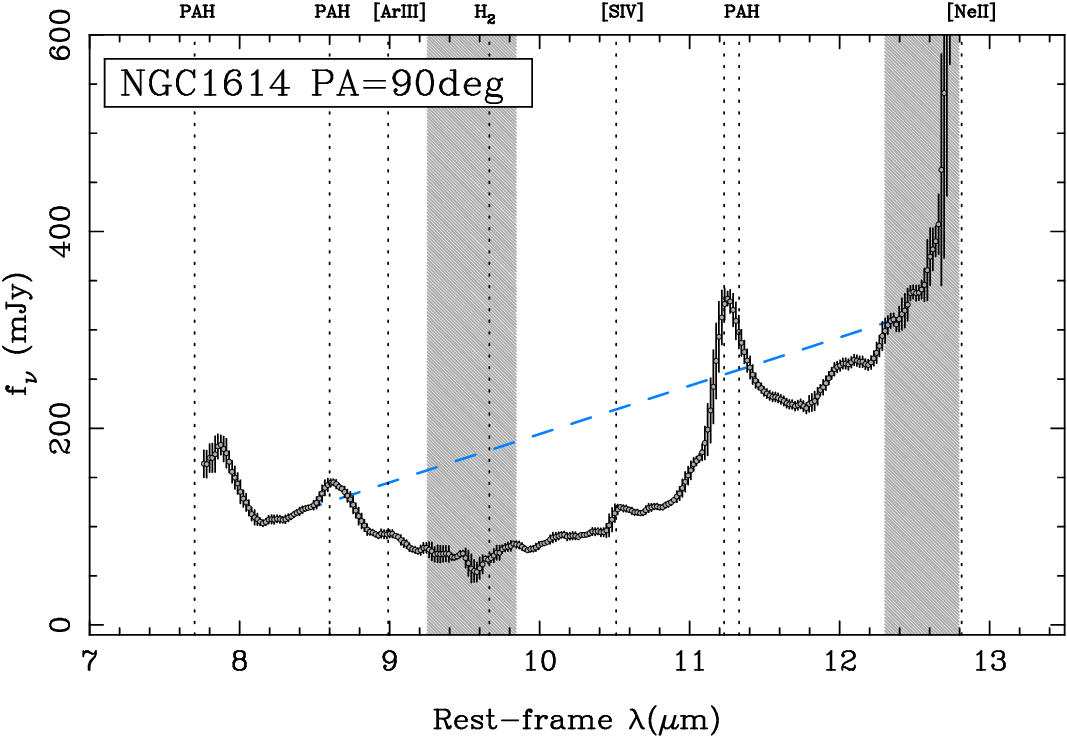}}}
\hspace{0.5cm}
\resizebox{0.35\hsize}{!}{\rotatebox[]{-90}{\includegraphics{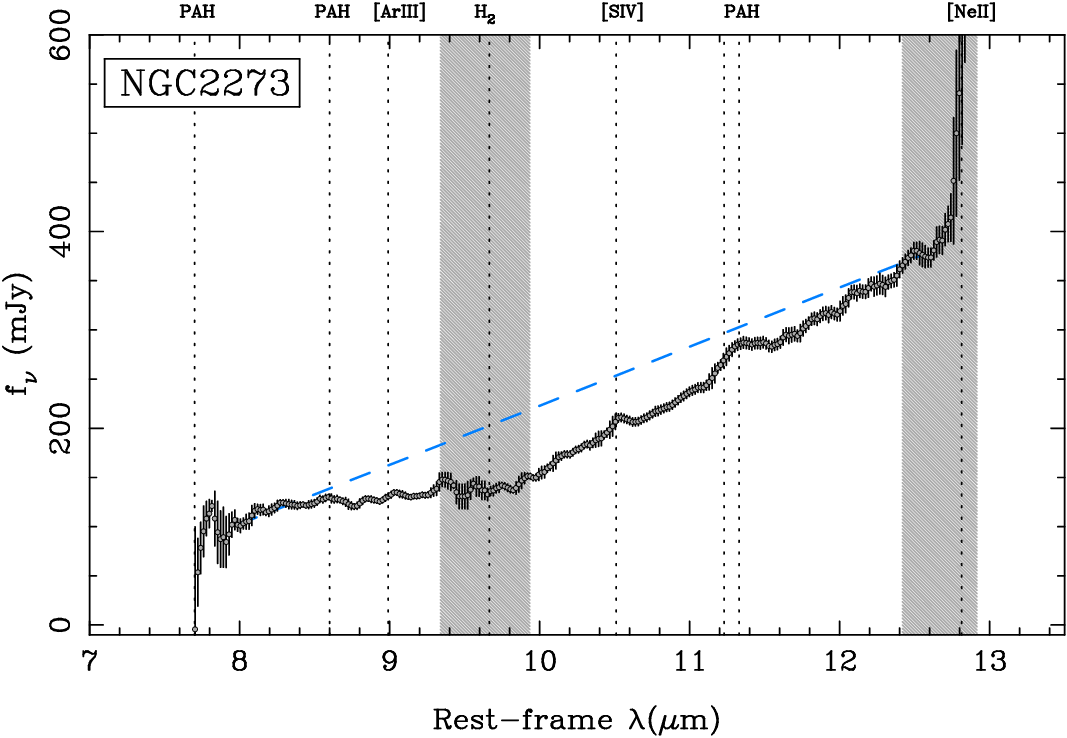}}}

\vspace{-0.8cm}
\resizebox{0.35\hsize}{!}{\rotatebox[]{-90}{\includegraphics{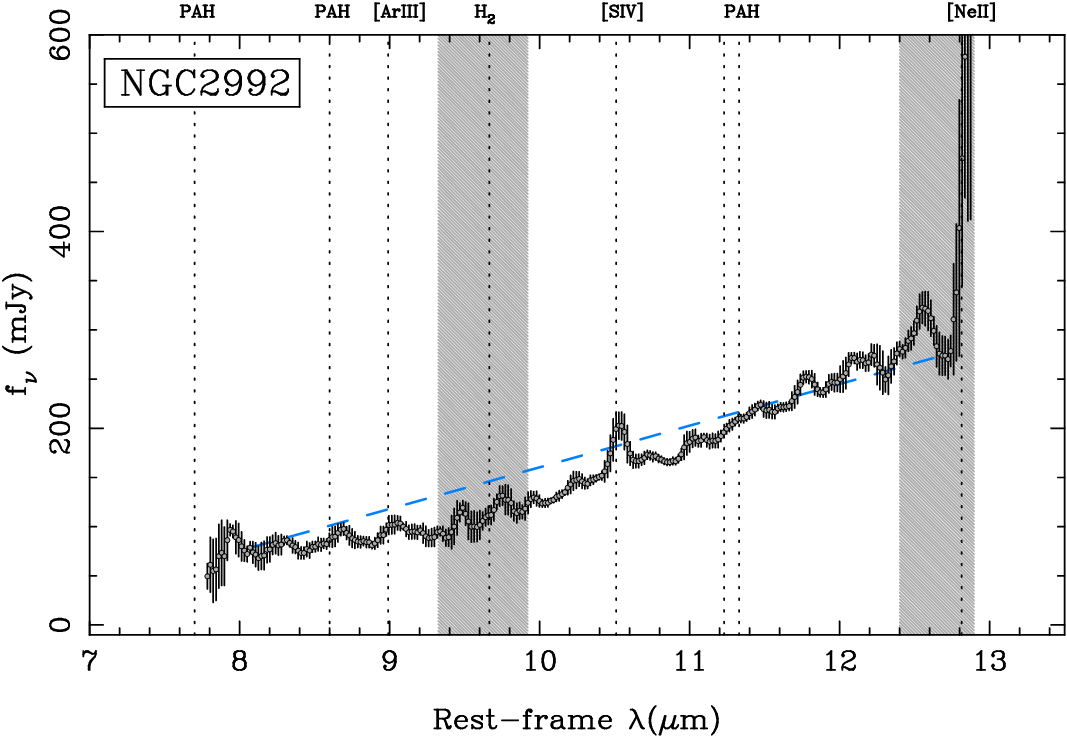}}}
\hspace{0.5cm}
\resizebox{0.35\hsize}{!}{\rotatebox[]{-90}{\includegraphics{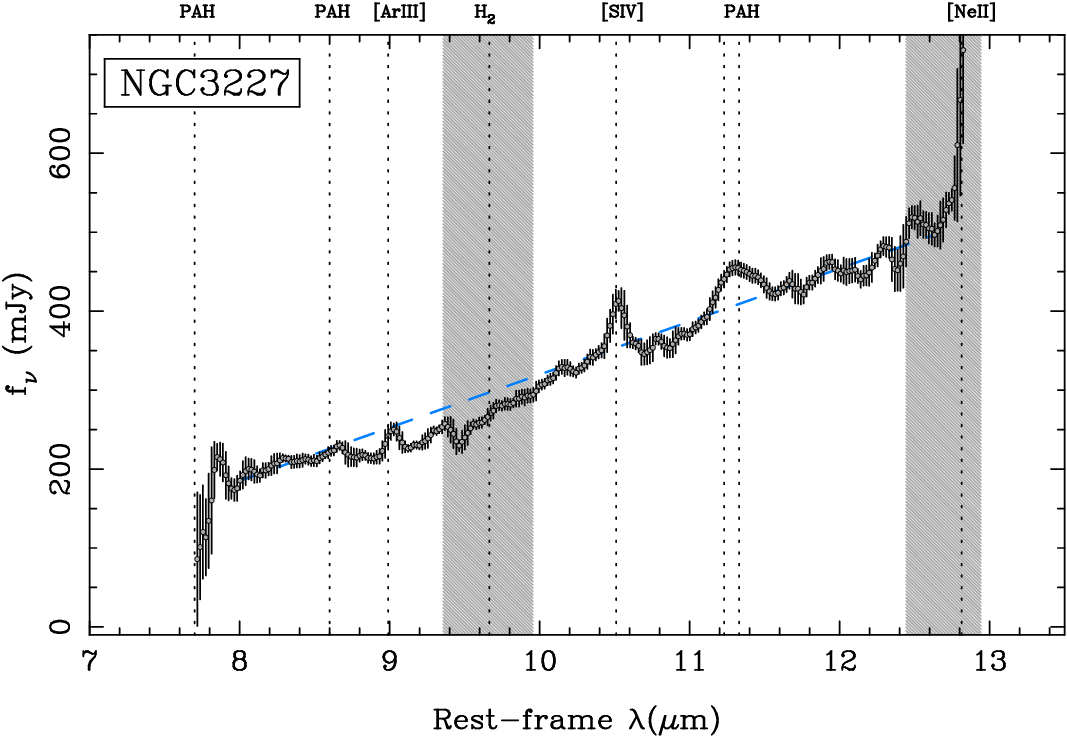}}}

\vspace{-0.8cm}
\resizebox{0.35\hsize}{!}{\rotatebox[]{-90}{\includegraphics{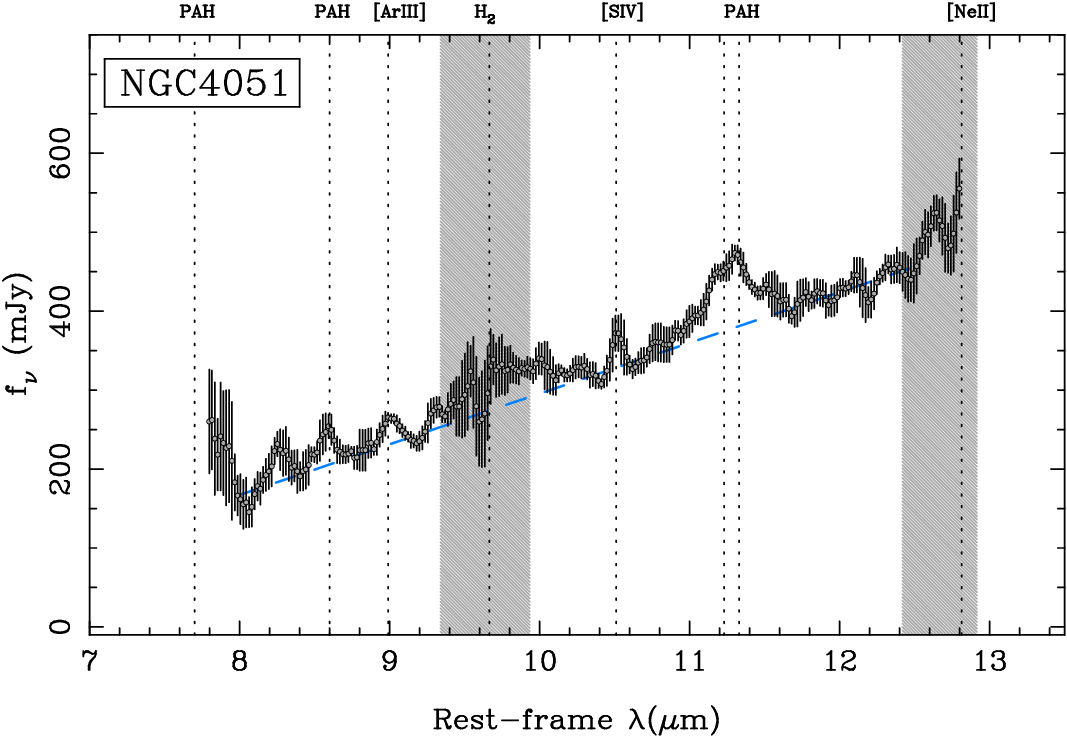}}}
\hspace{0.5cm}
\resizebox{0.35\hsize}{!}{\rotatebox[]{-90}{\includegraphics{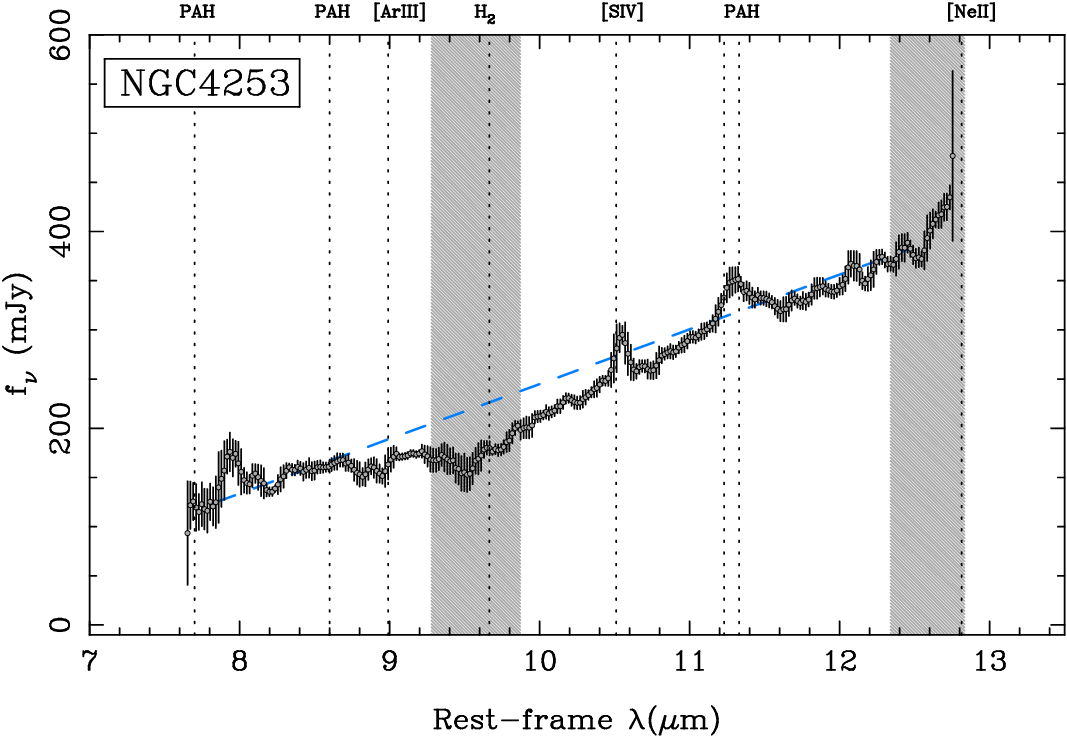}}}

\vspace{-0.8cm}
\resizebox{0.35\hsize}{!}{\rotatebox[]{-90}{\includegraphics{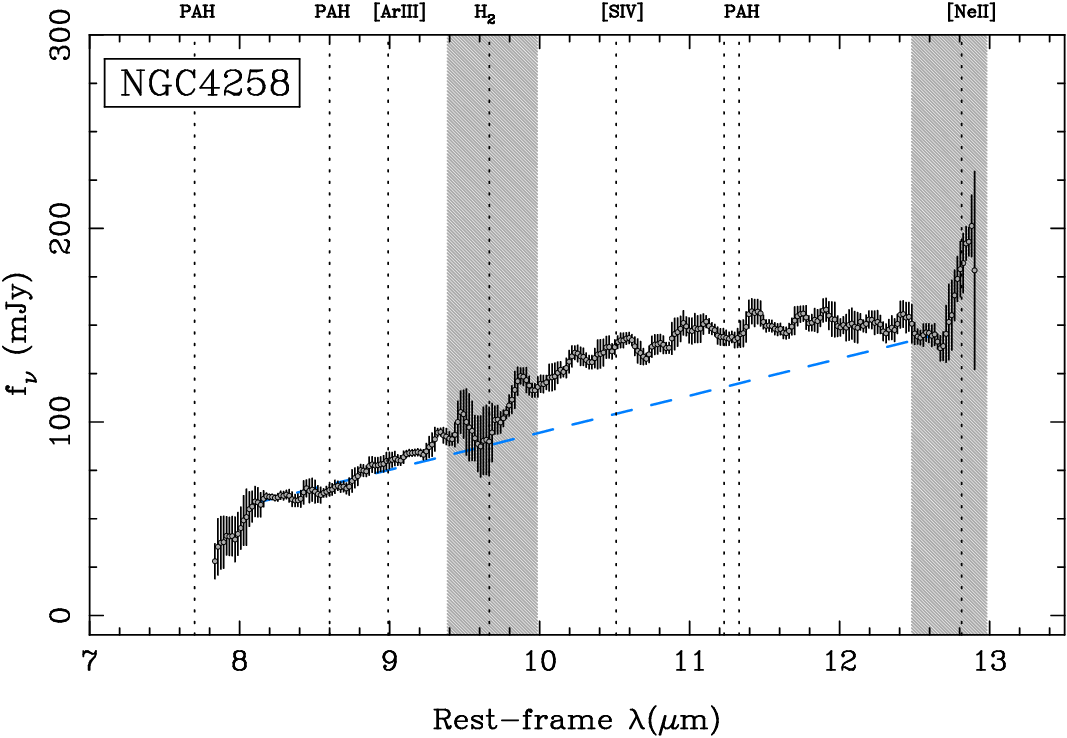}}}
\hspace{0.5cm}
\resizebox{0.35\hsize}{!}{\rotatebox[]{-90}{\includegraphics{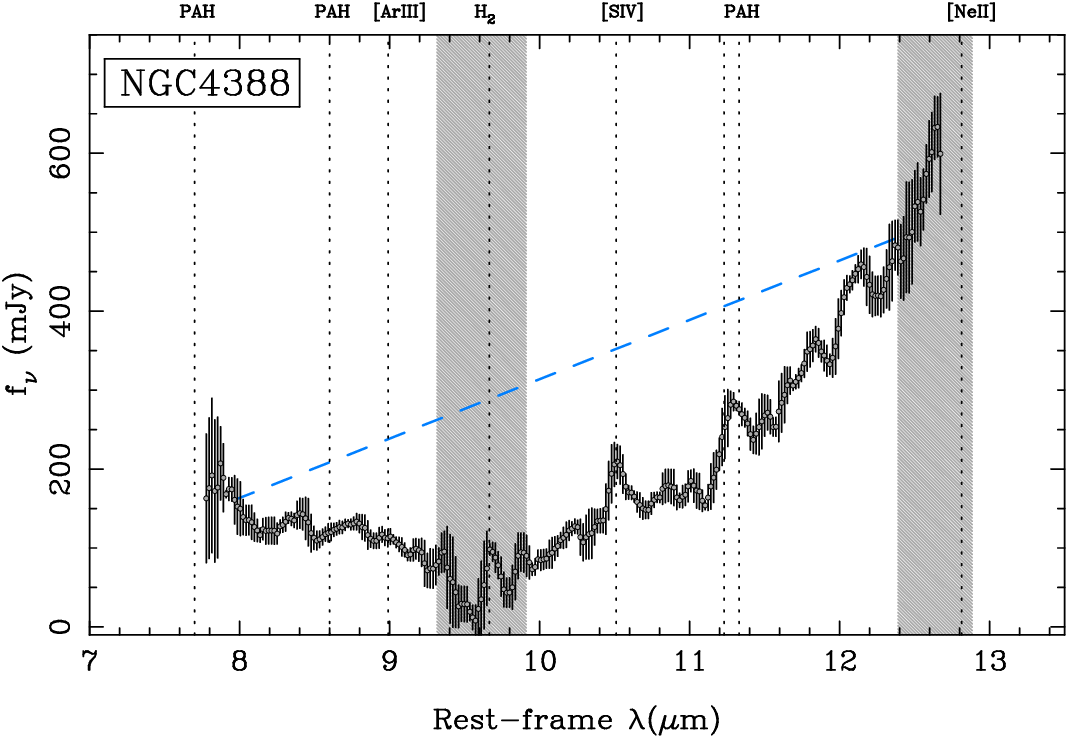}}}

\vspace{-0.9cm}
\caption{Continued.}
    \end{figure*}

\setcounter{figure}{4}
\begin{figure*}
\resizebox{0.35\hsize}{!}{\rotatebox[]{-90}{\includegraphics{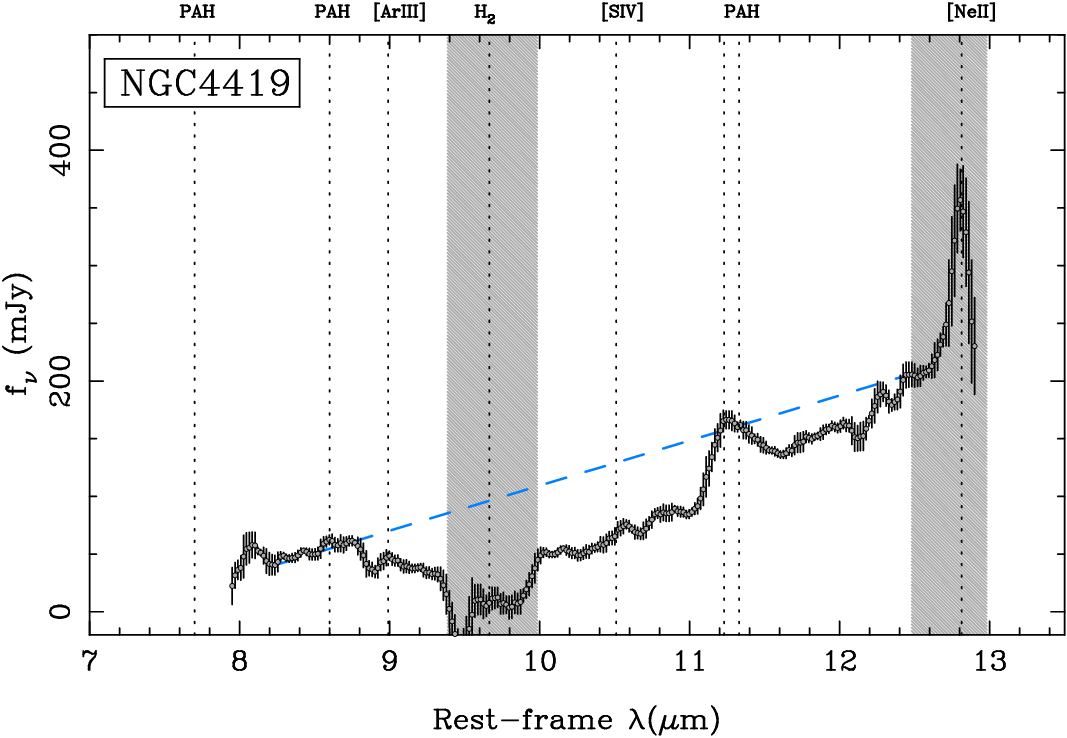}}}
\hspace{0.5cm}
\resizebox{0.35\hsize}{!}{\rotatebox[]{-90}{\includegraphics{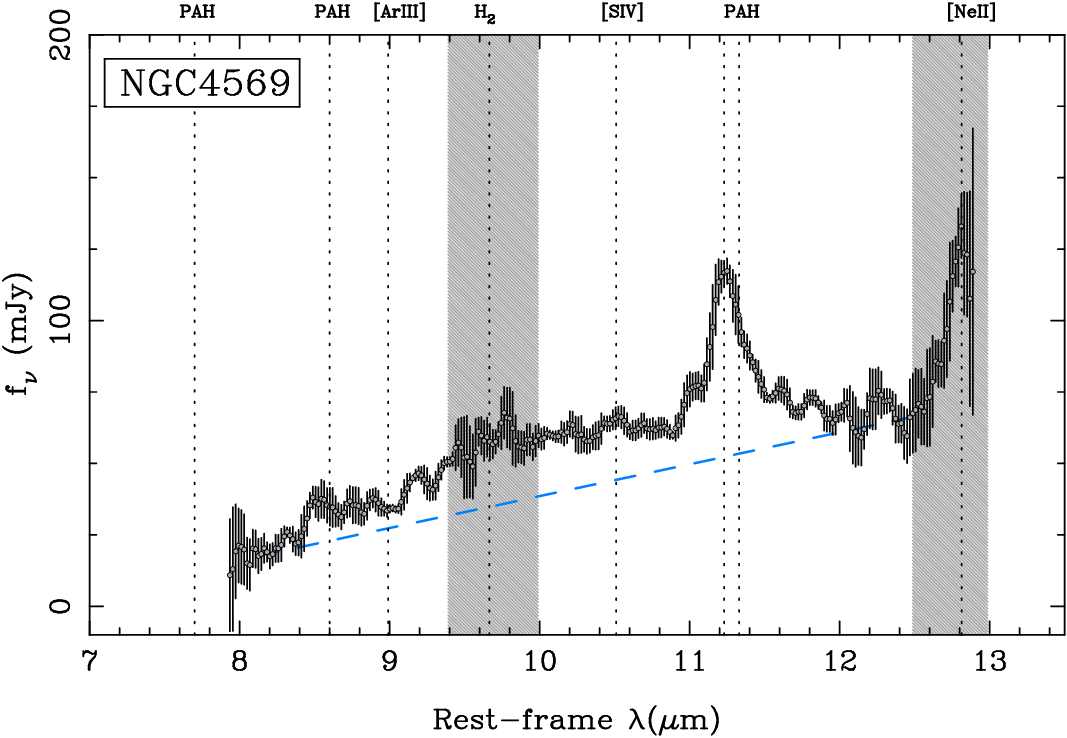}}}

\vspace{-0.8cm}
\resizebox{0.35\hsize}{!}{\rotatebox[]{-90}{\includegraphics{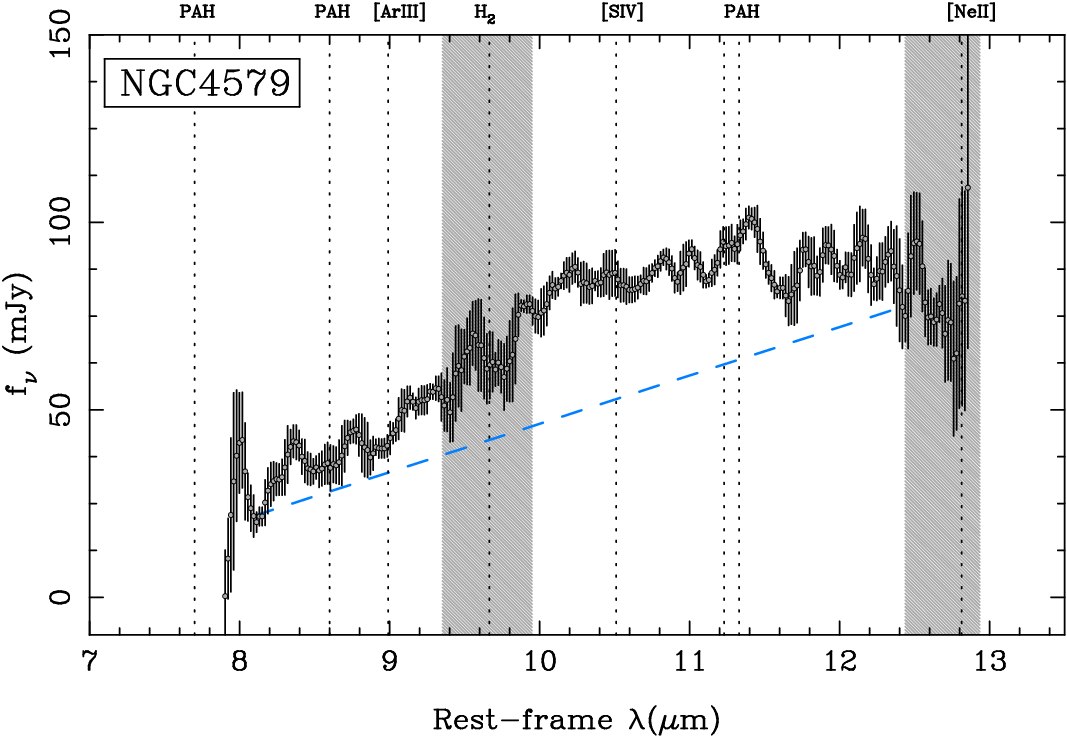}}}
\hspace{0.5cm}
\resizebox{0.35\hsize}{!}{\rotatebox[]{-90}{\includegraphics{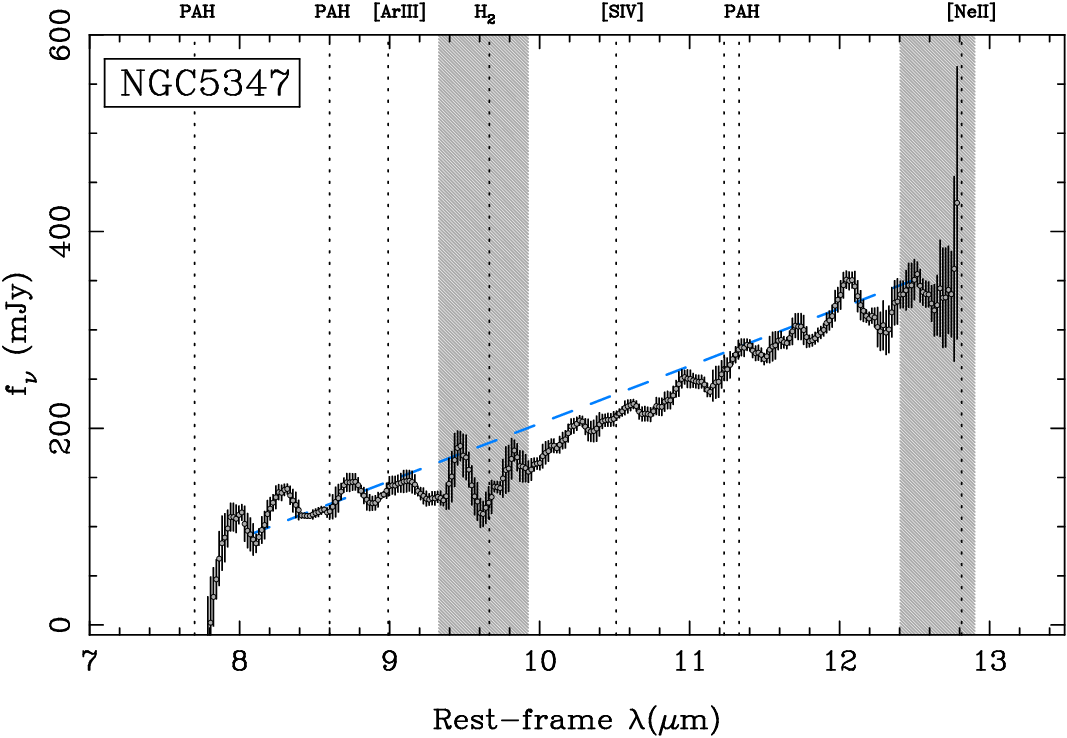}}}

\vspace{-0.8cm}
\resizebox{0.35\hsize}{!}{\rotatebox[]{-90}{\includegraphics{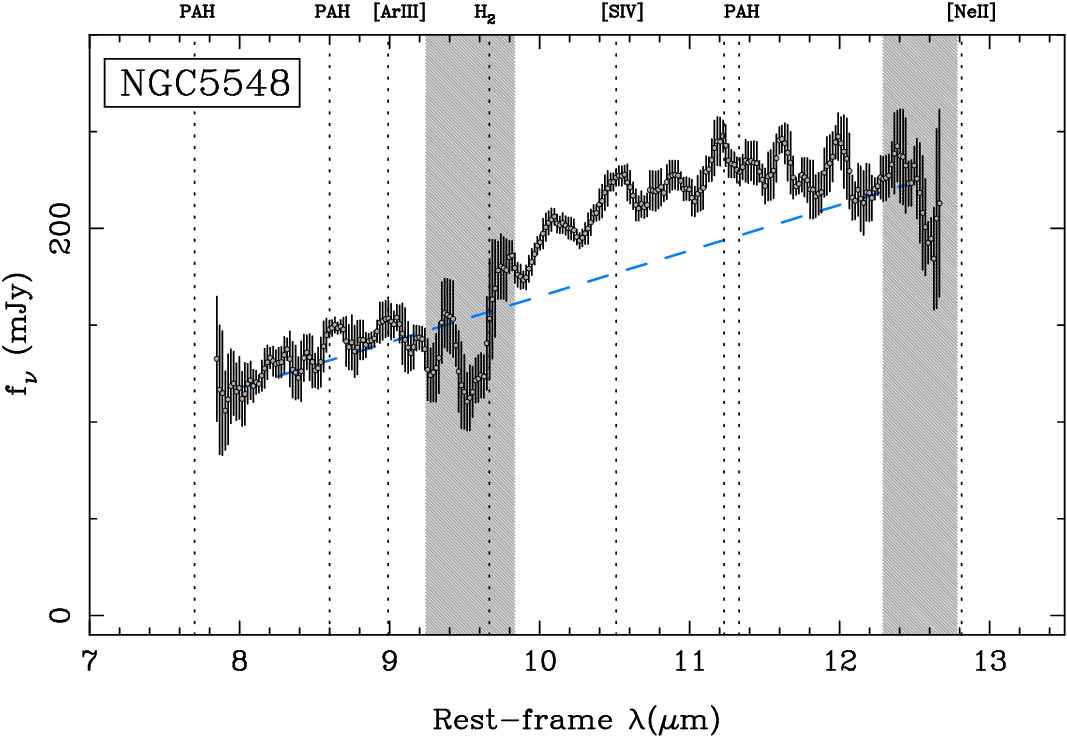}}}
\hspace{0.5cm}
\resizebox{0.35\hsize}{!}{\rotatebox[]{-90}{\includegraphics{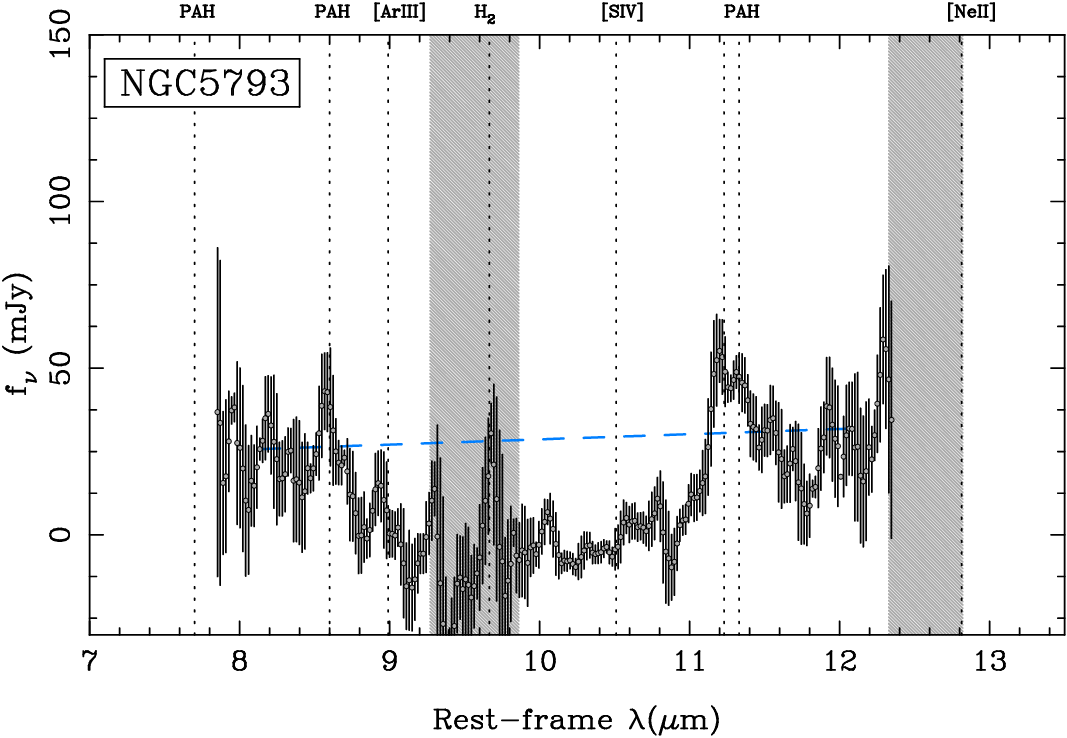}}}

\vspace{-0.8cm}
\resizebox{0.35\hsize}{!}{\rotatebox[]{-90}{\includegraphics{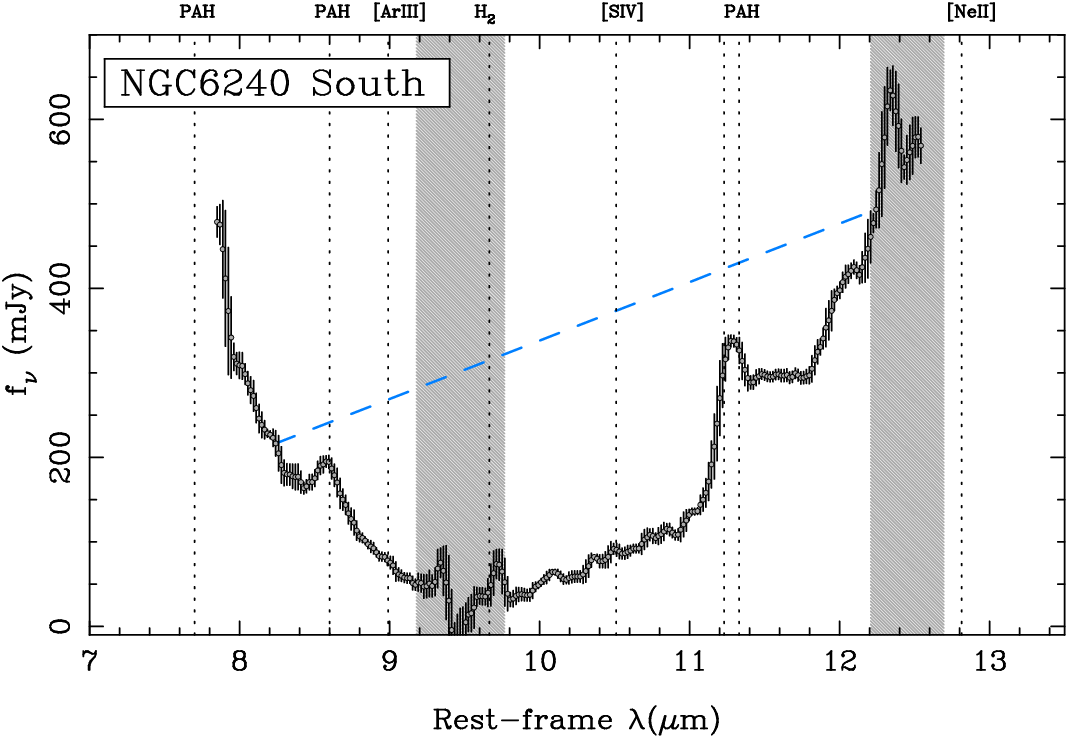}}}
\hspace{0.5cm}
\resizebox{0.35\hsize}{!}{\rotatebox[]{-90}{\includegraphics{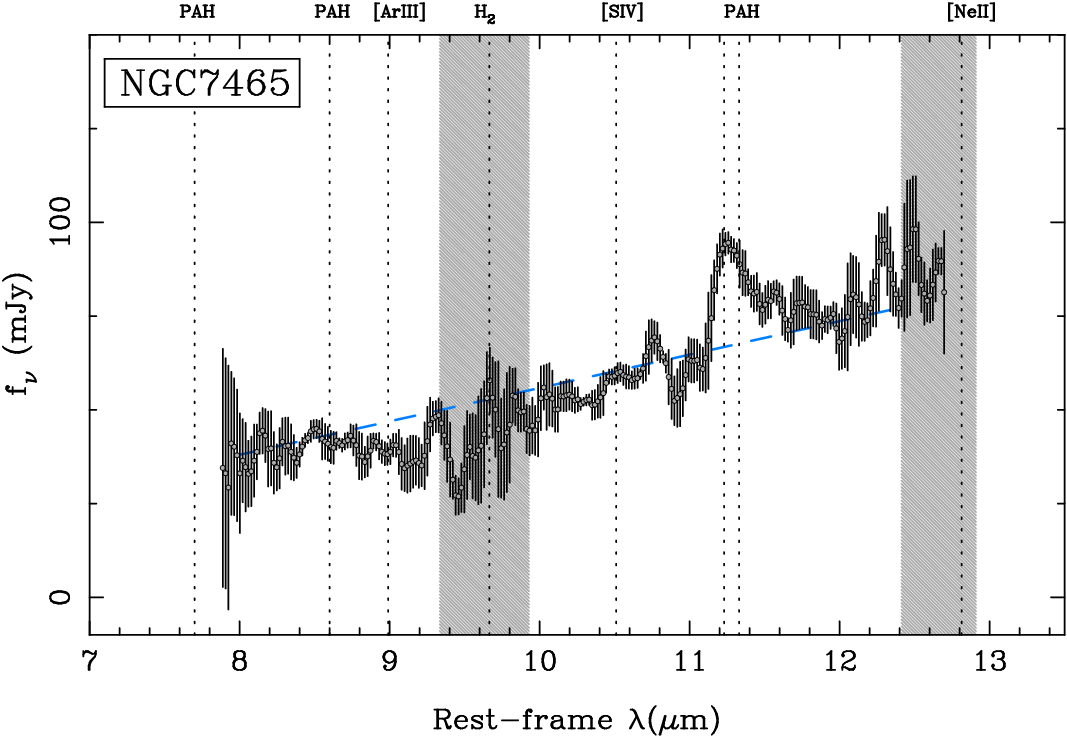}}}

\vspace{-0.8cm}
\resizebox{0.35\hsize}{!}{\rotatebox[]{-90}{\includegraphics{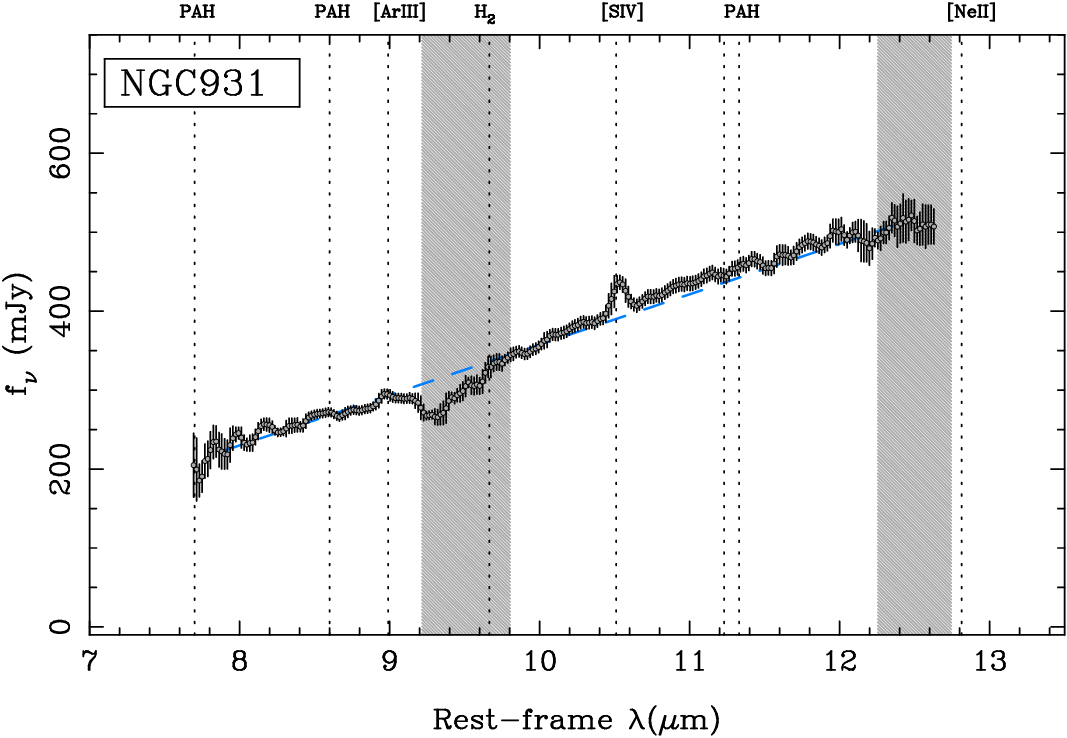}}}
\hspace{0.5cm}
\resizebox{0.35\hsize}{!}{\rotatebox[]{-90}{\includegraphics{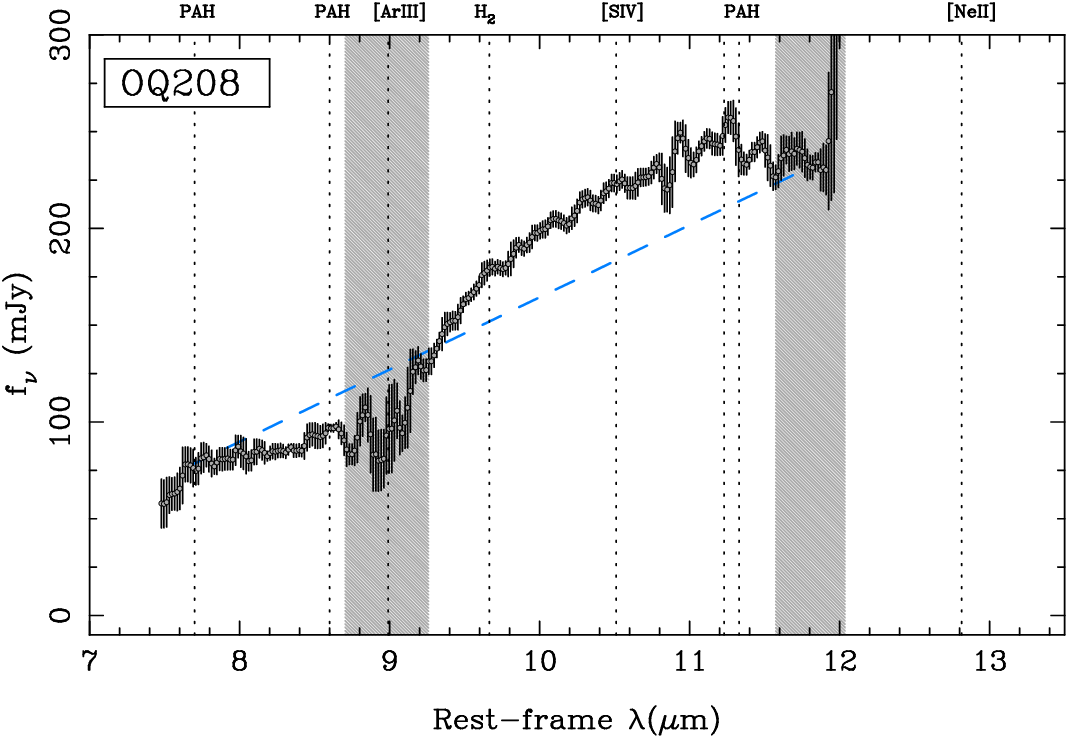}}}

\vspace{-0.9cm}
\caption{Continued.}
    \end{figure*}

\setcounter{figure}{4}
\begin{figure*}
\resizebox{0.35\hsize}{!}{\rotatebox[]{-90}{\includegraphics{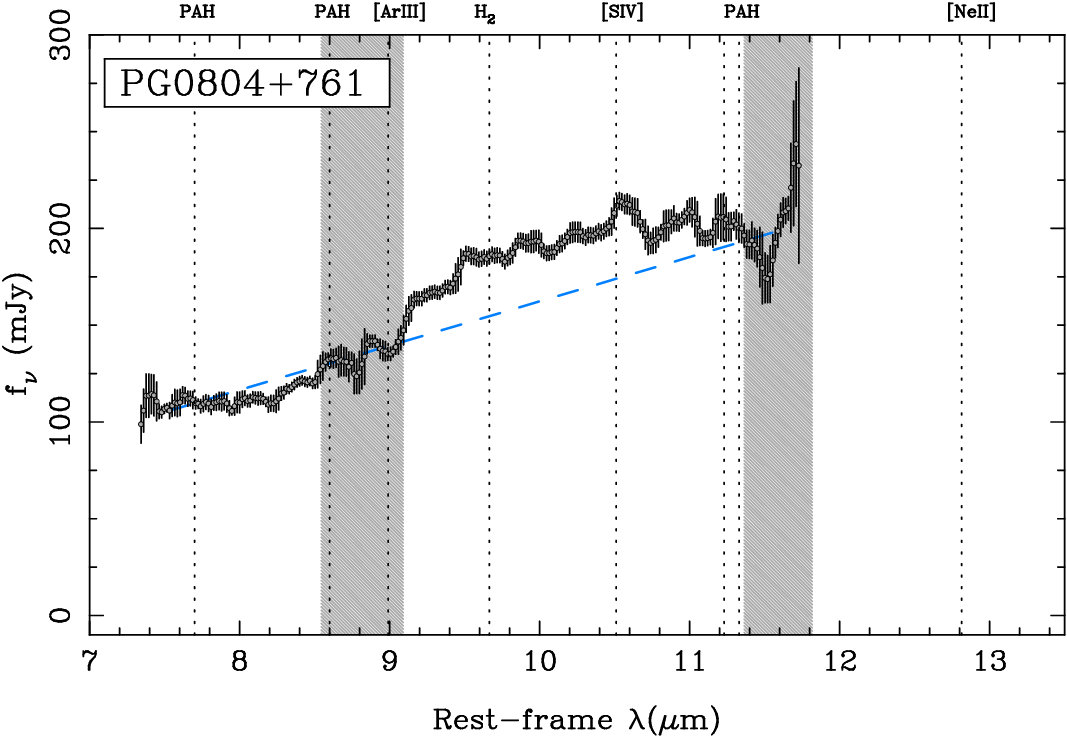}}}
\hspace{0.5cm}
\resizebox{0.35\hsize}{!}{\rotatebox[]{-90}{\includegraphics{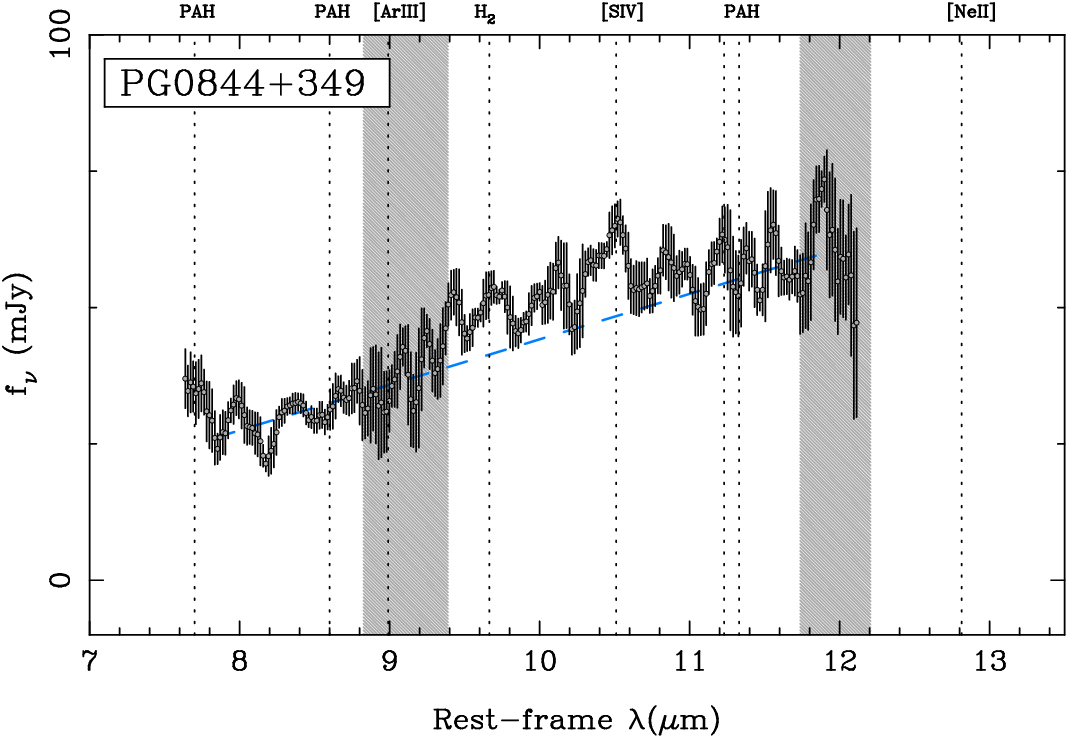}}}

\vspace{-0.8cm}
\resizebox{0.35\hsize}{!}{\rotatebox[]{-90}{\includegraphics{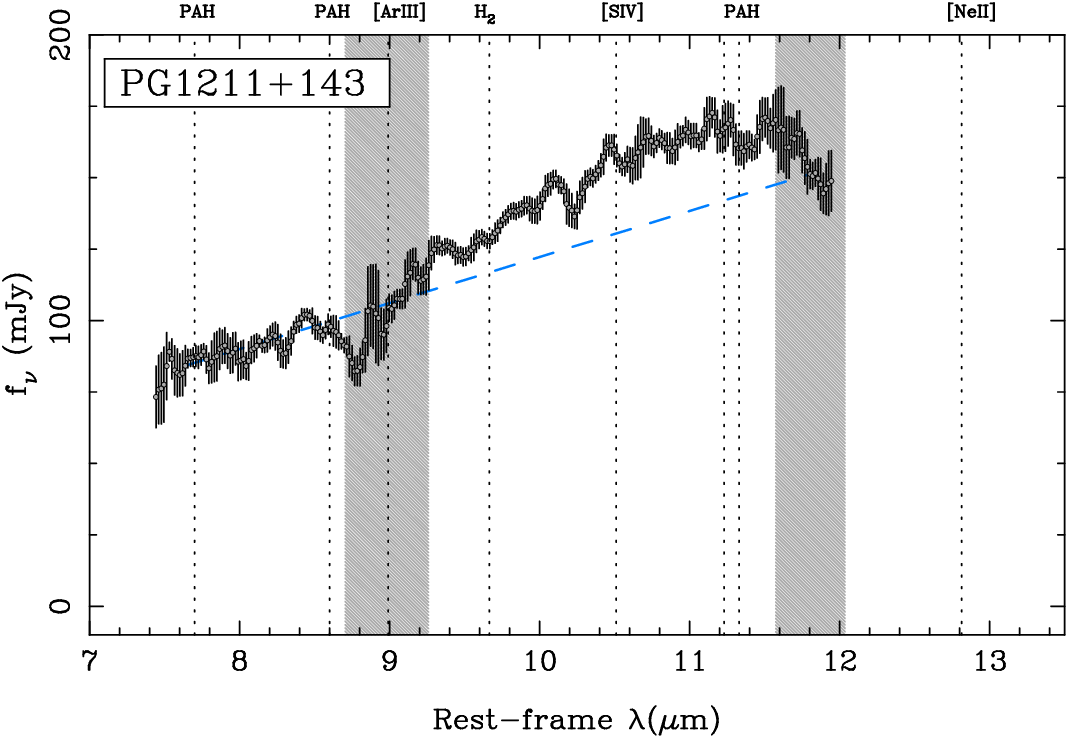}}}
\hspace{0.5cm}
\resizebox{0.35\hsize}{!}{\rotatebox[]{-90}{\includegraphics{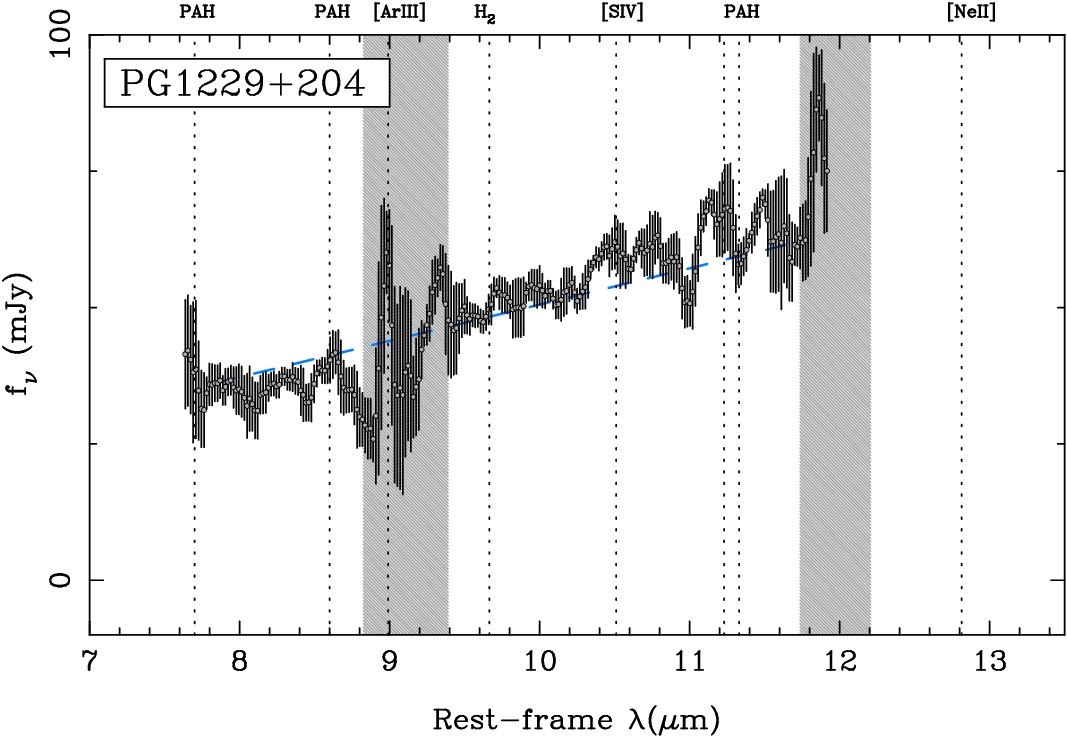}}}

\vspace{-0.8cm}
\resizebox{0.35\hsize}{!}{\rotatebox[]{-90}{\includegraphics{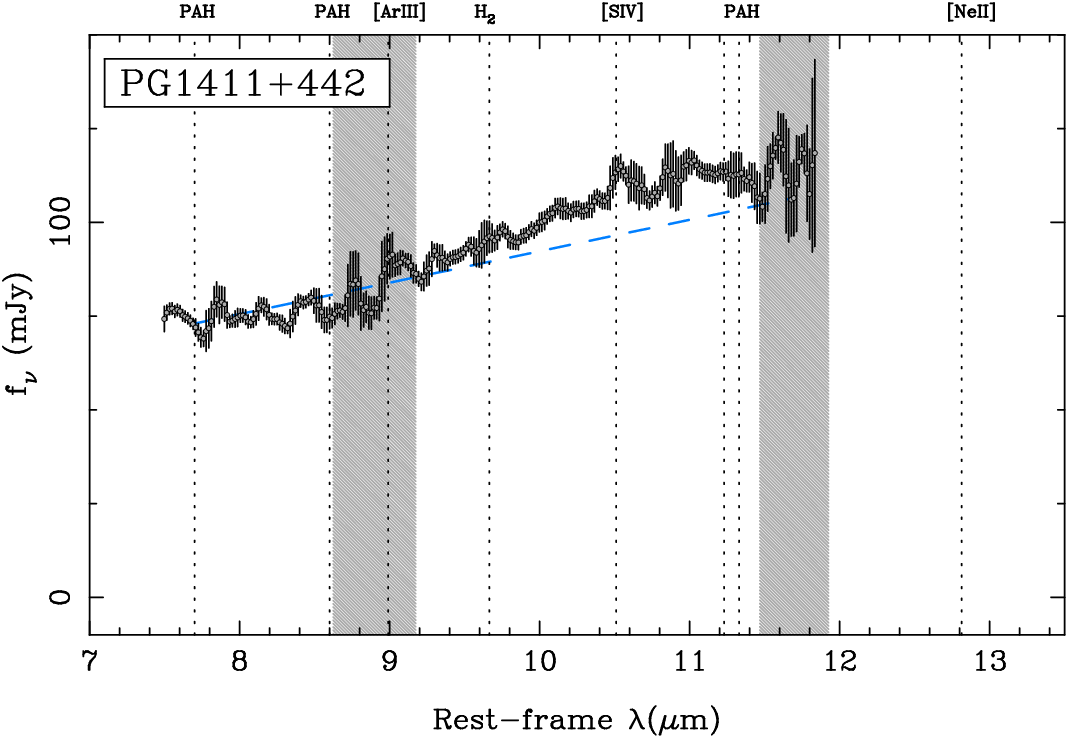}}}
\hspace{0.5cm}
\resizebox{0.35\hsize}{!}{\rotatebox[]{-90}{\includegraphics{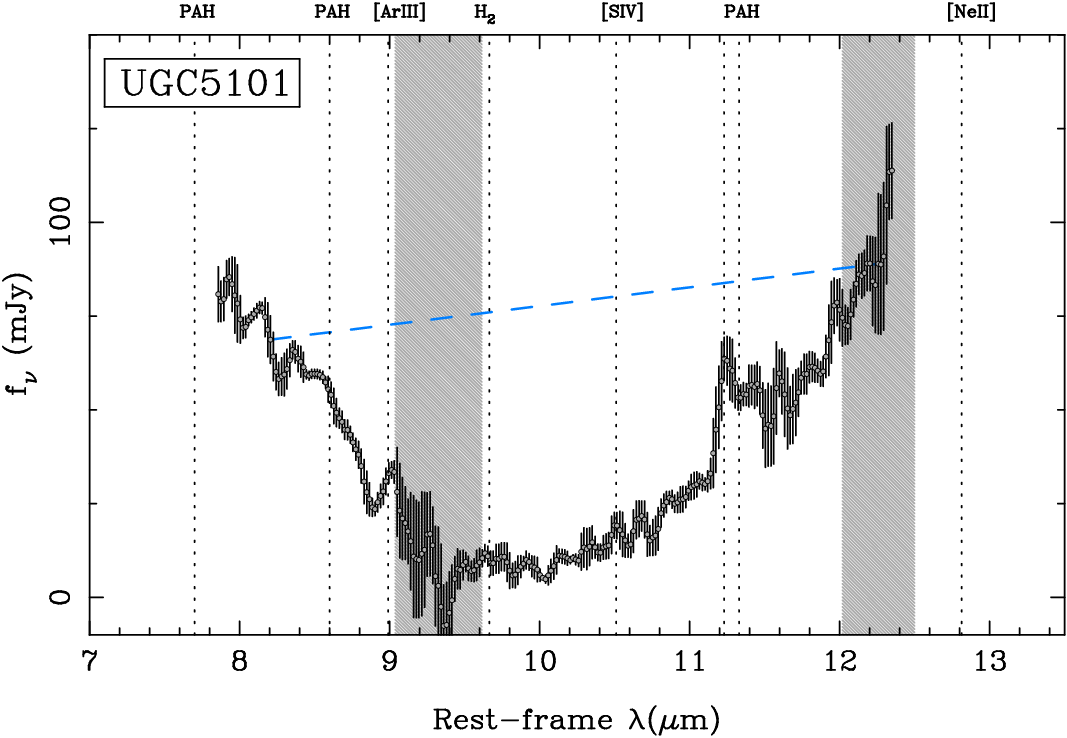}}}

\vspace{-0.9cm}
\caption{Continued.}
    \end{figure*}

Additionally, the typical errors associated with the photometric
calibration in the 
$N$-band window 
are usually estimated to be approximately $10-15\%$. For
our large data set we can compare the aperture photometry done on the
science images (i.e., those presented in Section~\ref{sec:imaging})
and on the shorter integration time acquisition
images taken for the spectroscopy. Since in the majority of cases the galaxy
images were taken on different nights, this comparison can give an
estimate of the uncertainties in the photometric calibration. To make
this comparison we chose the fluxes measured within a
2\,arcsec-diameter aperture to avoid the added uncertainty of the point source
correction for point-like sources while keeping the uncertainties due
to the background subtraction low. We present the comparison in
Fig.~\ref{fig:comparisonphot}.  In this figure we only show the
errors on the aperture photometry on the acquisition images
because the sky-subtracted image backgrounds have a higher standard
deviation due to the shorter integration times. 

As can be seen from
this figure, the agreement in the photometry is excellent. The most
discrepant point in Fig.~\ref{fig:comparisonphot} is Mrk~1014 for
which the flux density measured on the CanariCam acquisition image is
approximately twice that on the science image. Comparison with the
nuclear photometry of this source presented in \cite{Asmus2014}
indicates that the flux measured on the CanariCam science image is the correct
one. We cannot, however, rule out variability. The inset of  Fig.~\ref{fig:comparisonphot} shows the
distribution of the variation in the photometry for each galaxy,
computed as the difference between the two measurements divided by the
average of the two. The standard deviation of this distribution
(excluding Mrk~1014) is 11\%. We also marked in Fig.~~\ref{fig:comparisonphot} as
  star symbols those CanariCam science observations  
affected by pattern noise or had distorted PSFs (see
Table~\ref{table:imaging_log}, last column). We do not see a clear
trend for the most discrepant photometric points to be related to
these issues.

 A detailed comparison of our
  nuclear fluxes  with other works presenting 
photometry for large number of AGNs
\citep[e.g.,][]{Gorjian2004,Asmus2014} is not straightforward because of the
different $N$-band filters  and methods used to determine the nuclear
fluxes as well as possible variability in the mid-IR.

We measured the size of the nuclear regions before rotating and
smoothing the GTC/CanariCam $8.7\,\mu$m images
(Section~\ref{sec:imaging}) by fitting a moffat function to the
nuclear emission.  The measured
sizes (FWHM) of the nuclear mid-IR emission in arcseconds for the nuclear 
regions of the AGN are listed in Table~\ref{table:aperture_photometry}.  
For those galaxies with nuclear FWHMs similar to those of their
corresponding standard stars and no clear diffuse extended emission
we also estimated the unresolved nuclear fluxes. To do so, we used the
fluxes measured through the smallest aperture and estimated the
aperture correction for the total flux using the standard star
observations. We note that the unresolved nuclear fluxes estimated
using this method (see Table~\ref{table:aperture_photometry}) are likely
to be slightly overestimated as there is always a 
small fraction of resolved emission from the galaxy even in the
smallest aperture. We will present a detailed analysis of the
CanariCam unresolved emission of AGN using PSF-scaling photometry in
Garc\'{\i}a-Bernete et al. (2016, in preparation).

\subsection{Extraction and flux calibration of the
  spectra}\label{sec:spectraextraction}  
Before we extracted the 1D spectra we compared the measured nuclear
sizes of the galaxies with the FWHM 
of their corresponding standard stars. We used this comparison to
decide the type of extraction. The second column of
Table~\ref{table:spectral_measurements} specifies whether the spectrum was 
extracted as a point source or as an extended source. In the latter case we
give the extraction aperture in arcseconds. In the case of point
source extraction, {\sc redcan} uses an extraction aperture that
increases with wavelength to account for the decreasing diffraction-limited
angular
resolution and  performs an additional correction for
slit losses. For those galaxies with double nuclei detected in the
mid-IR, we only present in this work that corresponding to the
brightest nucleus.

For the nuclear spectra extracted as point sources we compared the
flux density $8.7\,\mu$m in the 1D flux-calibrated spectra 
with the unresolved fluxes estimated 
from the $8.7\,\mu$m CanariCam images (see
Table~\ref{table:aperture_photometry}). In most cases the $8.7\,\mu$m
fluxes agreed to within 25\%. When the discrepancy was large we scaled the
extracted 1D spectra to the imaging data unresolved fluxes estimated at
$8.7\,\mu$m (Section~\ref{sec:aperturephotometry}). In the
case of the extracted spectra for NGC~1614 along the two slit PA
(north-south, PA=0\,degrees and east-west, PA=90\,degrees),
we simulated the slit on the acquisition image to flux calibrate the
spectra. In the cases where there were more than one repetition or
spectroscopic observations taken on different nights we combined the
individual flux calibrated spectra.

\begin{table*}
 \centering
  \caption{Spectral measurements.}\label{table:spectral_measurements}  

  \begin{tabular}{lcccccccc}
 \hline
Galaxy & Extraction & size & $\lambda_1$ & $\lambda_2$ & $\alpha_{\rm MIR}$ & 
$S_{\rm Si}$ & $f_\nu (12\,\mu{\rm m})$ & EW(11.3$\mu$m PAH)\\
    &         & (pc) & ($\mu$m) & ($\mu$m) & & & (mJy) & ($\mu$m)\\ 
 \hline
3C273 & point &1376       & 7.0&  10.8& $-0.66  \pm0.12$& $ -0.02
\pm0.02$&  $317   \pm6$ & no\\
3C382 & point &  555 & 7.6&  11.9& $-1.48  \pm0.34$& $  0.15  \pm0.04$&  $123    \pm7$\\
3C390.3 & point &  541& 7.8&  11.8& $-0.45  \pm0.39$& $  0.12  \pm0.07$&  $130   \pm12$\\
IRAS~08572+3915 North & point& 570 & 7.7&  12.1& $ 0.66  \pm0.18$& $ -4.41  \pm0.85$&  $ 76   \pm 9$\\
IRAS~13197$-$1627  & point& 178& 8.0&  12.5& $-2.38  \pm0.07$& $ -0.38  \pm0.02$&  $859  \pm33$\\
IRAS~13349+2438 & point& 991 & 7.5&  11.6& $-1.34  \pm0.16$& $  0.11
\pm0.02$&  $585   \pm20$ & no\\
IRAS~14348$-$1447 South & point&786 & 7.5&  11.3& $ 2.98  \pm1.20$& $
-0.88  \pm0.67$&  $ 20   \pm12$ & no\\
IRAS~17208$-$0014 & 1\,arcsec& 809 & 7.6&  12.2& $ 1.39  \pm0.78$& $ -4.07   $      &  $ 43   \pm7$ & $0.56 \pm 0.07$ \\
Mrk~3 & point&   137  & 8.0&  12.3& $-2.67  \pm0.32$& $ -0.43  \pm0.06$&  $375  \pm30$\\
Mrk~231 & point&  420 & 7.8&  12.3& $-1.84  \pm0.11$& $ -0.84  \pm0.02$& $1592  \pm32$\\
Mrk~463 East & point& 499   & 7.7&  11.9& $-1.22  \pm0.15$& $ -0.53  \pm0.04$&  $460  \pm11$ & $<0.1$\\
Mrk~478 & point& 750 & 7.4&  11.7& $-1.25  \pm0.22$& $ -0.08  \pm0.02$&  $ 96   \pm 5$ & no\\
Mrk~841 & point&    367& 7.9&  12.1& $-2.28  \pm0.52$& $ -0.05  \pm0.03$&  $206   \pm 4$\\
Mrk~1014 & point& 1396 & 7.1&  10.6& $-2.78  \pm0.37$& $  0.17
\pm0.09$&  $ 92  \pm6$ & no\\
Mrk~1066 & point&  116& 8.0&  12.4& $-1.95  \pm0.20$& $ -1.01  \pm0.14$&  $185    \pm6$ & $0.34\pm 0.02$\\
Mrk~1073 & point&  230 & 8.1&  12.2& $-0.23  \pm0.28$& $ -0.76  \pm0.12$&  $ 91    \pm6$  & $0.08\pm0.02$\\
Mrk~1210 & point&  146 & 8.0&  12.4& $-2.62  \pm0.18$& $ -0.25  \pm0.03$&  $545   \pm13$\\
Mrk~1383 & point&  817 & 7.5&  11.5& $-1.87  \pm0.15$& $ -0.07
\pm0.02$&  $112    \pm3$ & no\\
NGC~931 & point&   161 & 7.9&  12.4& $-1.78  \pm0.25$& $  0.00  \pm0.03$&  $486   \pm 10$\\
NGC~1194 & point&  132& 7.9&  12.5& $-0.40  \pm0.19$& $ -0.94  \pm0.07$&  $196  \pm 4$\\
NGC~1275 & point& 173  & 8.0&  12.3& $-3.20  \pm0.18$& $  0.13  \pm0.03$& $1109   \pm22$\\
NGC~1320 & point&   85 & 8.2&  12.3& $-2.43  \pm0.39$& $ -0.18  \pm0.10$&  $393   \pm20$\\
NGC~1614 (PA=90\,deg)& 2\,arcsec& 131& 8.2&  12.3& $-2.63  \pm0.10$& $ -0.89  \pm0.10$&  $226    \pm9$ & $0.39\pm 0.02$ \\
NGC~1614 (PA=0\,deg) & 2\,arcsec& 131& 8.2&  12.3& $-2.69  \pm0.22$& $ -0.87  \pm0.06$&  $238    \pm5$ & $0.29 \pm 0.01$\\
NGC~2273 & point&  64& 8.0&  12.6& $-2.85  \pm0.16$& $ -0.39  \pm0.04$&  $315  \pm8$    & $0.033 \pm 0.005$\\
NGC~2992 & point&   90& 8.1&  12.7& $-2.75  \pm0.24$& $ -0.19  \pm0.10$&  $239    \pm8$\\
NGC~3227 & point&   51& 8.0&  12.6& $-2.12  \pm0.15$& $ -0.07  \pm0.04$&  $462    \pm9$   & $0.065 \pm 0.005$ \\
NGC~4051 & point&  32 & 8.0&  12.5& $-2.18  \pm0.37$& $  0.25  \pm0.09$&  $408   \pm13$   & $0.095 \pm 0.009$\\
NGC~4253 & point&   141& 7.8&  12.5& $-2.25  \pm0.44$& $ -0.23  \pm0.03$&  $342  \pm11$   & $0.060\pm 0.006$\\
NGC~4258 & point&  22 & 8.1&  12.6& $-1.96  \pm0.30$& $  0.16  \pm0.07$&  $150   \pm 5$\\
NGC~4388 & point&   43& 8.0&  12.5& $-2.47  \pm0.50$& $ -1.11  \pm0.20$&  $344   \pm19$   & \\
NGC~4419 & point&   43& 8.3&  12.5& $-3.52  \pm0.25$& $ -0.77  \pm0.14$&  $190   \pm 5$    & $0.21 \pm 0.01$ \\
NGC~4569 & 1\,arcsec&   82 & 8.1&  12.5& $-3.00  \pm0.87$& $  0.53  \pm0.07$&  $ 68   \pm 3$    & $0.30 \pm 0.02$\\
NGC~4579 & point&     43 & 8.1&  12.4& $-2.75  \pm0.61$& $  0.40  \pm0.07$&  $ 93   \pm 7$\\
NGC~5347 & point& 87 & 8.1&  12.5& $-2.94  \pm0.42$& $ -0.27  \pm0.04$&  $307   \pm 10$\\
NGC~5548 & point&  181 & 8.0&  12.5& $-1.36  \pm0.33$& $  0.13  \pm0.15$&  $220   \pm15$\\
NGC~5793 & 2\,arcsec& 126 & 8.1&  12.1& $-0.65  \pm1.06$& $ -0.39  $       &  $ 14   \pm 3$ & $\sim 1$ \\
NGC~6240 South & point& 247 & 8.2&  12.2& $-2.05  \pm0.21$& $ -1.49  \pm0.39$&  $297   \pm 7$ & $0.27 \pm 0.02$\\
NGC~7465  & point& 55 & 8.0&  12.4& $-1.38  \pm0.98$& $  0.04  \pm0.24$&  $ 74   \pm 4$  & $0.16 \pm 0.02$\\
OQ~208 & point&  731  & 7.7&  11.8& $-2.45  \pm0.27$& $  0.17
\pm0.02$&  $247   \pm 5$ & no\\
PG~0804+761 & point&  922 & 7.5&  11.6& $-1.46  \pm0.14$& $  0.19
\pm0.02$&  $203   \pm 4$ & no\\
PG~0844+349 & point&621 & 7.8&  12.0& $-2.00  \pm0.64$& $  0.28  \pm0.06$&  $ 57   \pm11$\\
PG~1211+143 & point& 772& 7.6&  11.8& $-1.29  \pm0.21$& $  0.12
\pm0.03$&  $171   \pm 8$ & no\\ 
PG~1229+204 & point& 615& 7.8&  11.8& $-1.36  \pm0.36$& $  0.11  \pm0.06$&  $ 68   \pm 9$\\
PG~1411+442 & point& 840& 7.7&  11.7& $-0.98  \pm0.23$& $  0.07
\pm0.03$&  $116   \pm 3$ & no\\
UGC~5101    & 1\,arcsec& 755& 8.2&  12.3& $-0.74  \pm0.55$& $ -1.88  \pm0.35$&  $ 46   \pm11$ & $0.21 \pm 0.03$\\

 \hline
\end{tabular}

{\it Notes}. The second column indicates the type of spectral
extraction, while the third column is the physical size of the slit
for unresolved sources or the
extraction aperture for extended sources. $\lambda_1$ and $\lambda_2$ are the rest-frame wavelengths
between which we fitted the continuum as well as the end points to
compute the mid-IR spectral index $\alpha_{\rm MIR}$. The nuclear
$12\,\mu$m flux densities are in observed frame and the quoted errors only include the dispersion
of the spectra (see Section~\ref{sec:spectraanalysis})
but not the additional $10\%$ photometric calibration
uncertainty. In the last column 'no' indicates those nuclei for
  which the rest-frame spectral range does not cover the $11.3\,\mu$m PAH
  feature or 
  the continuum next to the $11.3\,\mu$m PAH feature.
\end{table*}

\subsection{Analysis of the spectra}\label{sec:spectraanalysis}  
Figure~\ref{fig:Spectra} shows the fully reduced and flux calibrated 1D CanariCam spectra of the 45 AGN
in the ESO/GTC program. We smoothed the 1D CanariCam 
spectra by using a moving average of five spectral points. We computed the errors
as the standard deviation of the flux densities in these narrow wavelength bins. 
We note that this approach overestimates
the errors of the emission lines and PAH features.
We mark in this figure the most important emission lines and PAH features.

We measured the mid-IR spectral index in the 
CanariCam spectra, defined as $f_\nu \propto \nu^{\alpha_{\rm MIR}}$,
using the flux ratios at the approximate end points of the spectra. These wavelengths are
typically 8 and $12.5\,\mu$m (rest-frame) for the Seyferts and LLAGN, and
7.5 and $12\,\mu$m for the more distant AGN (see Table~\ref{table:spectral_measurements}).
For each galaxy   we chose  them visually  to avoid end wavelength points strongly
affected by low atmospheric  transmission.

The strength of the silicate feature is computed as $S_{\rm Si}= \ln
(f_{\rm cont}/f_{\rm feature}$), where $f_{\rm cont}$ is the continuum
at the wavelength of the feature and $f_{\rm feature}$ is the flux density
of the feature, which we evaluated at rest-frame $9.7\,\mu$m. The only exception was
NGC~4419 for which we evaluated them at $10\,\mu$m to avoid the bad ozone residual
in the spectrum at $9.7\,\mu$m rest-frame (see Fig.~\ref{fig:Spectra}). We 
fitted the continuum as a straight line between the same wavelengths
used to measure the spectral index.

In nuclei with strong PAH emission it is
difficult to estimate the continuum given the relatively narrow
spectral range attained from the ground. In those cases we placed the
blue end of the continuum in between the 7.7 and $8.6\,\mu$m PAH 
features (see e.g., NGC~6240 South or NGC~1614 in
Fig.~\ref{fig:Spectra}). Similarly, for AGN with deep
silicate absorption features it is difficult to measure the
underlying continuum and therefore the
resulting measured strength could be significantly underestimated.

To estimate the uncertainties in the measured spectral index we performed Montecarlo
simulations allowing the flux densities at  the end wavelengths to vary within their estimated errors. We then
computed the average and the standard deviation of these simulations
that are listed in Table~\ref{table:spectral_measurements}. For the strength of
the silicate feature we again performed Monte Carlo simulations allowing the flux of the
feature to vary within the measured errors to estimate the uncertainty in the measurement. 
For NGC~5793 and IRAS~17208$-$0014  we were not able to estimate
the uncertainty in the strength of the silicate feature
because the uncertainties of the flux density at  $9.7\,\mu$m allow for negative values.

Those AGN
in our sample  with the silicate feature in emission do not show the
peak at $9.7\,\mu$m but rather at longer wavelengths.
From Fig.~\ref{fig:Spectra}
we can see that the peak is usually at
$\lambda_{\rm rest}>10\,\mu$m, as also found by e.g. \cite{Thompson2009} and
\cite{Hatziminaoglou2015} from {\it Spitzer}/IRS spectroscopy.
However, we measure the strength of the silicate feature at
$9.7\,\mu$m rest-frame for the comparison with clumpy torus models in 
Section~\ref{sec:spectralindex}. We also measured the
strength of the  silicate feature in the Spitzer/IRS spectra
  and found that they were consistent within the uncertainties
  with our CanariCam values. The only exception
  was NGC~4051, for which the IRS value is consistent with zero
  or sligthly in emission \citep[see also][]{Wu2009}.

For those AGN with bright  $11.3\,\mu$m PAH feature emission, 
we measured the
equivalent width (EW) of the feature 
following the method described by \cite{HernanCaballero2011} and
\cite{Esquej2014}. Briefly, we fitted a local continuum at 
 $11.25\,\mu$m by interpolating
between two narrow bands on both sides of the feature
($10.75-11.0\,\mu$m and $11.65-11.9\,\mu$m). To obtain the EW we 
divided the flux of the feature integrated in the spectral 
range $11.05-11.55\,\mu$m by the fitted continuum and
corrected it for the missing flux \citep[see][for full
details]{HernanCaballero2011}. 
We estimated the uncertainties in the EW as the dispersion
around the measured fluxes and EWs in 100 Monte Carlo simulations of the
original spectrum with random noise distributed as the rms of the
CanariCam spectrum. We note that PAH fluxes, and thus EW, obtained with a
local continuum are lower (typically by a factor of two) than
those using a continuum fitted over a broad spectral range
\citep[e.g. with {\sc pahfit, see}][]{Smith2007}.

In  Table~\ref{table:spectral_measurements} we list for each galaxy
the rest-frame wavelengths used to compute the spectral index, the
spectral index and the strength of the silicate feature with the corresponding
uncertainties, the observed nuclear $12\,\mu$m flux densities from the
spectra and for those
nuclei with clear $11.3\,\mu$m PAH emission, the EW of the feature. 
We also list in this table the physical
sizes probed by the slit width, that is, the physical scale
corresponding to 
0.52\,arcsec for point source spectra, and the optimized extraction aperture
for extended sources. The median physical sizes probed by the
CanariCam spectroscopy are approximately 90\,pc for the Seyferts and LLAGN
and 640\,pc for the rest of the sample.

Finally, we estimated the nuclear rest-frame $12\,\mu$m
luminosities of our sample of AGN from the CanariCam 1D flux-calibrated 
spectra. For the most distant sources
  in our sample, these 
  were extrapolated from the fitted continuum to the observed
  CanariCam spectra (see above).

\section{Nuclear Mid-IR spectroscopic properties of local AGN}\label{sec:results}

\begin{figure}

\hspace{0.2cm}
\resizebox{0.92\hsize}{!}{\rotatebox[]{-90}{\includegraphics{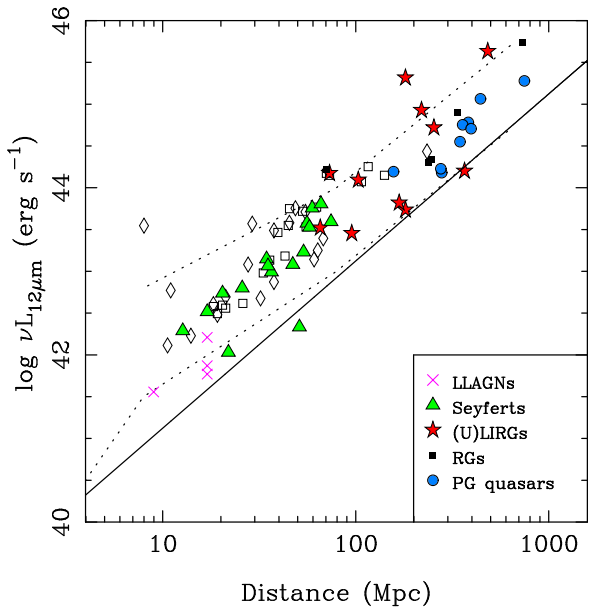}}}

\caption{Nuclear rest-frame monochromatic $12\,\mu$m
  luminosity against the luminosity distance for the 45 AGN in the
  ESO/GTC large program. The color symbols indicate the different AGN
  classes in our sample. The diamonds and squares are AGN (mostly Seyferts and
  LIRGs)  with Gemini/T-ReCS spectroscopy from
  Gonz\'alez-Mart\'{\i}n et al. (2013) 
and VLT/VISIR spectroscopy from H\"onig et al. (2010),
respectively. The dotted 
lines delineate 
  approximately the  location of the AGN in the mid-IR imaging atlas  of
  Asmus et al. (2014). The solid line shows a  constant flux density of
  45\,mJy at $12\,\mu$m (see Section~\ref{sec:12micronemission}). } 
\label{fig:lum12vsdistance}
    \end{figure}

\subsection{Nuclear $12\,\mu$m emission}\label{sec:12micronemission}
The nuclear mid-IR continuum of AGN and in particular the monochromatic
$12\,\mu$m nuclear emission are believed to be a good proxy for the AGN
luminosity. This is based on the good correlation found between the
AGN hard X-ray (absorption corrected) luminosity and the $12\,\mu$m
emission over four orders of magnitude 
\citep{Levenson2009,Gandhi2009,Asmus2011,Mason2012}. Moreover, these works found
that the $12\,\mu$m nuclear
emission of type 1 and type 2 AGN is not significantly
different \citep[although see][]{Yan2015,Burtscher2015}. This has been explained in the context
of the mild anisotropy of the mid-IR emission vs. e.g. the viewing angle 
predicted by clumpy torus models and smooth dusty torus models
\citep[see
e.g.][]{Horst2008, Nenkova2008,Levenson2009,Hoenig2010models,Feltre2012,Yan2015}.

It is interesting to compare the nuclear $12\,\mu$m luminosities of
our sample with the  sub-arcsecond mid-IR imaging compilation of 253
AGN presented by \cite{Asmus2014}.  This way we can determine if our
spectroscopic sample is representative of the local population of
mid-IR {\it bright} AGN. In Fig.~\ref{fig:lum12vsdistance} we plot
the nuclear 
rest-frame monochromatic $12\,\mu$m luminosity against the  distance for the 45 AGN in the
  ESO/GTC large program. This figure can be
compared directly with figure~13 of \cite{Asmus2014},  
although  we note that they used a slightly different cosmology
($H_0=63\,{\rm km \,
  s}^{-1}\,{\rm Mpc}^{-1}$). We show in our figure the approximate location of 
the Asmus et al. (2014) AGN with unresolved $12\,\mu$m emission
as dotted lines. We also show the line of constant $12\,\mu$m flux density
of 45\,mJy. This value is expected from the approximate flux limit
of our sample of 25\,mJy at $8.7\,\mu$m for spectroscopy  and the median spectral 
index of the sample $\alpha_{\rm MIR}=-2$ (see Table~\ref{table:spectral_measurements}). 
The comparison with the imaging compilation of \cite{Asmus2014}
demonstrates that, when compared to the imaging atlas, our spectroscopic
atlas at a given distance is not biased towards the most luminous mid-IR AGN.

The nuclear $12\,\mu$m luminosities of our
  sample cover over  four orders of
magnitude. According to the mid-IR vs. X-ray correlation, our sample
would probe four orders of magnitude of AGN luminosity, as also shown
in Fig.~\ref{fig:sample}.  The PG
quasars, (U)LIRGs and RGs have a median $\log (\nu L_{12\mu{\rm m}}/{\rm
  erg \,s}^{-1})=44.3$, whereas the Seyferts and 
LLAGN have $\log (\nu L_{12\mu{\rm m}}/{\rm erg \,s}^{-1})=43$. We
note that two LLAGN in our sample (NGC~4419   
and NGC~4569)  show bright PAH emission in their nuclear CanariCam spectra
(see Fig.~\ref{fig:Spectra} and also Section~\ref{sec:PAH}). Thus
their $12\,\mu$m luminosities might be
dominated by emission related to star formation activity as found in a few LLAGN \citep[see
e.g.][]{Mason2012,Asmus2014,GonzalezMartin2015}.   

We also plot in Fig.~\ref{fig:lum12vsdistance} a sample of AGN (mostly
Seyferts and LIRGs) with sub-arcsecond resolution Gemini/T-ReCS
spectroscopy taken from the work of
\cite{GonzalezMartin2013} and  VLT/VISIR spectroscopy from 
\cite{Hoenig2010}. The Seyfert galaxies observed in our
ESO/GTC large program occupy a similar region in this diagram as the
previously published sub-arcsecond mid-IR spectroscopy. Notably, however, our ESO/GTC
large program also extends the existing sub-arcsecond mid-IR
spectroscopy to more  
luminous and distant AGN, namely, ULIRGs \citep[see
also][]{Soifer2002}, PG quasars, and RGs.

\subsection{Mid-IR spectral index and the $9.7\,\mu$m silicate feature}\label{sec:spectralindex}
In this section we discuss two quantities that provide a simple 
description of the ground-based 
mid-IR spectrum of an AGN, namely the spectral index and the
strength of the silicate feature (see Section~\ref{sec:spectraanalysis}).

\begin{figure*}

\hspace{0.2cm}
\resizebox{0.4\hsize}{!}{\rotatebox[]{-90}{\includegraphics{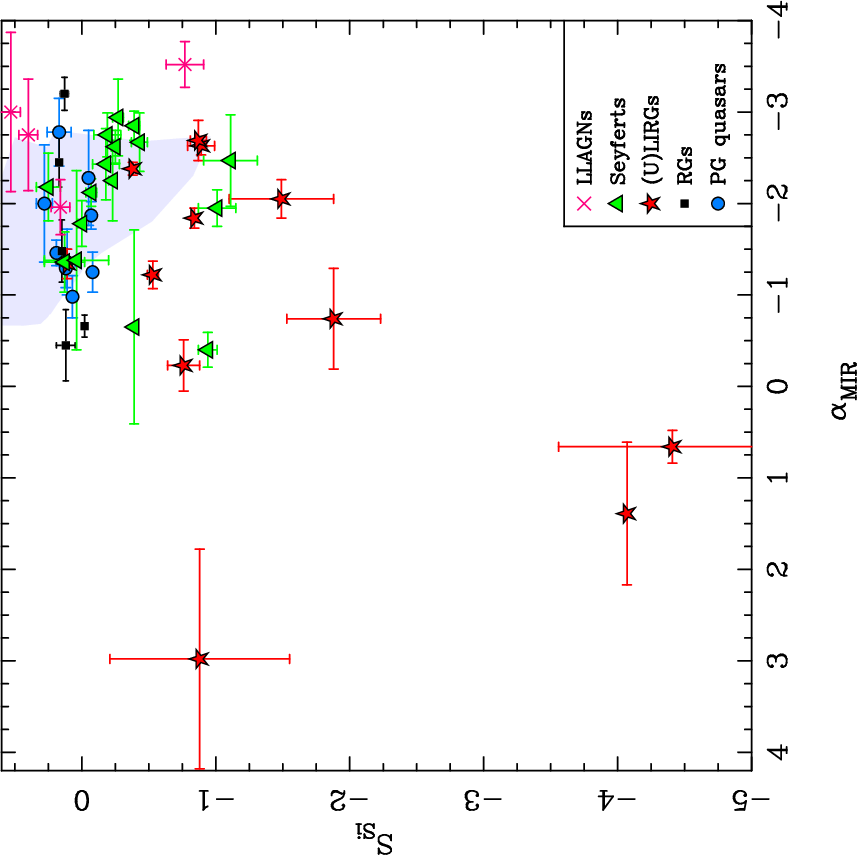}}}
\resizebox{0.4\hsize}{!}{\rotatebox[]{-90}{\includegraphics{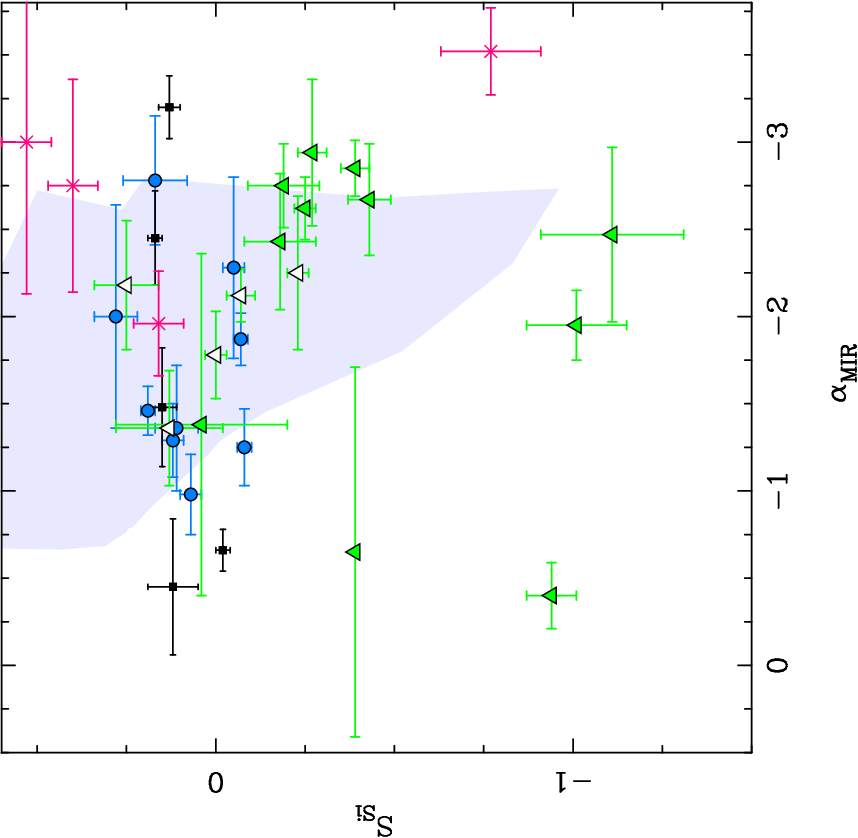}}}

\caption{{\it Left panel}. Nuclear $\alpha_{\rm MIR}$  against the strength of the silicate
  feature where positive numbers for $S_{\rm Si}$ indicate that the
  feature is in emission and negative numbers in absorption. The
  shaded area shows the approximate
  region covered in this diagram by a set of clumpy
  torus models shown in figure~12 of H\"onig et al. (2010).
  {\it Right panel}. Same as left but zooming to the region covered by the torus models.
  Seyfert 1s and 1.5s are plotted as white triangles
and Seyfert 1.8s, 1.9s, and 2s as green triangles. We do not show the U(LIRG) nuclei.}
\label{fig:alphavssilicatefeature}
    \end{figure*}

In the cases of relatively
featureless spectra the spectral index gives an estimate of the
shape of the AGN mid-IR continuum, which can be compared with clumpy
torus model predictions \citep[see below and
also][]{RamosAlmeida2014torus,Hoenig2010}.  
The mid-IR spectral indices of the Seyfert galaxies, PG quasars, and
RGs in our sample range  
between   $\alpha_{\rm MIR}=-0.23$ and
$\alpha_{\rm MIR}=-3.20$ (see Table~\ref{table:spectral_measurements} and
Fig.~\ref{fig:alphavssilicatefeature}). They are similar to those measured for 
other Seyfert nuclei using 
sub-arcsecond 8 to $18\,\mu$m imaging 
data \citep{RamosAlmeida2011} and  VLT/VISIR spectroscopy
\citep[$\alpha_{\rm MIR}=-1.65\pm0.44$ and $\alpha_{\rm
  MIR}=-2.07\pm0.54$ for type 1s and 2s, respectively,
see][]{Hoenig2010}.  \cite{Buchanan2006}, on the other hand, 
found from {\it Spitzer}/IRS spectroscopy (slit widths between 3.6 and
10.7\,arcsec) that only 
30\% of their $12\,\mu$m selected sample showed a clear power-law
component with spectral indices (measured between 5 and
$20\,\mu$m) -0.9 to -2.3. This small fraction is due to an important
contribution of  star formation and/or diffuse dust emission within the IRS slits. 

Four (U)LIRG nuclei  in our sample show
considerably flatter mid-IR 
spectral indices ($\alpha_{\rm MIR}>-0.5$) than the typical values of
other AGN. This is in part 
due to the presence of  strong PAH emission at 7.7, 8.6, and
$11.3\,\mu$m, which makes it 
difficult to get the intrinsic shape of the AGN continuum. Also in
cases of nuclei with deep silicate feature 
absorption (e.g., IRAS~08572+3915 North,
IRAS~17208$-$0014, UGC~5101), the
dust heating source appears to be so embedded that determining the intrinsic
mid-IR spectral shape might not be  possible unless we
perform a spectral decomposition and modeling of the AGN component
\citep[see e.g.][]{MartinezParedes2015}.

In terms of the silicate feature we can see from
Fig.~\ref{fig:Spectra} that most PG quasars and RGs
as well as three LLAGNs, the ULIRG/Sy1 IRAS~13349+2438  and some Sy1
nuclei  show the feature clearly in emission. This is similar to 
results obtained with {\it Spitzer}/IRS spectroscopy \citep[see
e.g.][]{Shi2006,Mason2012}. On 
sub-arcsecond scales most Seyfert nuclei show the feature in
moderate absorption or slight emission \citep[see also
  e.g.][]{Mason2006,Mason2009,Hoenig2010,GonzalezMartin2013,AAH2011,AAH2014}.
Nuclei in our sample with deep silicate features ($S_{\rm Si}<<-1$) are  
local (U)LIRG indicating that the AGN are likely embedded
and obscured by dust not directly
associated with the AGN
\citep{Levenson2007,AAH2011,AAH2013,GonzalezMartin2013}.

\cite{RamosAlmeida2014torus} showed that the 
nuclear mid-IR spectra of AGN can be used to constrain the number
of clouds and optical depth of the clouds of the \cite{Nenkova2008}
clumpy torus models.   \cite{Hoenig2010} used the mid-IR spectral
index $\alpha_{\rm MIR}$ together with the strength of the silicate feature to
also constrain the properties of the \cite{Hoenig2010models} torus models.
After fixing the width of the torus and the viewing angle,
\cite{Hoenig2010} used this diagram to constrain the 
number of clouds and radial distribution of the clouds in their models.
They  demonstrated however that the mid-IR nuclear
spectra do not provide information of the viewing angle of the
torus, as also verified by 
\cite{RamosAlmeida2014torus}.

Fig.~\ref{fig:alphavssilicatefeature} can be compared directly with
figure~12 of \cite{Hoenig2010} and in the right panel we zoom into the
region covered by a subset of the clumpy models of  
\cite{Hoenig2010models}. For a range of mid-IR spectral
indices between approximately $\alpha_{\rm MIR}=-0.6$ and $\alpha_{\rm
MIR}=-2.5$
their models predict silicate features in emission or relatively
shallow with values of the silicate strength  approximately $S_{\rm
  Si}=-1$ to $S_{\rm Si} =0.5$. 
This behaviour implies that the observed nuclear mid-IR spectra of
most of the (U)LIRG nuclei in our sample cannot be reproduced simply
with clumpy torus models. It has been showed that they require
foreground dust screen or a deeply embedded source
\citep{Levenson2007,AAH2013,Mori2014,MartinezParedes2015}. 
 The mid-IR spectrum of two (U)LIRG nuclei (IRAS~13197$-$1627  and IRAS~13349+2438)
 would be in principle reproduced by
 clumpy torus models as they show a relatively featureless continuum
 with no deep silicate feature.  

 In Fig.~\ref{fig:alphavssilicatefeature} (right  panel) we note that
 the PG quasars and the RGs occupy a narrow stripe around $S_{\rm Si}\sim 0$
 with $\alpha_{\rm MIR}$ ranging between -0.5 and -3, and would be in principle
 well reproduced by clumpy torus models. However, for the RGs there might be
 a non negligible contribution from synchrotron emission to the mid-IR
 spectrum. For PG quasars it has been shown that the full mid-IR spectra
with the silicate feature in emission cannot always be fully reproduced with the 
Nenkova et al. (2008) clumpy torus models alone and need extra components
\citep[see][and Mart\'{\i}nez-Paredes et al. 2016 in preparation]{Mor2009}.
\cite{Mason2013}
found a similar result for the LLAGN NGC~3998 and showed that an optically
thin dust model reproduced better the overall infrared emission of this
AGN while producing the silicate feature in emission.

The nuclear mid-IR spectral index and silicate features of
most Seyfert nuclei also fall 
in the region covered by the \cite{Hoenig2010models} clumpy
torus models.  There is a tendency for
the type 1s to have on average flatter $\alpha_{\rm MIR}$ than type 2s, as already shown
by \cite{Hoenig2010}. On the other hand, the Seyfert nuclei showing deeper silicate
feature than predicted by the models are type 2s and suggest the
presence of extended dust components not related to the torus
\citep[see also][]{AAH2011,GonzalezMartin2013}. Although this
subset of clumpy torus
models plotted in this figure
does not produce spectral indices as steep as $\alpha_{\rm MIR}=-3$,
it is likely due to the reduced parameter space (e.g., fixed
angular width of the torus, optical depth of the clouds, and range of
viewing angles) plotted in their
figure. Indeed, figure~8 of \cite{Hoenig2010models} shows that
steep mid-IR indices can be produced for instance for
viewing angles $i=90\,$deg
(equatorial view). Also, the steep mid-IR continuum of NGC~2992 ($\alpha_{\rm 
  MIR}=-2.8$) is reproduced by the 
Nenkova et al. (2008) clumpy torus models \citep {GarciaBernete2015}. 
In a series of future papers we will present detailed comparisons between the observed spectra of
  Seyfert nuclei and PG quasars and predictions from clumpy torus models.

 The silicate features seen in emission are rather muted and appear to peak at wavelengths longer than $10\,\mu$m, as found in Spitzer/IRS spectra of type 1 Seyferts and QSOs \citep{Hao2005}.  The spectra shown here have similar profiles, which may arise through optical depth effects in radiative transfer \citep{Nikutta2009} or increased mean silicate grain sizes or from additional contributions from crystalline silicates such as enstatites which peak at longer wavelengths than the amorphous silicates typically seen in the ISM.  \cite{Spoon2006} have  argued that many ULIRGs contain significant levels of crystalline silicates in their absorption spectra and \cite{Kemper2011}
  suggest that the crystalline grains may be produced in the circumnuclear environment of the AGN.

\subsection{Emission from the $11.3\,\mu$m PAH feature}\label{sec:PAH}
The presence of PAH features in the nuclear regions of AGN can be used
to trace the nuclear star formation activity. In particular  PAH
emission appears to be well suited to probe recent star formation as
they trace  the emission of both  B stars and O stars
\citep{Peeters2004,DiazSantos2010}. Although many AGN show $11.3\,\mu$m PAH
emission in their mid-IR spectra, these features appear weaker than those observed in star-forming galaxies
\citep{Roche1991}. However, \cite{Esquej2014} did not find
evidence of strong suppression of the $11.3\,\mu$m PAH emission in the
nuclear regions of Seyferts in the
Revised Shapley-Ames (RSA) catalog. The low EWs of
the $11.3\,\mu$m PAH feature in AGN have been interpreted as due to 
dilution by the presence of a strong AGN continuum rather than PAH
destruction \citep{AAH2014,RamosAlmeida2014}. 

Seven Seyferts (Mrk~1066, NGC~2273,
NGC~3227, NGC~4051, NGC~4253,  NGC~5793, and NGC~7465) in the
ESO/GTC sample show
clear $11.3\,\mu$m PAH emission in their nuclear spectra (see
Fig.~\ref{fig:Spectra}), although as found in other Seyfert nuclei 
 the EW of the 
feature is lower than in star forming galaxies
\citep{GonzalezMartin2013,Esquej2014,AAH2014}. The nearly 50\% detection 
rate for the Seyferts in our sample is similar to that of Seyferts in
the RSA sample 
\citep{Esquej2014}. This again suggests that at least the carriers of the $11.3\,\mu$m PAH
feature do not get completely destroyed in the vicinity of Seyfert-like AGN at
typical distances from the AGN of 45\,pc in our sample and as close as 10\,pc,
as shown by \cite{Esquej2014}.

Two of the LLAGN in our sample (NGC~4569 and NGC~4419) show bright nuclear
PAH emission, whereas the other two (NGC~4579 and NGC~4258) do
not and present the silicate feature in emission. At least for
NGC~4569 the detection of bright $11.3\,\mu$m PAH emission might be
associated with the presence of massive young stars in the nuclear
region \citep{Maoz1998}. Although this is a
small sample of LLAGN, this variety of nuclear mid-IR spectral shapes
agrees with the findings of \cite{Mason2012} from sub-arcsecond
resolution mid-IR 
imaging data. They concluded that the  nuclear mid-IR emission of LLAGN
can be produced by several mechanisms (dust heated by AGN, dust heated
by star formation, synchrotron emission) and might depend on the
Eddington ratio of the AGN and the 
radio loudness. In any case, for the two LLAGN with PAH emission we can
conclude that they have undergone star formation activity in the 
recent past and that the molecules responsible for the PAH features
can survive at distances as close to the 
nucleus as 20\,pc. The latter is 
well understood in the context of the survival of the PAH  molecules
at lower AGN luminosities even if they are closer to the AGN
\citep[see details in ][]{Esquej2014,AAH2014}.

\begin{figure}

\hspace{0.2cm}
\resizebox{0.95\hsize}{!}{\rotatebox[]{-90}{\includegraphics{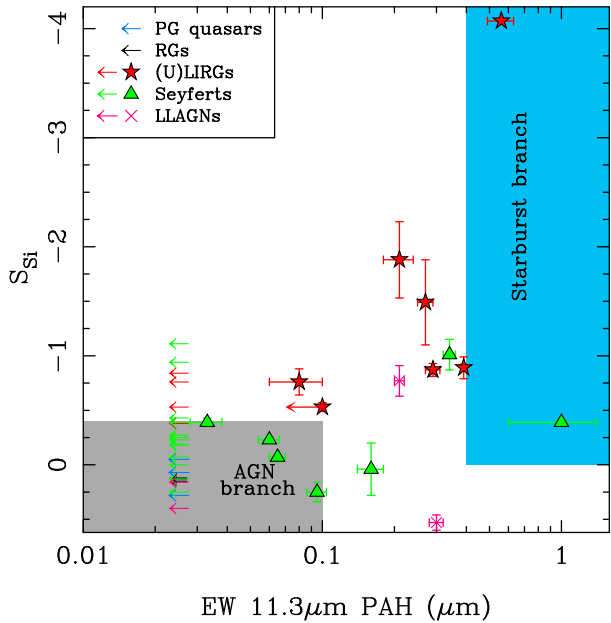}}}

\caption{Nuclear 
EW of the $11.3\,\mu$m PAH feature against the strength of the silicate
  feature for the ESO/GTC large program AGN. For those
  AGN without a clear detection of the $11.3\,\mu$m PAH feature, the
  EW is plotted as an upper limit at the level of $\sim 0.03\,\mu$m. We
  plot the approximate location of the AGN and the starburst branches
  based on Hern\'an-Caballero \& Hatziminaoglou (2011).}
\label{fig:EWvssilicatefeature}
    \end{figure}

The nuclei of several (U)LIRGs in our sample:
UGC~5101 \citep{MartinezParedes2015},
NGC~6240 South, Mrk~1073, IRAS~17208$-$0014 \citep{AAH2014},  and NGC~1614
\citep{PereiraSantaella2015} show clear
nuclear $11.3\,\mu$m PAH emission, whereas 
Mrk~463 East has a tentative detection. This means that in these (U)LIRGs
the AGN does
not completely dominate the nuclear (typical physical scales
between 50 and 800\,pc) mid-IR emission  \citep[see
also e.g.][]{Soifer2002,Mori2014,AAH2014,MartinezParedes2015}. This is
in contrast 
with the nuclei of the PG quasars and RGs, which have similar $12\,\mu$m
nuclear luminosities (see Fig.~\ref{fig:lum12vsdistance}),
but no bright  nuclear PAH emission, at least in
terms of the EW of the feature. 

As can be seen from Table~\ref{table:spectral_measurements}, the EW of
the $11.3\,\mu$m PAH 
feature of the AGN in our sample tend to be smaller than those
measured from {\it Spitzer}/IRS spectra
\citep[see][]{HernanCaballero2011}
for the same type of object. This is well known 
\citep[see figure~11 of][]{AAH2014} and is understood in
terms of an increased AGN contribution relative to that of star
formation within the CanariCam (and
 other mid-IR instruments on 8-10\,m-class telescopes) slit
when compared to the almost 10 times wider IRS slit. It is only when
there is a strong nuclear and highly concentrated starburst that the  
EW of the $11.3\,\mu$m PAH measured from CanariCam spectra would be
similar to that of the IRS spectra.

In Fig.~\ref{fig:EWvssilicatefeature} we compare the EW of the
$11.3\,\mu$m PAH feature with the strength of the silicate
feature for the nuclear emission of our sample of AGN. These
diagrams have been used in the literature to 
distinguish between AGN-dominated and starburst-dominated sources. 
As discussed by \cite{HernanCaballero2011}, using the $11.3\,\mu$m PAH
feature instead of the $6.2\,\mu$m PAH feature has a number of
advantages. First it avoids the water ice absorption feature in the wings
of the $6.2\,\mu$m PAH feature  in deeply obscured
sources. Second, the spectral range required is narrower and in our
case is accessible with ground-based
telescopes for nearby AGN. In this diagram, the {\it Spitzer}/IRS starburst
branch is nearly vertical with values of the EW typically between 0.4
and $1\,\mu$m. The AGN-dominated sources lie in a
horizontal branch with EW typically of less than $0.1\,\mu$m, and the
silicate feature in slight emission or absorption. 

The nuclear mid-IR emission of  PG quasars, RG and 
most of the Seyfert nuclei in our sample appears to be dominated by
AGN emission, based on Fig.~\ref{fig:EWvssilicatefeature}.  On the
physical scales probed by the CanariCam spectra 
only the nuclear mid-IR emission of the Seyfert NGC~5793 and the
ULIRG IRAS~17208$-$0014 would appear to be
dominated by star 
formation. Most of the (U)LIRG nuclei in our sample \citep[see
also][]{Soifer2002} and a few Seyfert nuclei appear
to be composite sources, that is, 
these sources  are examples with different degrees of
contribution from the AGN and nuclear star formation  to the observed nuclear
emission. 


\subsection{Fine structure lines}\label{sec:finestructure}
The most conspicuous mid-IR fine structure lines that can be observed
from the ground are [Ne\,{\sc ii}] at $12.81\,\mu$m and [S\,{\sc iv}]
at $10.51\,\mu$m.  The [Ne\,{\sc ii}]$12.81\,\mu$m  line has a
relatively low excitation potential and is mostly related to star
formation activity in AGN \citep[see e.g.][]{Roche1991,
  PereiraSantaella2010}. Unfortunately, [Ne\,{\sc ii}]$12.81\,\mu$m 
 falls very close to the edge of the CanariCam spectra and it
is only seen or party seen in some of the nearest galaxies in our
sample (e.g., NGC~2273, NGC~2992, NGC~4419, NGC~4569, see
Fig.~\ref{fig:Spectra}). 

The [S\,{\sc iv}]$10.51\,\mu$m fine structure line has an intermediate excitation potential and
therefore it has been observed both in AGN and star forming galaxies. 
\cite{Dasyra2011} detected this line in half of a large sample of AGN
using {\it Spitzer}/IRS spectra and showed that it is a good tracer of
the narrow line region emission in AGN. However, this line has also
been observed
in H\,{\sc ii} regions, for instance in M101 \citep{Gordon2008} and star
forming regions in local LIRGs \citep{PereiraSantaella2010mapping}. 

As can be seen from Fig.~\ref{fig:Spectra}, in general the [S\,{\sc
  iv}]$10.51\,\mu$m  line is not bright in the
AGN in our sample and is only clearly detected in approximately
one-third of them. These are mostly Seyfert nuclei both
type 1 and 2 \citep[for detections of this line at sub-arcsecond
  resolution see also][]{Hoenig2008,DiazSantos2010,GonzalezMartin2013}
and also in some (U)LIRG nuclei such as IRAS~13197$-$1627,
NGC~1614 and Mrk~463 East and
the PG quasar
Mrk~841. Being a faint line with the added complication that it is inside
the $9.7\,\mu$m broad silicate, it is likely that the non-detection in some
CanariCam spectra is due to the limited signal-to-noise ratios in that
part of the spectra.

\section{Summary}\label{sec:summary}
We have presented mid-IR Si-2 ($\lambda_{\rm c}=8.7\,\mu$m) imaging
and $7.5-13\,\mu$m 
spectroscopy ($R=\lambda / \Delta
\lambda \sim  175$)  of a sample of 45 local AGN obtained with
GTC/CanariCam through an ESO/GTC large program (ID 182.B-2005, PI
Alonso-Herrero). The sample
includes 4 LLAGN and 16 Seyfert nuclei at a median distance of 35\,Mpc. It
also contains 5 RG, 11 (U)LIRGs, and 9 PG quasars at a median distance
of 254\,Mpc. The ESO/GTC large program observations together with CanariCam
GT observations are part of a mid-IR survey of 
local AGN aimed at the study of the obscuring material around active
nuclei. The ESO/GTC large program imaging and spectroscopic
observations were obtained under sub-arcsecond resolution conditions
with a median value of 0.3\,arcsec (FWHM).

The goal of this work was to present a brief overview of the mid-IR
spectroscopic properties of the nuclear regions of the AGN in the
ESO/GTC sample. The nuclear $12\,\mu$m luminosities of these
AGN  cover more than four orders of magnitude
$\nu L_{12\mu{\rm m}}\sim 3\times 10^{41}-10^{46}\,{\rm 
  erg \,s}^{-1}$. We summarize 
our main results as follows,

\begin{enumerate}
\item We demonstrated that in terms of the nuclear $12\,\mu$m 
  luminosity the sample is representative of the local population of mid-IR
  emitting AGN, based on the comparison with the \cite{Asmus2014} imaging
  atlas. The $12\,\mu$m luminosities of
  LLAGN and Seyfert nuclei are similar to those of other
mid-IR  spectroscopic observations obtained on sub-arcsecond
resolution. The RGs, PG quasars and (U)LIRG nuclei in our ESO/GTC
large program  expand the existing sub-arcsecond resolution spectroscopy 
to more luminous and distant AGN. 
The CanariCam 0.52\,arcsec-width slit probes typical physical regions of 93\,pc
in the LLAGN and Seyferts, and 640\,pc in the rest of the sample.

\item We measured mid-IR spectral indices for the Seyfert nuclei, RG,
  LLAGN, 
  and PG quasars in the range $\alpha_{\rm MIR}=-0.2$ to $\alpha_{\rm
    MIR}=-3$, which are similar to those of other Seyfert nuclei
  observed on sub-arcsecond resolution. Some (U)LIRG nuclei show
  flatter mid-IR indices, which are in part due to the presence of PAH emission
  and/or deep silicate absorption. 

\item We found that on sub-arcsecond scales most PG quasars and RGs as well as
three LLAGNs, and one ULIRG/Sy1 nuclei show the silicate feature in
emission. The majority of  
Seyfert nuclei show the feature in moderate absorption
or slight emission.  Most of the nuclei in our sample
with deep silicate features ($S_{\rm Si} << - 1$) are local LIRG and
ULIRG and suggests that their AGN are likely embedded
and obscured by dust not directly associated with the AGN.

\item We used a simple diagram comparing 
  the spectral index $\alpha_{\rm MIR}$ and the 
strength of the silicate feature $S_{\rm Si}$ to show that
  most PG quasars, RGs, LLAGN, and Seyfert nuclei lie in the region
  covered by the clumpy torus models of \cite{Hoenig2010models}.  However,
  some RGs and PG quasars might require extra components
  to explain their mid-IR emission.

\item We detected clear $11.3\,\mu$m PAH emission in the nuclear
  regions of 40\% of the AGN in
    our sample where the spectral range allowed a measurement, namely  7
  Seyferts, 5 (U)LIRGs, and
  2 LLAGN. Based on an  EW($11.3\,\mu$m PAH) vs. $S_{\rm Si}$ diagram,  the
mid-IR emission of PG quasars, RG and half of the Seyfert
nuclei in our sample appears to be dominated by AGN emission, whereas only 
one ULIRG and one Seyfert nucleus would be mostly powered 
 by star
formation. Most (U)LIRG nuclei and a few Seyfert
nuclei in our sample appear to be composite sources.
\end{enumerate}

\section*{Acknowledgments}

We thank an anonymous referee for suggestions that helped improve
  the paper. We are extremely grateful to the GTC staff for their
constant and enthusiastic support to obtain the ESO/GTC large
program observations presented in this paper.  We thank Rhys
  Poulton for help with some of the spectroscopic analysis  presented
  in this paper.

A.A.-H. acknowledges financial support from the Spanish Ministry of Economy and Competitiveness (MINECO) under the 2011 Severo Ochoa Program MINECO SEV-2011-0187. A.A.-H. and
A.H.-C. acknowledge financial support from the Spanish Ministry of Economy and 
Competitiveness through grant AYA2012-31447, which is party funded by the  
FEDER program, P.E. from grant AYA2012-31277, and L.C. from grant
AYA2012-32295.  C.R.A. acknowledges
financial support from the Marie Curie Intra European Fellowship
within the 7th European Community Framework Programme
(PIEF-GA-2012-327934). N.A.L. and R.E.M. are supported by the Gemini
Observatory, which is 
operated by the Association of Universities
for Research in Astronomy, Inc., on behalf of the international Gemini
partnership of
Argentina, Australia, Brazil, Canada, Chile, and the United States of
America. C.P. acknowledges support from UTSA to help enable this research.

Based on observations made with the GTC,
installed in the Spanish Observatorio del Roque de los Muchachos of the
Instituto de Astrof\'{\i}sica de Canarias, in the island of La Palma.
This research has made use of the NASA/IPAC Extragalactic Database
(NED) which is operated by JPL, Caltech, 
under contract with the National Aeronautics
and Space Administration.

\label{lastpage}

\end{document}